\documentclass[pra,reprint,superscriptaddress,amsmath,amssymb,aps,floatfix]{revtex4-1}
\usepackage[utf8]{inputenc}
\usepackage{hyperref}
\usepackage{graphicx}
\usepackage{comment}
\usepackage{braket}

\usepackage{color}
\usepackage[normalem]{ulem} 	
\definecolor{myblue}{rgb}{0.4, 0.3, 0.7}
\definecolor{purple}{rgb}{0.63,0,1}

\definecolor{dark-green}{rgb}{0,0.4,0.1}
\definecolor{dark-gray}{rgb}{0.4,0.4,0.4}

\definecolor{pink}{rgb}{1,0,0.9}

\newcommand{\change}[1]{{#1}}
\newcommand{\Tr}{\operatorname{Tr}}

\newcommand{\subfigref}[2]{\hyperref[#1]{\ref*{#1}(#2)}}

\begin{document}
\title{
    Corner transfer matrix renormalization group approach\\ in the zoo of Archimedean lattices
    }
\author{I.V. Lukin}
\email{illya.lukin11@gmail.com}
\affiliation{Karazin Kharkiv National University, Svobody Square 4, 61022 Kharkiv, Ukraine}
\affiliation{Akhiezer Institute for Theoretical Physics, NSC KIPT, Akademichna 1, 61108 Kharkiv, Ukraine}

\author{A.G. Sotnikov}
\email{a\_sotnikov@kipt.kharkov.ua}
\affiliation{Karazin Kharkiv National University, Svobody Square 4, 61022 Kharkiv, Ukraine}
\affiliation{Akhiezer Institute for Theoretical Physics, NSC KIPT, Akademichna 1, 61108 Kharkiv, Ukraine}

\date{\today}

\begin{abstract}
    We develop a new methodology to contract tensor networks within the corner transfer matrix renormalization group approach for a wide range of two-dimensional lattice geometries.
    We discuss contraction algorithms on the example of triangular, kagome, honeycomb, square-octagon, star, ruby, square-hexagon-dodecahedron, and dice lattices.
    As benchmark tests, we apply the developed method to the classical Ising model on different lattices and observe a remarkable agreement of the results with the available from the literature.
    The approach also shows the necessary potential to be applied to various quantum lattice models in a combination with the wave-function variational optimization schemes.
\end{abstract}

\maketitle

\section{Introduction}

Tensor-network (TN) methods are powerful nonperturbative approaches to describe both classical and quantum systems on the lattice. 
For the up-to-date reviews on this topic, we direct the reader to Refs.~\cite{Tensor_networks_intro_1, tensor_networks_intro_2, tensor_network_intro_3}. The general idea of the TN approach is to reformulate the problem of interest (e.g., computation of the partition function or a search for the ground state of the quantum many-body Hamiltonian) in terms of a contraction of a large number of tensors connected with each other through a certain lattice-network structure.  The problem is now reformulated as a contraction of these networks of tensors. While one can perform contraction of the one-dimensional network exactly, for the two-dimensional (2d) and higher-dimensional tensor networks it is imperative to employ approximate approaches. 

The contraction of the 2d tensor networks can be realized by means of the transfer matrix approaches~\cite{Haegeman_2017_ED_TM, VUMPS} (see also Refs.~\cite{yang2023efficient3D, vanderstraeten2018_3d_frustrated, nishinoPEPS1, Gendiar2003, gendiar2004estimation} for the three-dimensional analysis) or with various tensor renormalization groups~\cite{TNR, TRG}.  Another approach is the corner transfer matrix renormalization group (CTMRG). 

Corner transfer matrices (CTM) originally appeared \change{as a set of equations in Refs.~\cite{CTM_intro, CTM_intro_2} and also in the integrable model context \cite{Baxter1976, Baxter1977, baxter1981rogers}. They were later adapted to the efficient numerical renormalization group method---CTMRG~\cite{nishino1996corner}}.   The applications of CTM and CTMRG approaches include the computation of properties of infinite projected entangled pair state (iPEPS) wave functions~\cite{CTMRG_for_iPEPS_1, CTMRG_for_iPEPS_2, orus2012_corner_tensors}, variational optimization of iPEPS wave functions and gradient summations~\cite{CTMRG_variatopnal_iPEPS}, excited states on the top of the iPEPS wave function~\cite{ExcitedStates},  hybrid approaches of CTMRG and tensor renormalization group~\cite{CTMRG_for_TRG}, and series expansions~\cite{chan2012series}. The original approach was developed for the square-lattice tensor network and was later generalized to the hyperbolic lattices~\cite{hyperbolic_1, hyperbolic_2, hyperbolic_3, hyperbolic_4, hyperbolic_6, hyperbolic_7, hyperbolic_8}.
Recently, CTMRG was generalized and applied to the honeycomb-lattice quantum and frustrated classical systems~\cite{Lukin2023ctmrg_honeycomb, nyckees2023critical}.  

In this study, we introduce a further generalization of the CTMRG approach to other relevant lattice geometries. We discuss details and perform the necessary benchmark tests of the methodology for triangular, kagome, honeycomb (two-site unit cell), square-octagon, star, ruby, square-hexagon-dodecahedron, and dice lattices.  The conceptual framework employed to derive these CTMRG approaches is potentially generalizable to many other lattice geometries, which are not covered in this work, e.g., the Shastry-Sutherland, maple-leaf, or square-kagome lattice. 

In principle, one can deal with all the mentioned lattices in the framework of CTMRG by using the coarse-graining mapping to the square lattice with subsequent application of the most general CTMRG scheme to the latter. This strategy was followed previously in the studies of the iPEPS wave functions on various lattices~\cite{Corboz_honeycomb, Corboz_Kagome, Corboz_Shastry_Sutherland} and it is implemented in the recent libraries for variational iPEPS optimization~\cite{Hasik_Kagome, rizzi2023varipeps}. We believe that the majority of our results can be obtained with this square-lattice-mapping methodology as well. However, the recent study~\cite{nyckees2023critical} shows that the CTMRG approach tailor-made for the lattice and its respective symmetries becomes more efficient in terms of the computational cost and necessary CTM bond dimensions. Also, within our approach, one can directly access relevant physical quantities, such as the corner (entanglement) spectra, which are not easily obtained in the mapping-based schemes. Finally, the variational iPEPS optimization usually requires allocation of rather large memory resources to track full CTMRG convergence. The application of the minimal CTMRG algorithm can largely reduce these memory requirements.

The paper is organized as follows. In Sec.~\ref{sec:Algorithms} we discuss essential details of CTMRG algorithms for each lattice geometry. \change{The most essential parts here are the triangular and kagome lattices. Other lattice geometries can be skipped during the first reading. } Section~\ref{sec:Benchmarks} is devoted to benchmark tests of the developed approach on the classical Ising model on different lattices. In Sec.~\ref{sec:Conclusions} we summarize our results and discuss outlook.

\section{Algorithms}\label{sec:Algorithms}

\begin{figure}
\includegraphics[width= \linewidth]{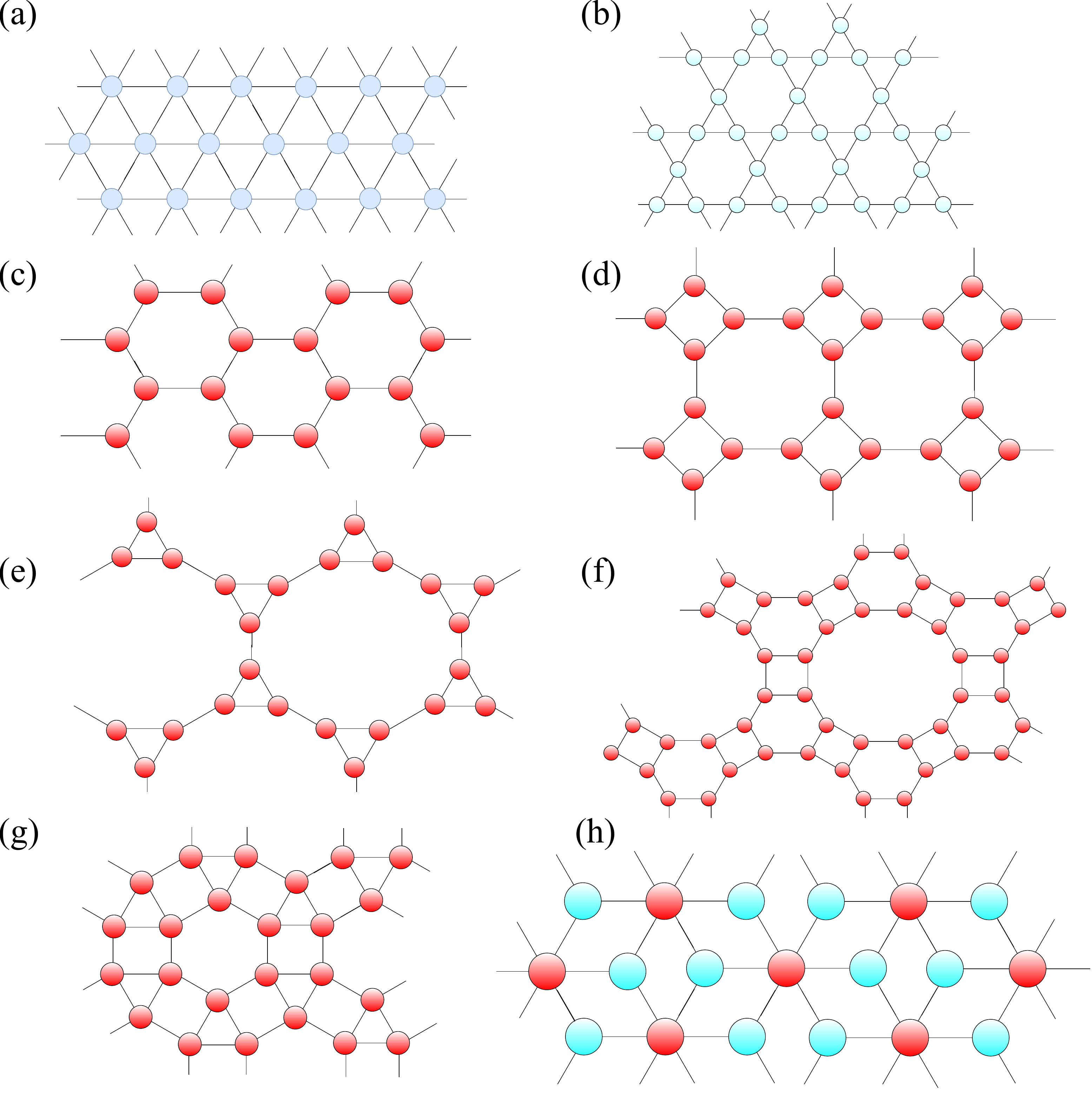} 
 \caption{\label{fig:Lattices}%
   Overview of different lattices for which we construct (generalize), discuss, and apply the CTMRG algorithms: (a) triangular, (b) kagome, (c) honeycomb, (d) square-octagon, (e) star, (f) square-hexagon-dodecahedron, (g) ruby, and (h) dice. }
\end{figure}

Before diving into specific details of each lattice geometry, let us point out the general strategy to construct the CTMRG environments and corresponding update rules. Briefly, the method can be expressed as follows: 
\begin{itemize}
    \item[(i)] define all unique boundary matrix-product states (bMPS) on the lattice and find how the individual tensors of the bMPS are updated during the absorption of the bulk tensors into the bMPS;
    \item[(ii)] define the corner matrices as intersections of different bMPS; 
    \item[(iii)] find the updates of corners from the updates of bMPS local tensors; 
    \item[(iv)] employ the corner tensors to define environments, which enable finding optimal truncations for the local bMPS tensors.  
\end{itemize}
In the following subsections, we illustrate and explain how this scheme can be applied to different lattice geometries. In particular, we develop and apply the CTMRG approach on triangular, kagome, honeycomb, square-octagon, star, square-hexagon-dodecahedron, ruby, and dice lattices, which are shown in Fig.~\ref{fig:Lattices}.  

All the proposed CTMRG methods have the $\chi^{3}$ scaling of the computational cost, where $\chi$ is the bond dimension of the CTM environments. The precise dependence of the cost on the bulk tensor network bond dimension $D$ varies from lattice to lattice (and also depends on the details of a particular realization, e.g., the application of iterative schemes as the randomized singular value decomposition or Lanczos method). 

\subsection{Triangular lattice}

\begin{figure}
\includegraphics[width= \linewidth]{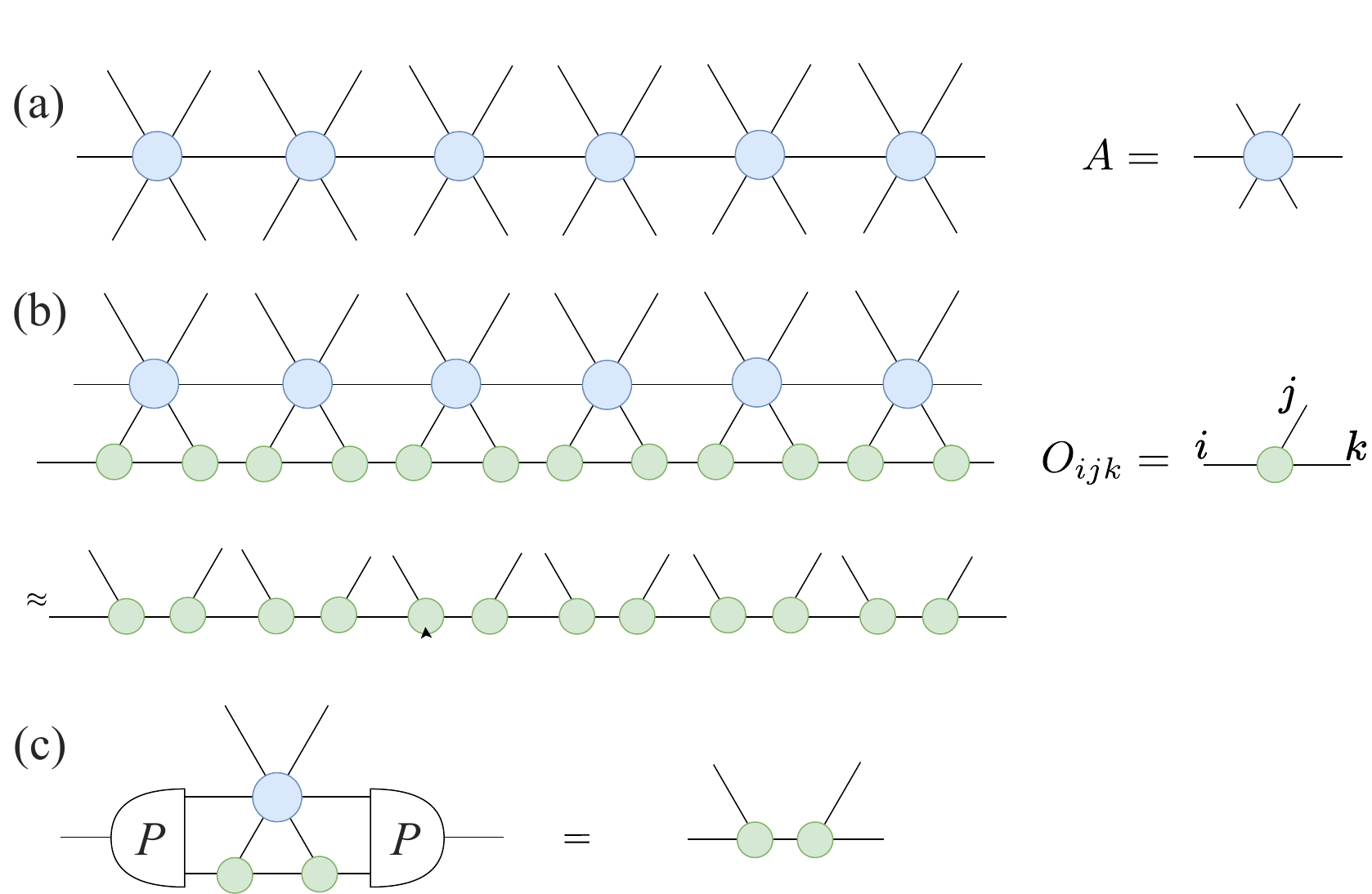} 
 \caption{\label{fig:Figure1}%
    Triangular lattice: (a) Construction of the transfer-matrix network and notation of the bulk rank-6 tensor $A$ (we omit its indices for simplicity); (b) definition of the boundary MPS (bMPS); (c) iterative update of  the individual tensors holding within bMPS. }
\end{figure}

In this subsection, we study the vertex model on the triangular lattice. The rank-6 tensor $A$, which is symmetric under rotations and arbitrary reflections, is placed on every lattice site, as in Fig.~\ref{fig:Figure1}(a). Our aim is to contract a tensor network consisting of these tensors. We begin our discussion with the boundary matrix product state and then arrive at the CTMRG construction. 

First, we assume that the leading eigenvector of the transfer matrix can be approximated by a translationally invariant matrix product state (MPS), which is shown in Fig.~\ref{fig:Figure1}(b). The boundary MPS (bMPS) consists of rank-3 tensors $O_{ijk}$, which are not symmetric with respect to any of the three indices. This form of the MPS can be deduced both from the translational symmetry of the transfer matrix and from its reflection invariance under mirror transformations. For the bMPS to be the eigenvector of the transfer matrix we can assume that the local condition shown in Fig.~\ref{fig:Figure1}(c) must hold. Here, we introduce the isometric projector $P$, which can be naturally obtained from the corner matrices, as we discuss below. 

\begin{figure}
\includegraphics[width= \linewidth]{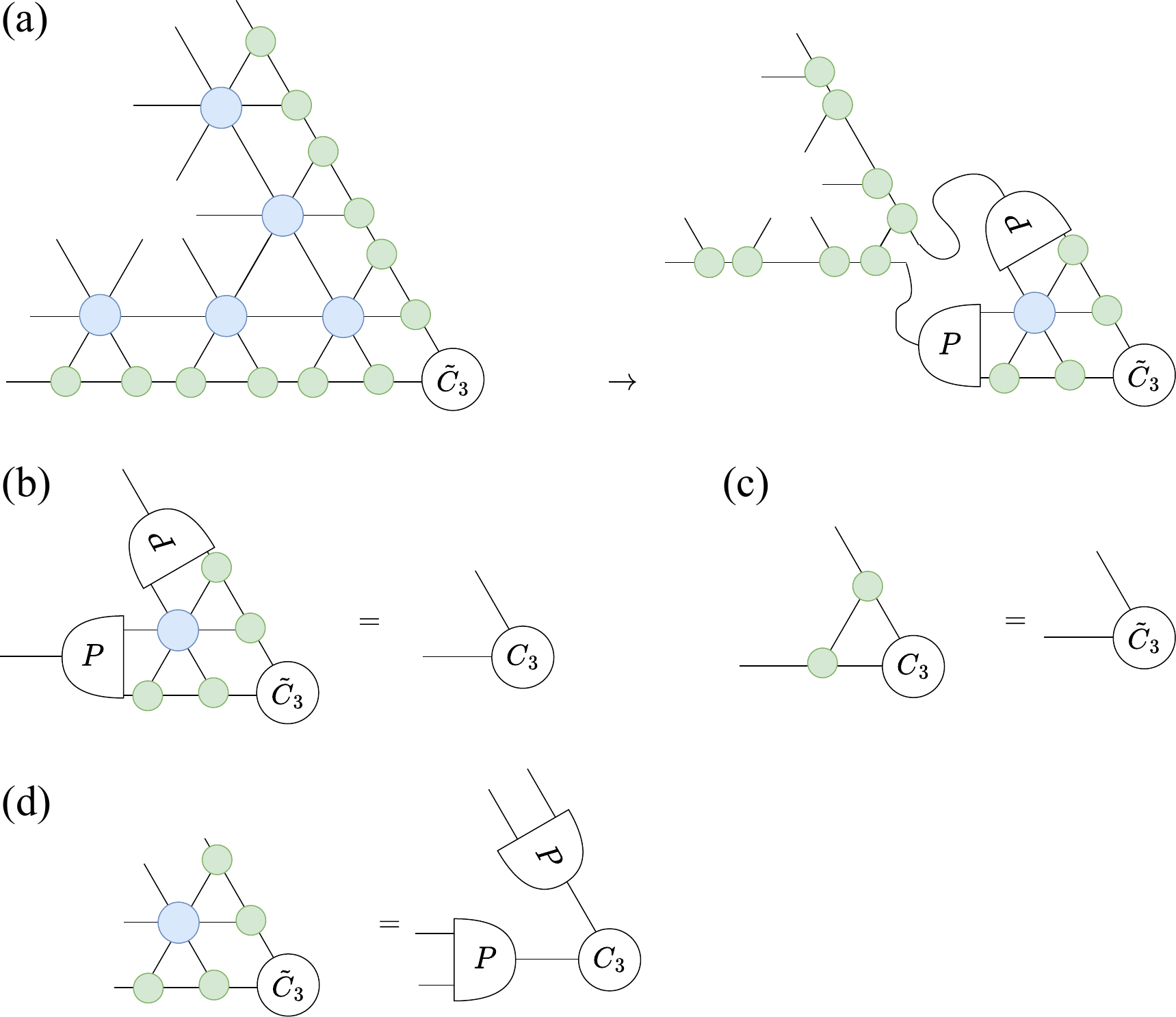} 
 \caption{\label{fig:Figure2}%
   Triangular lattice: (a) Definition of the corner matrix $\tilde{C}_{3}$ on the intersection of two boundary MPS and its update; (b) update procedure of the corner matrix $\tilde{C}_{3}$, which results in the new corner matrix $C_{3}$; (c) the second update step, which transforms $C_{3}$ back into $\tilde{C}_{3}$; (d) the projectors $P$ can be chosen from the truncated eigendecomposition of the matrix $C_{3}$. The matrix $C_{3}$ can be assumed to be always diagonal. }
\end{figure}

Next, we determine the corner tensors. To this end, we take two bMPS, which intersect as shown in Fig.~\ref{fig:Figure2}(a). It is natural to assume that on the intersection point of two bMPS, there must be a corner matrix $\tilde{C}_{3}$. From the symmetry of the problem, it follows that the matrix $\tilde{C}_{3}$ is symmetric in its indices. Let us now absorb one layer of bulk tensors $A$ into the bMPS. Such absorption leads to the update rule for the corner matrix $\tilde{C}_{3}$, which is shown in Figs.~\ref{fig:Figure2}(b) and \ref{fig:Figure2}(c). Note that the update consists of two steps [Figs.~\ref{fig:Figure2}(b) and \ref{fig:Figure2}(c)], and leads to the definition of the second corner matrix $C_{3}$. The reason for the appearance of the two different matrices lies in the presence of two different triangles (pointing up and down), while the bMPSs may intersect on both types of triangles. We see that the projectors $P$ naturally appear in the update rule for the matrix $C$. Besides that, the density matrix around the triangle is proportional to the third power of $C_{3}$: $\rho \propto C_{3}^{3}$. This leads us to the suggestion to choose the projectors $P$ in the way to diagonalize the matrix $C_{3}$ (note that this matrix is also symmetric). This step of determining the projector from the eigendecomposition of the matrix $C_{3}$ is shown in Fig.~\ref{fig:Figure2}(d). The eigendecomposition must be truncated according to the spectrum of the corner matrix back to the original bond dimension of the bMPS, which we denote as $\chi$.

\begin{figure}
\includegraphics[width= \linewidth]{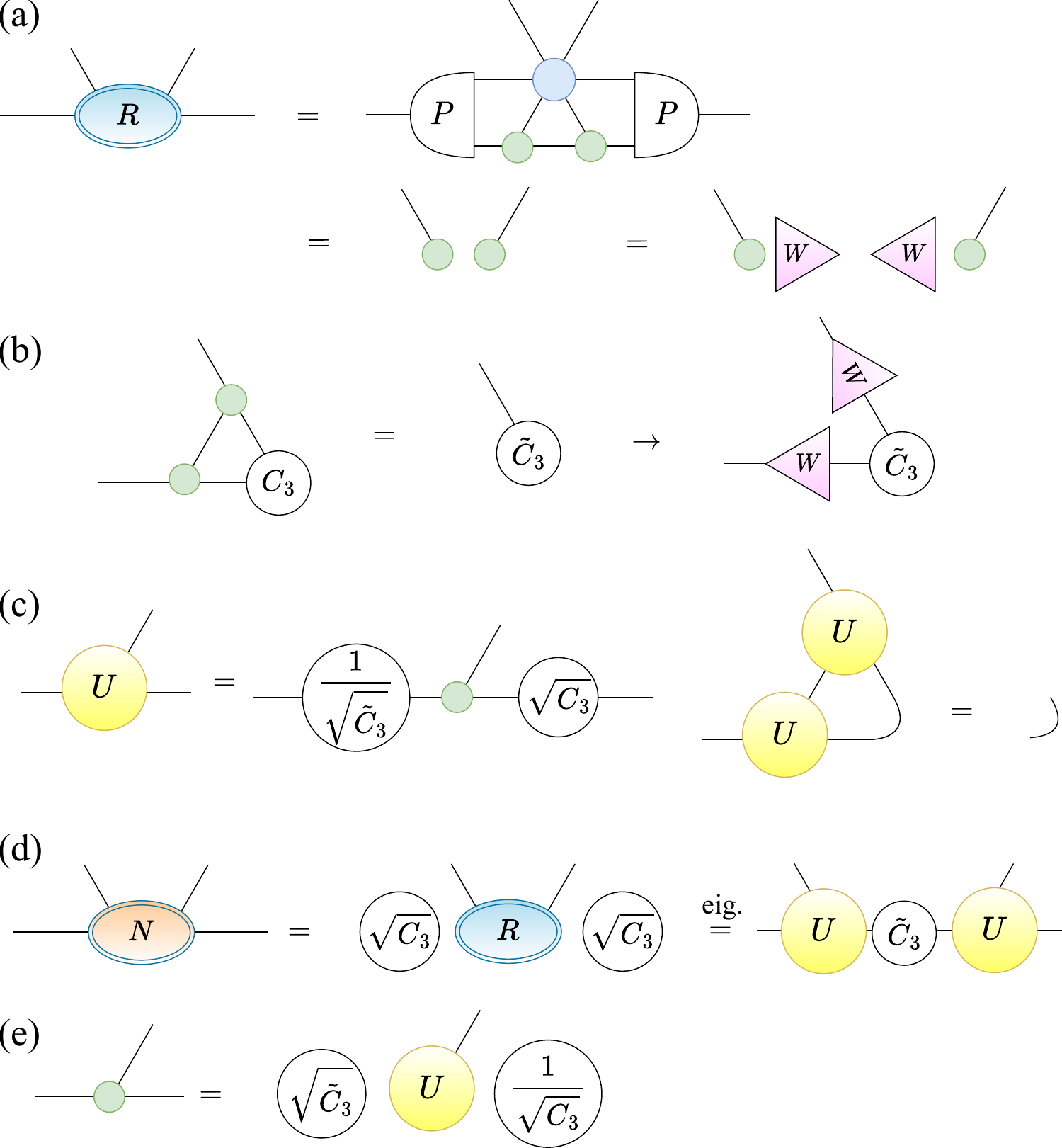} 
 \caption{\label{fig:Figure3}%
   Triangular lattice: (a) Definition of the tensor $R$, which is factorized into two tensors $O$, and the possible ambiguity (gauge choice) in the factorization due to the arbitrary orthogonal matrix $W$, which can be inserted in the factorized index. (b) Ambiguity in the factorization leads to the transformation of the corner matrix $\tilde{C}_{3}$. We can fix the gauge by the condition, that the corner matrix $\tilde{C}_{3}$ is also diagonal. (c) The definition of the new tensor $U$, which can be shown (from the point (b)) to be isometric. 
   (d) We can define the new tensor $N$ as the tensor $R$ weighted by the corner matrices. 
   The eigendecomposition of $N$ results in both isometry $U$ and the diagonal corner matrix $\tilde{C}_{3}$. 
   (e) The original bMPS tensor $O$ can be found from the isometry $U$ by the inverse transformation. This ends the update step for the tensor $O$ and simultaneously finds the updated corner matrix $\tilde{C}_{3}$. }
\end{figure}

The analysis of the corner matrices leads us to the natural definition of the projectors $P$. Still, this is not sufficient to complete the update of the bMPS tensor $O$, since the update step of the tensor $O$ shown in Fig.~\ref{fig:Figure1}(c) contains not only a projection step but also a factorization step. In principle, this factorization can be performed by the eigendecomposition or singular-value decomposition, but this procedure will be not unique, and it is not guaranteed that the truncation with this factorization remains optimal. The possible ambiguity in the factorization step of the tensor $R$ is shown in Fig.~\ref{fig:Figure3}(a): We can insert a pair of orthogonal matrices $W$ in the factorized index and reabsorb these orthogonal matrices back into the definition of the bMPS tensor $O$. Note also that the exact factorization leads to an increase of the factorized index bond dimension. We should now fix the gauge freedom in the factorization algorithm and also define the optimal truncation of the factorized index. The gauge transformation of the tensor $O$ results in the change of the corner matrix~$\tilde{C}_{3}$, as shown in Fig.~\ref{fig:Figure3}(b). Besides, the density matrix for the factorized index can be naturally cast in the form $\rho_{f} \propto \tilde{C}_{3}^{3}$. This leads us to the natural fix of the gauge freedom and simultaneously to the truncation choice: we should choose the matrix $W$ to diagonalize the corner matrix $\tilde{C}_{3}$ and truncate the factorized index according to the spectrum of the corner matrix.

Let us discuss how this factorization can be performed in practice. We assume now that the matrices $C_{3}$ and $ \tilde{C}_{3}$ are both diagonal. Then, the connection between these matrices, which is shown in Fig.~\ref{fig:Figure3}(b), can be used to define the new tensor $U$ [shown in Fig.~\ref{fig:Figure3}(c)], which is isometric. We can also define the tensor $N$, as a tensor $R$ (to be factorized), weighted by the square roots of the corner matrices $C_{3}$. Using the connection between the bMPS tensor $O$ and the isometry $U$ we arrive at the decomposition of the tensor $N$, which is shown in  Fig.~\ref{fig:Figure3}(d). Due to the isometricity of the tensor $U$ and the diagonal $\tilde{C}_{3}$, we can conclude that the decomposition of the tensor $N$ coincides with its eigendecomposition. The order of this derivation may now be reversed: first, we calculate $N$ from $R$; then we make an eigendecomposition of $N$ to find both $U$ and new corner matrix $\tilde{C}_{3}$; finally, we can use the connection between $O$ and $U, C_{3}, \tilde{C}_{3}$ to define the new bMPS tensor $O$.

To summarize, the full CTMRG loop runs as follows:
\begin{enumerate}
    \item Start from some initial tensor $O$ and the diagonal matrix $\tilde{C}_{3}$.
    \item Determine the projector $P$ and the corner matrix~$C_{3}$ from the tensors $O$, $A$, and $\tilde{C}_{3}$, as shown in Fig.~\ref{fig:Figure2}(d).
    \item Employ the projector $P$ to obtain the tensor~$R$, which is defined in Fig.~\ref{fig:Figure3}(a). Transform the tensor~$R$ into the tensor~$N$, as defined in Fig.~\ref{fig:Figure3}(d), and compute its truncated eigendecomposition to find the isometry $U$ and the new corner matrix $\tilde{C}_{3}$.
    \item Obtain the new bMPS tensor $O$ from the tensors~$U$, $C_{3}$, and $\tilde{C}_{3}$. 
    \item Return to the point 2 and repeat until convergence. Convergence can be measured in terms of diagonal elements of the corner matrices or in terms of some observables. 
\end{enumerate}

Note that these steps have some residual sign ambiguity in the eigendecompositions, which we fix by additional sign rule on the isometry elements. The eigendecompositions can also be performed with iterative methods, e.g., the Lanczos algorithm to reduce the computational cost. Finally, according to our observations, the matrices $C_{3}$ and $\tilde{C}_{3}$ converge in practice to almost identical values. 

\begin{figure}
\includegraphics[width= \linewidth]{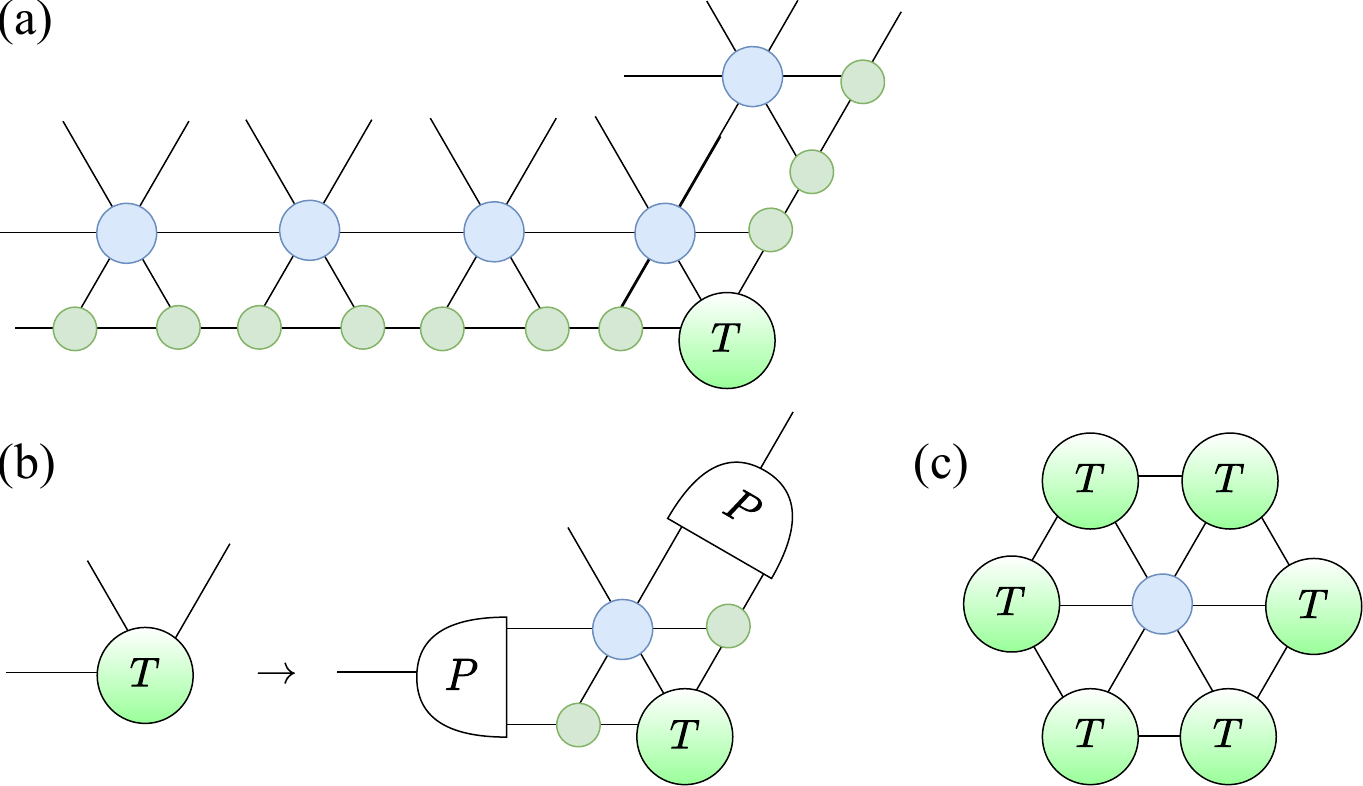} 
 \caption{\label{fig:Figure4}%
   Triangular lattice: (a) Definition of the corner tensor $T$, as the intersection of two bMPS with $2\pi/3$ angle; (b) update rule for the tensor $T$; (c) environment of the bulk tensor $A$ in terms of the corner tensor $T$.}
\end{figure}

The CTMRG construction on the triangular lattice can be augmented by the additional structure, in particular, by the tensor $T$, which is shown in Fig.~\ref{fig:Figure4}(a). This new tensor naturally appears at the intersection of two bMPS at the angle $2\pi/3$ and it is symmetric in its two bMPS indices. In Fig.~\ref{fig:Figure4}(b) we show the update rule for this tensor. This update rule can be run in parallel to the CTMRG loop, since the tensor $T$ does not appear in the main CTMRG procedure. In Fig.~\ref{fig:Figure4}(c) it is shown how the corner tensor $T$ can be used to define explicitly the $C_6$-symmetric environment for the bulk tensor $A$, which can be used for the calculation of observables, such as the magnetization. 

\subsection{Kagome lattice}\label{subsec:Alg.kagome}
\begin{figure}
\includegraphics[width= \linewidth]{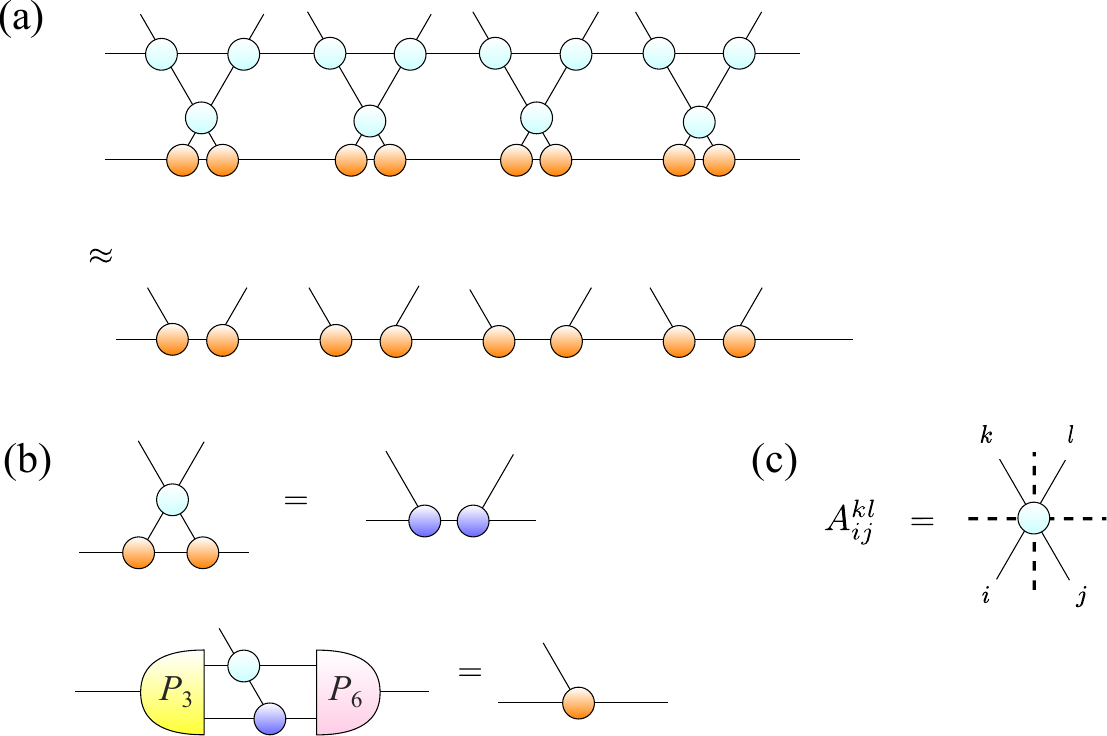} 
 \caption{\label{fig:Figure5}%
   Kagome lattice: (a) Definition of the boundary MPS on the kagome lattice and the structure of the transfer matrix. (b) Local conditions on the tensors of the bMPS, which guarantee  approximate holding of the above eigenvector condition. To fulfill this, we introduce the (isometrical) projectors $P_{3}$ and $P_{6}$ and determine them later from the corner matrices update rules. (c) The bulk tensor $A_{ij}^{kl}$ with the dashed lines indicating to the reflection symmetry axes. }
\end{figure}

After the triangular lattice, it is natural to construct the CTMRG on other lattices by applying the analogous procedure. Here, we extend the construction to the kagome lattice, which consists of triangles and hexagons. Hence, we need two types of corners, which correspond to the corners of hexagons and triangles.
We assume the kagome lattice to be filled with bulk tensors $A$ placed in all its nodes. Since we study here the isotropic CTMRG (and not its directional generalization), we should choose the tensor $A$ in the way that the tensor network is rotationally invariant under the rotations of the hexagons (which makes all hexagonal corners identical) and rotations of triangles (which leads to identical triangular corners). To ensure that these corner matrices are also symmetric, we also assume the reflection invariance of tensors both upon reflections in the hexagons and triangles. This leads to the following conditions on the tensor $A_{ij}^{kl} = A_{lk}^{ji} = A_{kl}^{ij} = A_{ji}^{lk}$. These symmetries are illustrated in Fig.~\ref{fig:Figure5}(c), where the reflection symmetry axes are shown by the dashed lines. 

Let us now describe the CTMRG algorithm on this lattice. First, we introduce the boundary MPS as the leading eigenvector of the transfer matrix on the kagome lattice. This bMPS is shown in Fig.~\ref{fig:Figure5}(a) together with the transfer matrix structure. The bMPS has the same form as in the case of the triangular lattice, but the transformation procedure is different. This is evident from the update rules of the local bMPS tensors $O$, which are shown in Fig.~\ref{fig:Figure5}(b). This update procedure consists of two steps: the first one involves factorization of the rank-4 tensor, while the second one uses two projectors $P_{3}$ and $P_{6}$. The first factorization step is performed analogously to the factorization in the triangular lattice case, while the projectors are determined according to the update rules of the hexagonal and triangular corner matrices. 

\begin{figure}
\includegraphics[width= \linewidth]{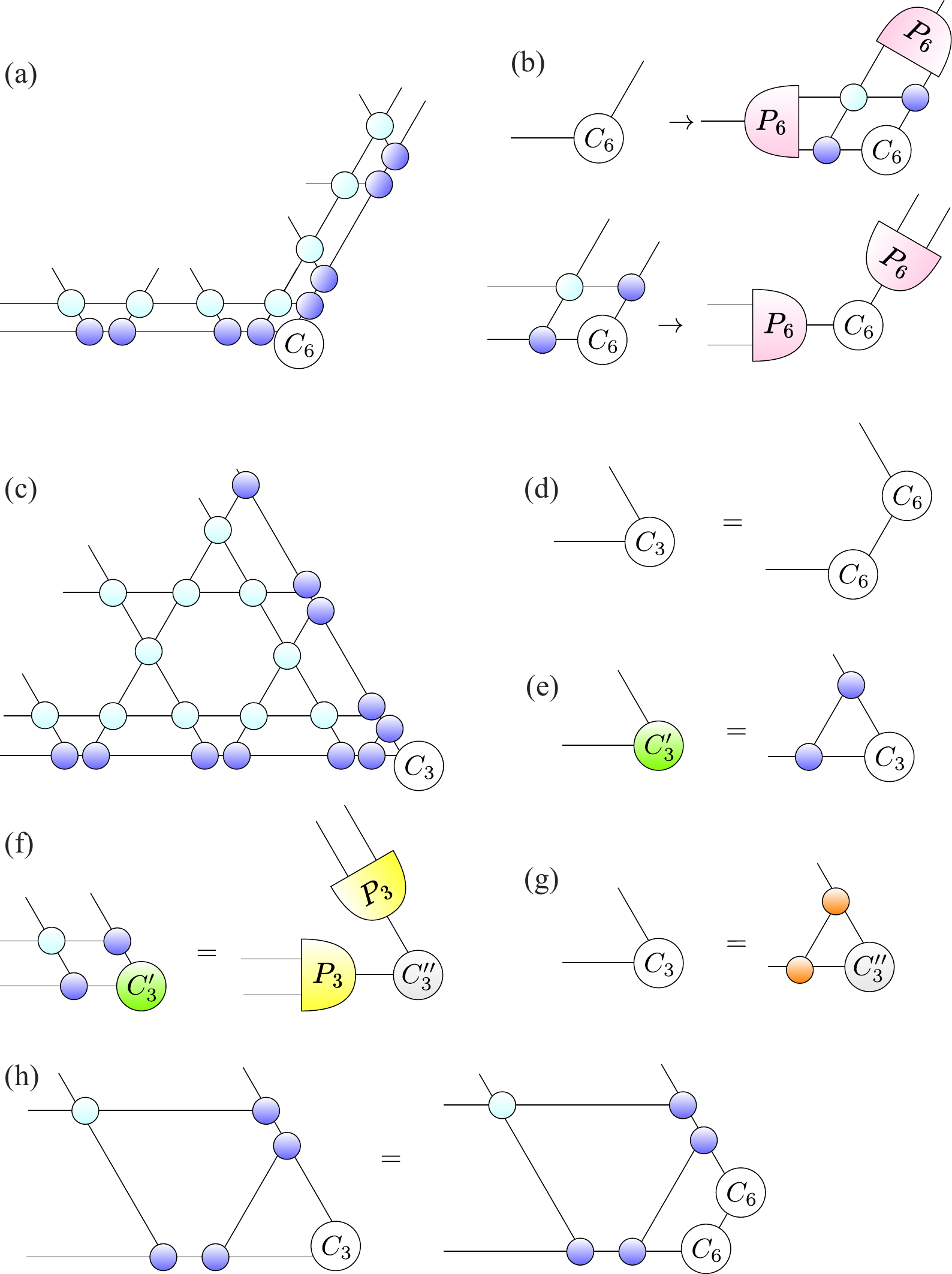} 
 \caption{\label{fig:Figure6}%
    Kagome lattice: (a) Definition of the corner matrix $C_{6}$ as an intersection of two bMPS with the angle $2\pi/3$ between them. (b) The update rule for the matrix $C_{6}$; the isometrical projector $P_{6}$ can be chosen in the way that the updated matrix $C_{6}$ will be diagonal. (c) The definition of the corner matrix $C_{3}$ as an intersection point between two bMPS with the angle $\pi/3$; (d) The matrix $C_{3}$ can be expressed as the square of the matrix $C_{6}$; (e) still, the matrix $C_{3}$ has a different update rule: here, we define the new corner matrix $C'_{3}$ (in green) from the matrix $C_{3}$; (f) yet another corner matrix $C''_{3}$ (gray) can be obtained from $C'_{3}$. The diagonalization of this matrix leads to the projector $P_{3}$. (g) One can arrive back to the $C_{3}$ matrix from $C''_{3}$. (h) The illustration of the equality in the point (d) between the matrices $C_{3}$ and $C_{6}^2$. It is always possible to change the crossing of two bMPS with the angle $\pi/3$ into two crossings with an additional bMPS, where the new crossings are at the angle $2\pi/3$. }
\end{figure}

Next, we define the corner matrices. The first corner matrix $C_{6}$ is shown in Fig.~\ref{fig:Figure6}(a) as an intersection of two boundary bMPS. From the update rules of the bMPS tensors it can be found that the $C_{6}$ matrix transforms according to the update rule in Fig.~\ref{fig:Figure6}(b). We can now choose the isometry $P_{6}$ to diagonalize the new $C_{6}$ matrix and truncate it back to its original bond dimension. One can perform truncation according to the spectrum magnitude. The definition of the second corner matrix $C_{3}$ is shown in Fig.~\ref{fig:Figure6}(c). This matrix is obtained as an intersection of the two bMPS with the angle $\pi/3$ between them. But such an intersection may be obtained in two different ways: We can either intersect two bMPS directly or we can intersect both bMPS with a third one (with angle $2\pi/3$), which results in the product of two $C_{6}$ matrices, as is shown in Fig.~\ref{fig:Figure6}(h). For the consistency between these two definitions of the intersection, we must conclude that $C_{3} = C_{6}^{2}$ (at least at the converged state of the CTMRG).

\begin{figure}
\includegraphics[width= \linewidth]{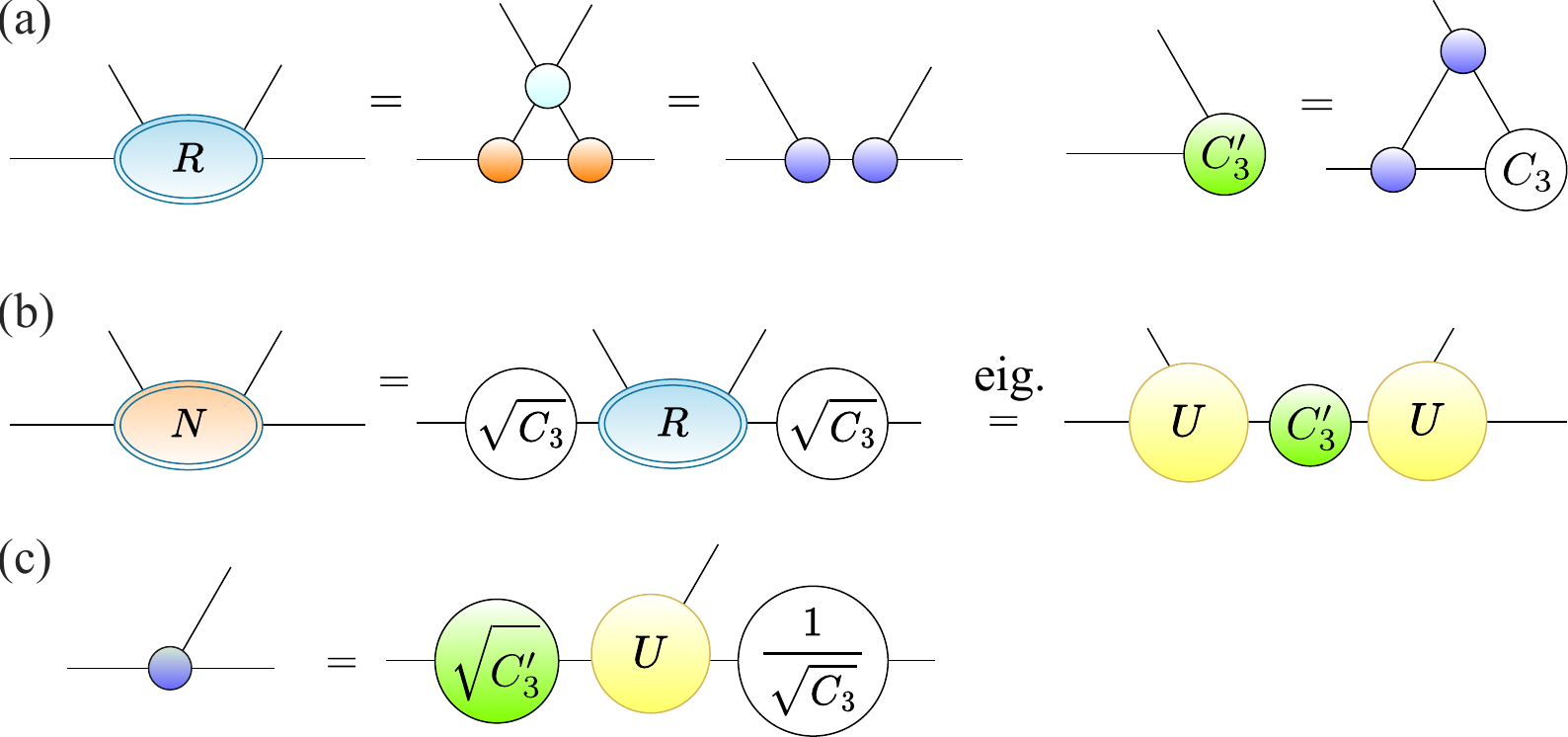} 
 \caption{\label{fig:Figure7}%
   Kagome lattice: (a) Definition of the new rank-4 tensor $R$ to be factorized and the update rule of the matrix ${C}'_{3}$. (b) The tensor $N$ as a weighted tensor $R$ and its eigendecomposition (eig.) from which we define the new matrix ${C}'_{3}$ and the isometry $U$. (c) The reverse transformation between the bMPS local tensor and the isometry $U$. }
\end{figure}

In spite of the given connection between the two matrices, they have rather different update rules. The update rules for the $C_{3}$ matrix are shown in Figs.~\ref{fig:Figure6}(e)--\ref{fig:Figure6}(g). Here, we additionally introduce the matrices $C'_{3}$ and $C''_{3}$.
The reason for three different corners lies in the fact that the bMPS can intersect with the angle $\pi/3$ in hexagons and in both types of triangles (pointing up and down). We can obtain the isometrical projector $P_{3}$ from the diagonalization of the matrix $C''_{3}$, while $C'_{3}$ can be used for the proper factorization, as shown in Fig.~\ref{fig:Figure5}(b). The details of this factorization are the same as in the case of triangular lattice and shown in Fig.~\ref{fig:Figure7}. 

We can now discuss the full CTMRG iteration loop. There is a certain ambiguity in the updates of the tensors. In particular, should the condition $C_{3} = C_{6}^{2}$ be enforced in every iteration, or should it only hold in the converged state? Below, we describe one of options to update the tensors, which we find rather fast and stable. The scheme is as follows:
\begin{enumerate}
    \item Initialize the bMPS tensor $O$ and matrix $C_{6}$. Determine the matrix $C_{3}=C_{6}^2$. 
    
    \item Find the rank-4 tensor $R$, as shown in Fig.~\ref{fig:Figure7}(a). Factorize it according to Figs.~\ref{fig:Figure7}(b) and \ref{fig:Figure7}(c) and obtain the matrix $C'_{3}$ and the updated bMPS tensor. 
    
    \item Determine $P_{3}$ from the eigendecomposition of $C''_{3}$. 
    
    \item Update $C_{6}$ and find the new projector $P_{6}$. 
    
    \item Use the projectors $P_{3}$ and $ P_{6}$ to update the bMPS tensor according to Fig.~\ref{fig:Figure5}(b). 
    
    \item Determine the new matrix $C_{3} = C_{6}^{2}$. 
    
    \item Return to the point 2 and repeat until convergence.
\end{enumerate}

Point 6 is the most controversial, since the  matrix $C_{3}$ can also be obtained from $C''_{3}$ [see Fig.~\ref{fig:Figure6}(g)]. But the update rule in Fig.~\ref{fig:Figure6}(g) does not necessarily lead to the diagonal matrix $C_{3}$ and does not enforce the connection between the matrices $C_{3}$ and $C_{6}$. We choose to employ the update rule in Fig.~\ref{fig:Figure6}(g) as a convergence check instead. For the converged CTMs, we observe that the two different ways to obtain $C_{3}$ agree. Note also that operating with $C_{6}$ leads to a higher precision.

\subsection{Remarks on the honeycomb lattice}\label{subsec:Alg.honeycomb}

In the previous work, we studied the variational iPEPS optimization on the honeycomb lattice~\cite{Lukin2023ctmrg_honeycomb}. In Ref.~\cite{nyckees2023critical}, an alternative formulation of the honeycomb lattice CTMRG was proposed. In this subsection, we discuss the derivation of both approaches and the consistency between them. We also show how the method can be naturally generalized to the two-site unit cells. 

\begin{figure}
\includegraphics[width= \linewidth]{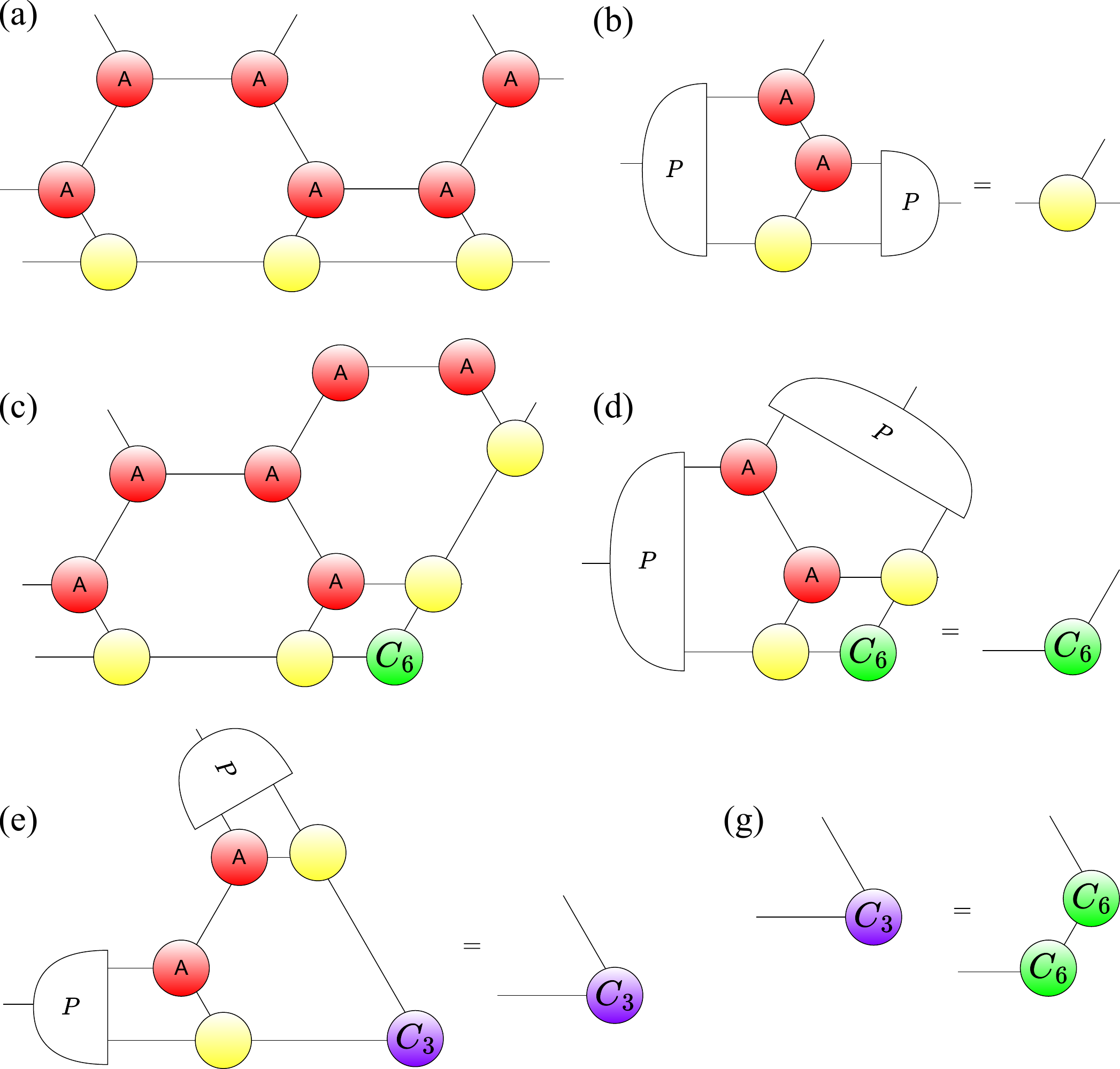} 
 \caption{\label{fig:Figure8}%
   Honeycomb lattice: (a) Definition of the transverse matrix and boundary MPS. (b) The update procedure for the bMPS tensors. (c) The definition of the tensor $C_{6}$. (d) The update procedure for the tensor $C_{6}$. (e) Definition and update for the corner matrix~$C_{3}$. (g) Consistency condition for the corner matrices $C_{3}$ and $C_{6}$.}
\end{figure}

First, we assume that the tensor network consists of identical rank-3 tensors $A$ with rotational and reflection invariance. In Fig.~\ref{fig:Figure8}(a) we show the transfer matrix and boundary MPS for this tensor network. The update for the local bMPS tensors is shown in Fig.~\ref{fig:Figure8}(b). Here, we introduce the isometrical projectors $P$, which we define from the updates of the corner matrices.

Next, let us study the intersection of two bMPS at the angle $\pi/3$. As in the case of the kagome lattice, we can define the corner matrix $C_{6}$ on the intersection point of two bMPS [see Fig.~\ref{fig:Figure8}(c)]. The update for the corner matrix $C_{6}$ is shown in Fig.~\ref{fig:Figure8}(d). As in the case of kagome lattice, it is also possible to define the corner matrices~$C_{3}$ (on the intersection point of two bMPS with the angle $2\pi/3$) with the update rule shown in Fig.~\ref{fig:Figure8}(e). For consistency of the observables calculation, the condition $C_{3} = C_{6}^{2}$ [shown in Fig.~\ref{fig:Figure8}(g)] must hold.

There are now two different ways to define the projectors $P$: we can choose them to either diagonalize the corner matrices $C_{6}$ or $C_{3}$. The iterations from the two methods can be different, but the final result should be independent of the scheme due to the consistency condition $C_{3} = C_{6}^{2}$. According to our observations, if we define the matrix $C_{3}$ from the condition $C_{3} = C_{6}^{2}$ for the converged $C_{6}$ tensor, then the $C_{3}$ matrix is the fixed point of the update in Fig.~\ref{fig:Figure8}(e), and the updates for different corner matrices are consistent. 

\begin{figure}
\includegraphics[width= \linewidth]{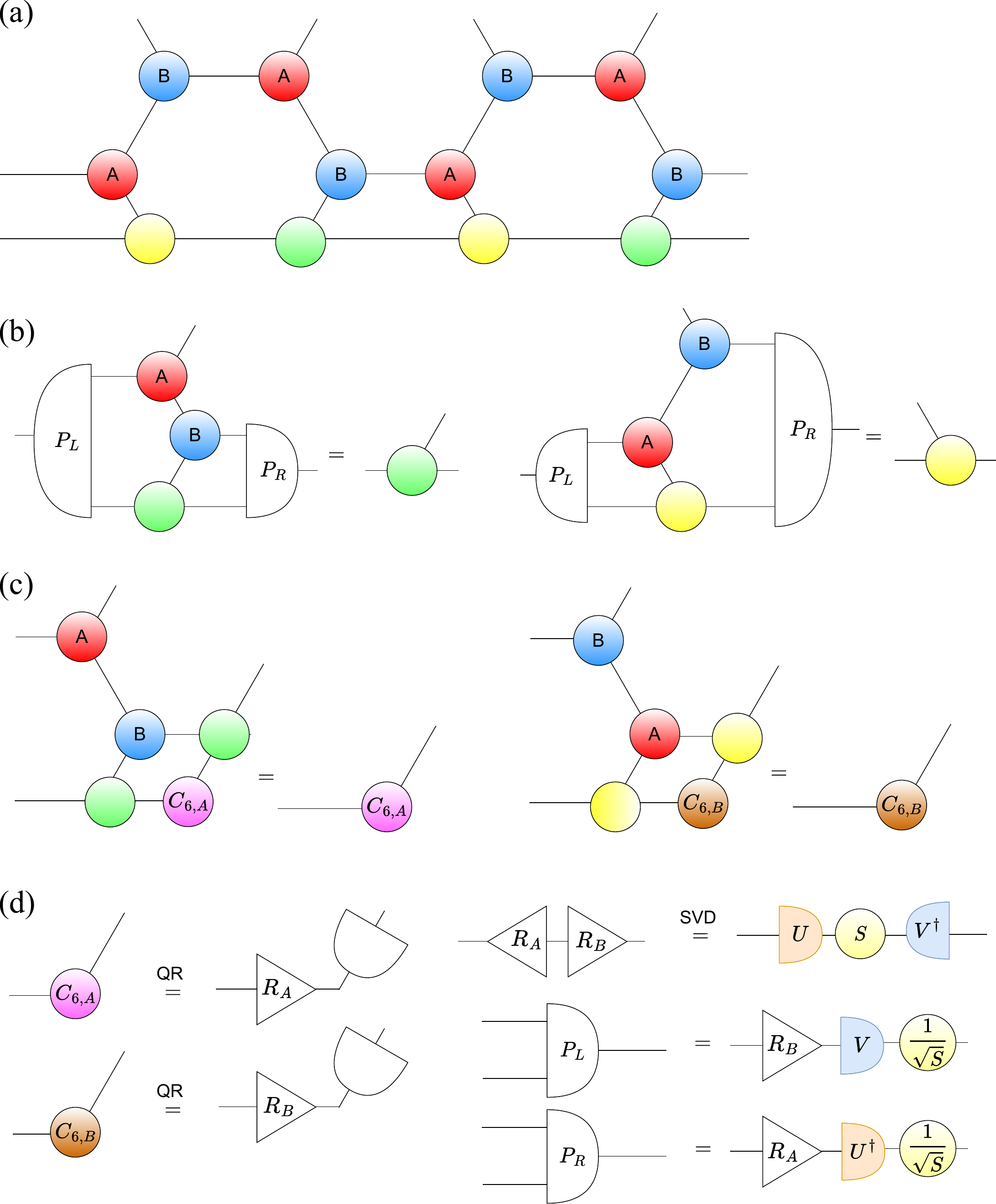} 
 \caption{\label{fig:Figure9}%
   Honeycomb lattice: (a) Definition of the two-site bMPS and transfer matrix for the case of two-site unit cell. (b) The update procedure for the bMPS tensors, where we insert the projectors $P_{L}$ and $ P_{R}$ ($P_{L}P_{R} = 1$). (c) The updates of the corner matrices $C_{6,A}$ and $ C_{6,B}$. (d) Biorthogonalization procedure (involving QR decomposition and SVD), which allows to define the projectors $P_{L}$ and $ P_{R}$ from the corner matrices.  }
\end{figure}

Next, let us generalize this construction to the larger unit cell. We assume that the tensor network on the bipartite honeycomb lattice consists of two different tensors $A$ and $B$, which are placed on two different sublattices. We additionally set that these tensors are invariant under rotations and reflections. Then, the boundary MPS consists of two types of tensors, as shown in Fig.~\ref{fig:Figure9}(a). The update rules for the boundary MPS tensors are shown in Fig.~\ref{fig:Figure9}(b). 
The key difference from the case with a one-site unit cell is that there are two different projection tensors $P_{L}$ and $P_{R}$, which are no longer isometric, but just biorthogonal, $P_{L} P_{R} = 1$. There are also two different corner matrices $C_{6}$, which are updated according to Fig.~\ref{fig:Figure9}(c) (projectors are not shown here, since the enlarged corner matrices will be used to construct projectors).

Now, we discuss the procedure to find biorthogonal projectors $P_{L}$ and $ P_{R}$ from the enlarged corner matrices. In case one tries to use the density matrix, then it becomes clear that the latter is no longer symmetric. In principle, one can use the singular-value decomposition (SVD) of the density matrix to define a projector in the isometrical form (according to the directional update), but, in practice, we observe that such a definition of projectors leads to the breaking of some consistency relations between the converged tensors. We choose instead to employ the biorthogonalization procedure, which was proposed in the context of CTMRG in Ref.~\cite{CTMRG_for_iPEPS_2}. The biorthogonalization procedure is shown in Fig.~\ref{fig:Figure9}(d), and defines two biorthogonal projectors $P_{L}$ and $P_{R}$. Note that the authors of Ref.~\cite{CTMRG_for_iPEPS_2} applied biorthogonalization to larger environments, but noted that the corners themselves can be used as a reduced form of the environment. The enlarged corner matrices $C_{6,A} \to P_{L} C_{6,A} P^{T}_{L}$ and $ C_{6,B} \to P_{R} C_{6,B} P^{T}_{R}$ are then projected back to the original bond dimension with the obtained projectors.  We observe that these reduced corner environments work well far from criticality, but at the vicinity of the phase transition, it may be necessary to use larger environments (with two or even three corner matrices) to find the proper projectors.

\subsection{Square-octagon lattice}

Square-octagon and star lattices can be reduced to square and honeycomb lattices. Still, it is interesting if the structure of these lattices allows for a more natural formulation of CTMRG. In this and the following subsections we show that this is indeed the case. 
\begin{figure}
\includegraphics[width= \linewidth]{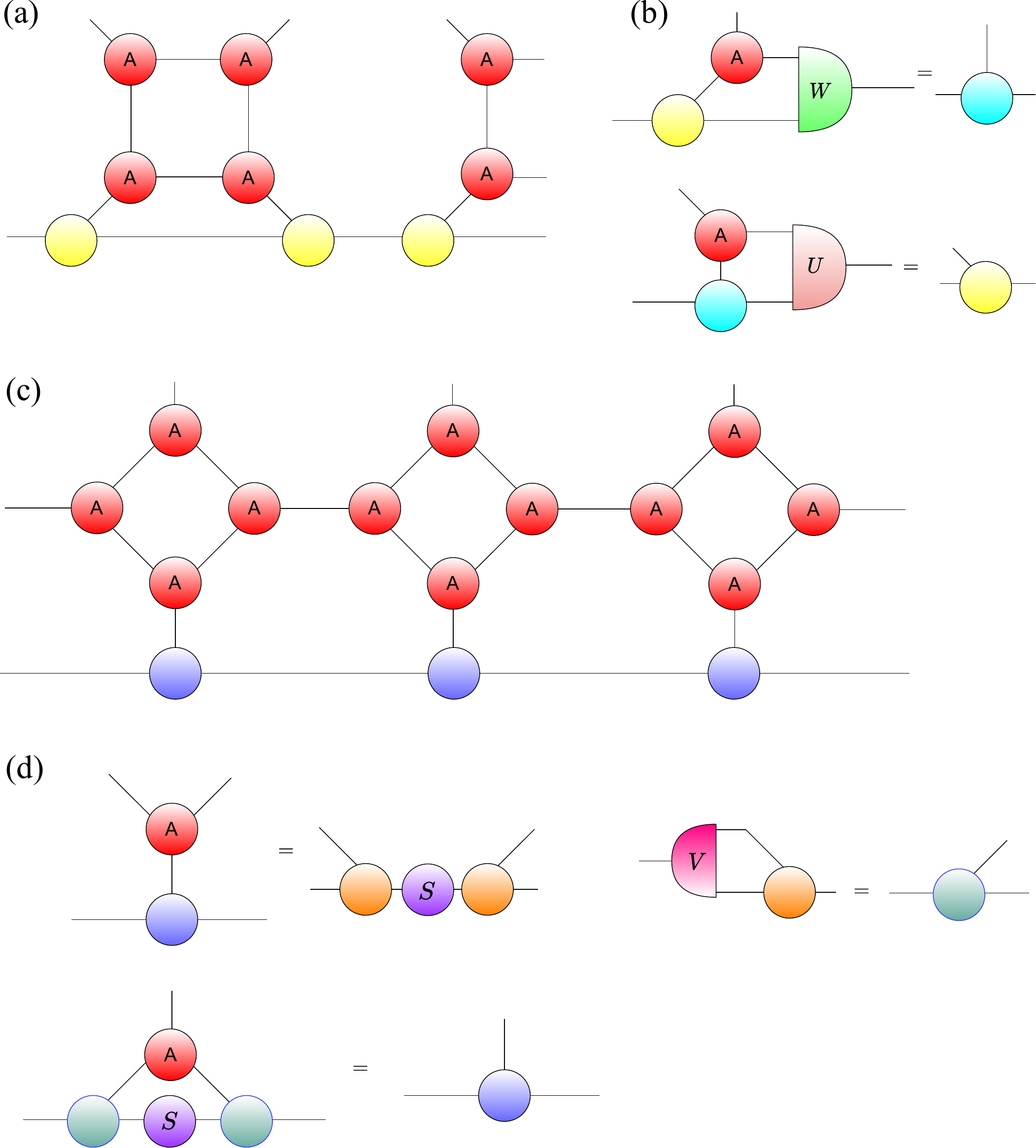} 
 \caption{\label{fig:Figure10}%
   Square-octagon lattice: (a) Definition of the first transfer matrix and the respective bMPS. (b) The local updates of bMPS tensors for the first type of transfer matrix. (c) The second type of transfer matrix on the square-octagon lattice and the corresponding bMPS. (d) The local update rules for the bMPS local tensors of the second type. }
\end{figure}

First, we assume that the tensor network consists of identical rank-3 tensors $A_{ijk}$ placed in the rotationally invariant manner on the nodes of the square-octagon lattice. We also set the tensors $A$ to be symmetric under the reflection of two last indices $A_{ijk} = A_{ikj}$, which correspond to the indices connecting tensors inside squares on the square-octagon lattice. 
We can now define the boundary MPS for this tensor network. In contrast to the previously discussed lattice geometries, the square-octagon lattice possesses two different kinds of bMPS, which are shown in Figs.~\ref{fig:Figure10}(a) and \ref{fig:Figure10}(c), respectively. These bMPS can intersect at the angle $\pi/4$. The local updates for the corresponding local bMPS tensors are shown in Figs.~\ref{fig:Figure10}(b) and \ref{fig:Figure10}(d). Here, we introduce three different isometrical projectors $U$, $W$, and $V$, and also a factorization step in Fig.~\ref{fig:Figure10}(d).

\begin{figure}
\includegraphics[width= \linewidth]{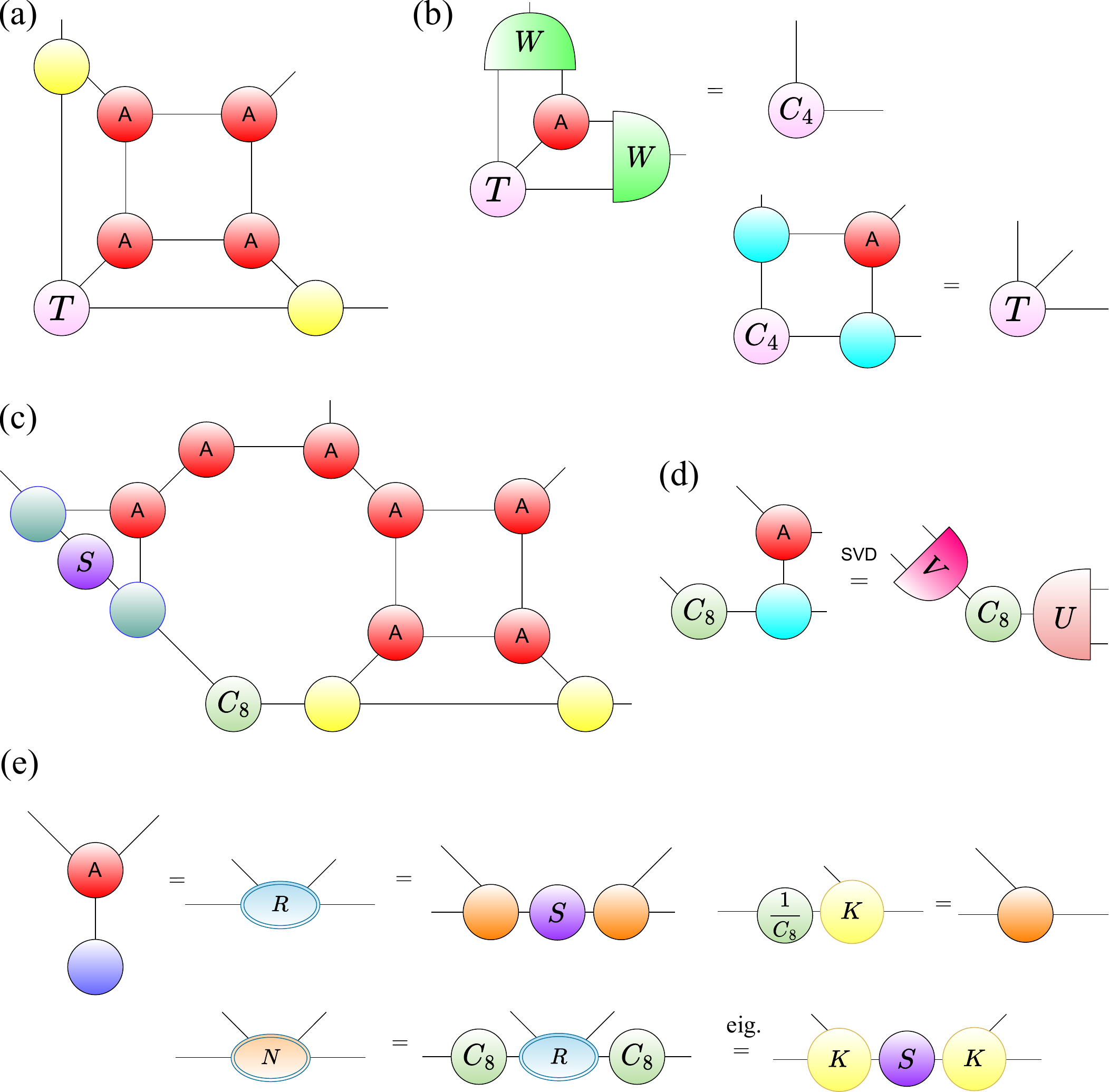} 
 \caption{\label{fig:Figure11}%
   Square-octagon lattice: (a) Intersection point of two bMPS of the first kind defines the corner tensor $T$ (symmetric upon reflections). (b) The update rule for the corner tensor $T$ and the corner matrix $C_{4}$. The isometry $W$ can be chosen to diagonalize the corner matrix $C_{4}$. (c) The corner matrix $C_{8}$ must be inserted at the intersection point of two different types of bMPS. (d) The update step of the corner matrix $C_{8}$ enlarges its dimensions, which can be then truncated back with the SVD decomposition, which also defines the projectors $U$ and $V$. (e) The factorization step of the update for the second bMPS. The factorization is performed in the same way as for the kagome lattice.}
\end{figure}

Similarly to other lattice geometries, we now introduce the corner matrices, which allow us to find the projectors $V$, $U$, and $W$. Let us take the intersection of the two bMPS of the first kind, which is shown in Fig.~\ref{fig:Figure11}(a) and results in the reflection-symmetric corner tensor~$T$. In Fig.~\ref{fig:Figure11}(b) we show the update steps for this tensor $T$, which is transformed into the corner matrix $C_{4}$. The latter matrix is internal to the squares of the square-octagon lattice and the density matrix for the square link can be defined as $\rho_{sq} = C_{4}^{4}$. Hence, we choose the isometry $W$ to diagonalize the matrix $C_{4}$ and truncate the index according to the magnitude of its eigenvalue spectrum. In Fig.~\ref{fig:Figure11}(b) we also show how the corner tensor $T$ can be obtained back from the corner matrix $C_{4}$. 

Next, let us discuss the intersection point of two bMPS of different types. This type of intersection is shown in Fig.~\ref{fig:Figure11}(c) and corresponds to the $C_{8}$ corner matrix (this is the corner matrix of the octagon angles). The update step of this matrix is shown in Fig.~\ref{fig:Figure11}(d). Note that after the update step, the corner matrix $C_{8}$ is not necessarily symmetric, hence, the simple eigendecomposition does not work. This is rather natural, since the corners of octagons in the square-octagon lattice are not symmetric under reflections. 

Let us specify how the density matrices connected by the internal links of octagons can be defined in terms of the matrices $C_{8}$. We have two types of links: those, which are shared between two octagons, and the links between an octagon and a neighbor square. These links correspond to two types of the density matrix. For the octagon-octagon link, the density matrix has a form $\rho_{o-o} = (C_{8} C_{8}^{T})^{4}$, while for the octagon-square links it is $\rho_{o-s} = (C_{8}^{T} C_{8})^{4}$. Both these matrices are symmetric and can be simultaneously diagonalized by the SVD of $C_{8}$, which is shown in Fig.~\ref{fig:Figure11}(d). 
This SVD naturally defines the isometric projectors $U$ and $V$, which are used in the bMPS tensor updates. 

Finally, we discuss the factorization step. It can be performed in nearly the same way, as in the case of the kagome lattice. The details of the factorization are shown in Fig.~\ref{fig:Figure11}(e). The matrix $R$ to be factorized is transformed into the matrix $N$ with two corner matrices $C_{8}$. The new matrix $N$ then naturally corresponds to the density matrix internal to the square $\rho \propto N^{4}$, and can be diagonalized and truncated according to its spectrum magnitude. In the final step, we apply the inverse of the diagonal matrix $C_{8}$.

As a final remark, let us note the following. First, in principle, we can perform only the iteration steps, which correspond to the bMPS of the first kind, since all the necessary projectors can be obtained from the tensors of the first bMPS. The second bMPS is auxiliary, but it can be used to compute certain observables more naturally. A situation with the auxiliary bMPS, which is not necessary for the performance of the algorithm, is a common trait of several lattices studied below. Note that the auxiliary bMPS may be not necessary for the simplest case of completely symmetric minimal unit cell discussed in this study; the larger unit cells will make it completely necessary to use both bMPS. Second, the factorization step here does not necessarily need the absorption of the eigenvalue spectrum $S$ into the bMPS tensors. This allows the method to work even when the matrix $S$ is not positive definite. Hence, the CTMRG on the square-octagon lattice is not plagued by the problems connected with positivity and can be applied to arbitrary bulk tensors~$A$, which respect the symmetry conditions.

\subsection{Star lattice}

The star lattice also has two different bMPS, as is the case for the square-octagon lattice. It is interesting that both bMPS can be converged independently from each other, but the simplest algorithm still couples them. First, we describe the main bMPS update, which is original for the star lattice, and then add the second bMPS (analogously to the honeycomb lattice) at the end of the subsection. 
\begin{figure}
\includegraphics[width= \linewidth]{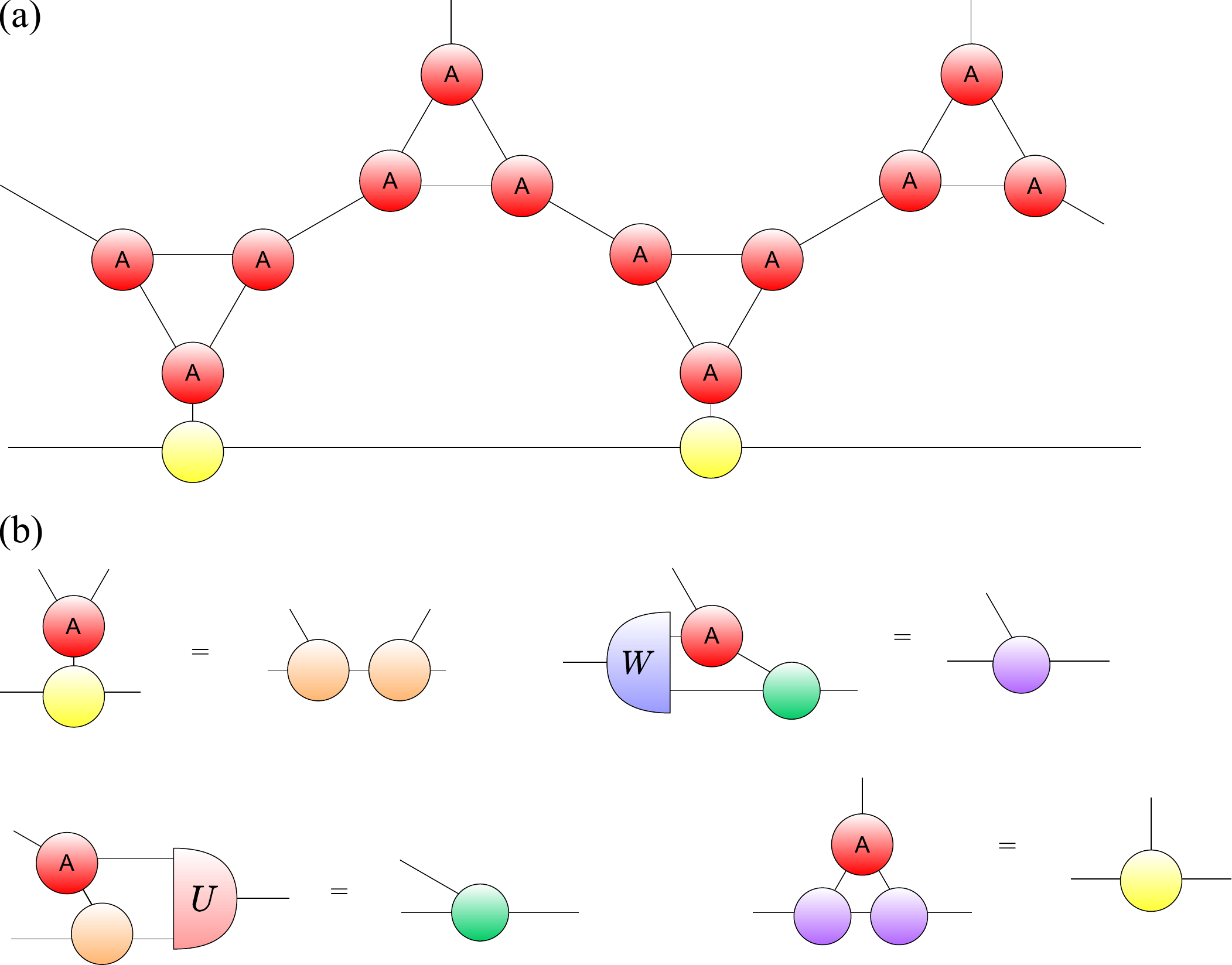} 
 \caption{\label{fig:Figure12}%
   Star lattice: (a) Definition of the first transfer matrix on the star lattice and the respective bMPS. (b) The local updates of bMPS tensors of the first type of the transfer matrix. }
\end{figure}

The structure of the transfer matrix and the corresponding bMPS are shown in Fig.~\ref{fig:Figure12}(a). The transfer matrix is considerably larger than the previously discussed cases. Hence, the local bMPS tensors have a larger number of update steps, which are illustrated in Fig.~\ref{fig:Figure12}(b). Here, we have a factorization step and two different isometric projectors $U$ and $V$. As in the previous cases, these projectors can be obtained from the corner matrices.  

\begin{figure}
\includegraphics[width= \linewidth]{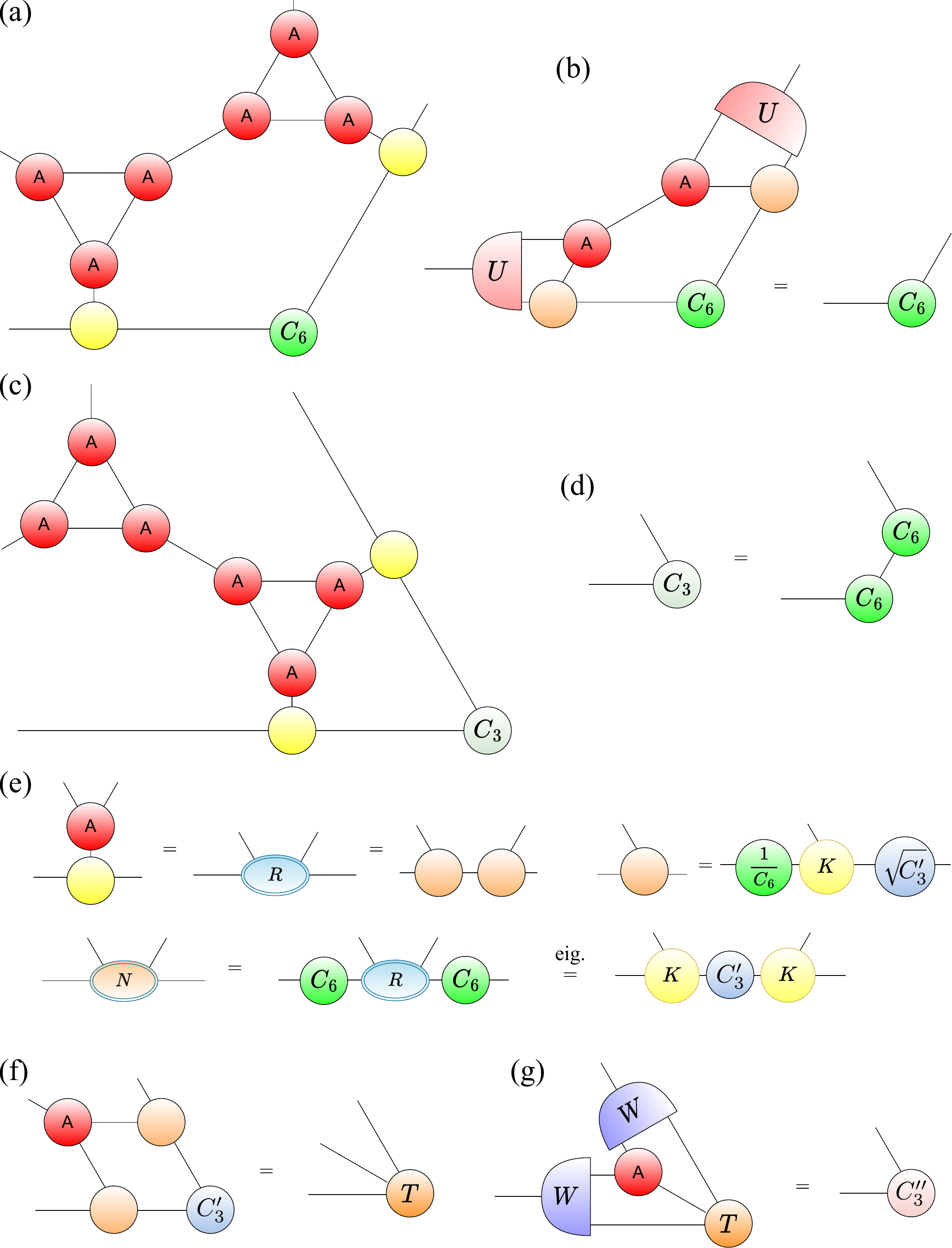} 
 \caption{\label{fig:Figure13}%
   Star lattice: (a) Definition of the corner matrix $C_{6}$ as an intersection point between two bMPS with the angle $\pi/3$. The density matrix on the dodecahedrons can be cast in the form $\rho_{d} = C_{6}^{6}$. (b) The update step of the corner matrix $C_{6}$. (c) The definition of the corner matrix $C_{3}$ as an intersection point between two bMPS with the angle $2\pi/3$. (d) The consistency condition between the corner matrices $C_{6}$ and $C_{3}$, similar to the kagome lattice. (e) The factorization step in the bMPS tensors update determining the matrix ${C}'_{3}$, which describes the internal correlations inside the triangles. (f) The update step for the matrix ${C}'_{3}$. (g) The final step of the update of ${C}'_{3}$, which results in the corner matrix ${C}''_{3}$. The isometrical projector $W$ is chosen to diagonalize the new corner matrix.
   }
\end{figure}

In principle, we can already introduce the corner matrix $C_{12}$, which appears at the intersection of two different types of bMPS. Still, if we restrict ourselves to the update of only one type of bMPS, then it is more natural to introduce only $C_{6}$ and various corner matrices $C_{3}$. The corner matrix $C_{6}$ appears at the intersection point of two bMPS with the angle $\pi/3$, as in Fig.~\ref{fig:Figure13}(a). The update rule for the corner matrix $C_{6}$ is shown in Fig.~\ref{fig:Figure13}(b). This step defines the isometrical projectors $U$, which can be chosen to diagonalize and truncate the new corner matrix~$C_{6}$.

Next, we introduce the corner matrix $C_{3}$, which is shown in Fig.~\ref{fig:Figure13}(c). Its update rules are very similar to the kagome lattice. First, as in the case of the kagome lattice, the consistency condition between the matrices $C_{6}$ and $C_{3}$ inside the dodecahedrons must hold. This consistency condition is illustrated in Fig.~\ref{fig:Figure13}(d). This natural decomposition of the corner matrix $C_{3}$ allows us to perform the factorization step in the same way, as for the kagome lattice: we show this in Fig.~\ref{fig:Figure13}(e). The factorization also results in the new corner matrix ${C}'_{3}$, which corresponds to the down-directed triangles. 

The following update steps are shown in Figs.~\ref{fig:Figure13}(f) and \ref{fig:Figure13}(g) and result in the additional corner matrix ${C}''_{3}$, which corresponds to the up-directed triangles. The diagonalization and truncation according to the eigenvalue spectrum magnitude of this matrix give us the last isometric projector $W$. One can also note an additional update step, which maps the matrix ${C}''_{3}$ back into the matrix $C_{3}$. However, we do not perform this step in the algorithm, since it can break the consistency condition in Fig.~\ref{fig:Figure13}(d) during the algorithm convergence. Instead, we use this mapping as a consistency check between the converged values of the tensors.

\begin{figure}
\includegraphics[width= \linewidth]{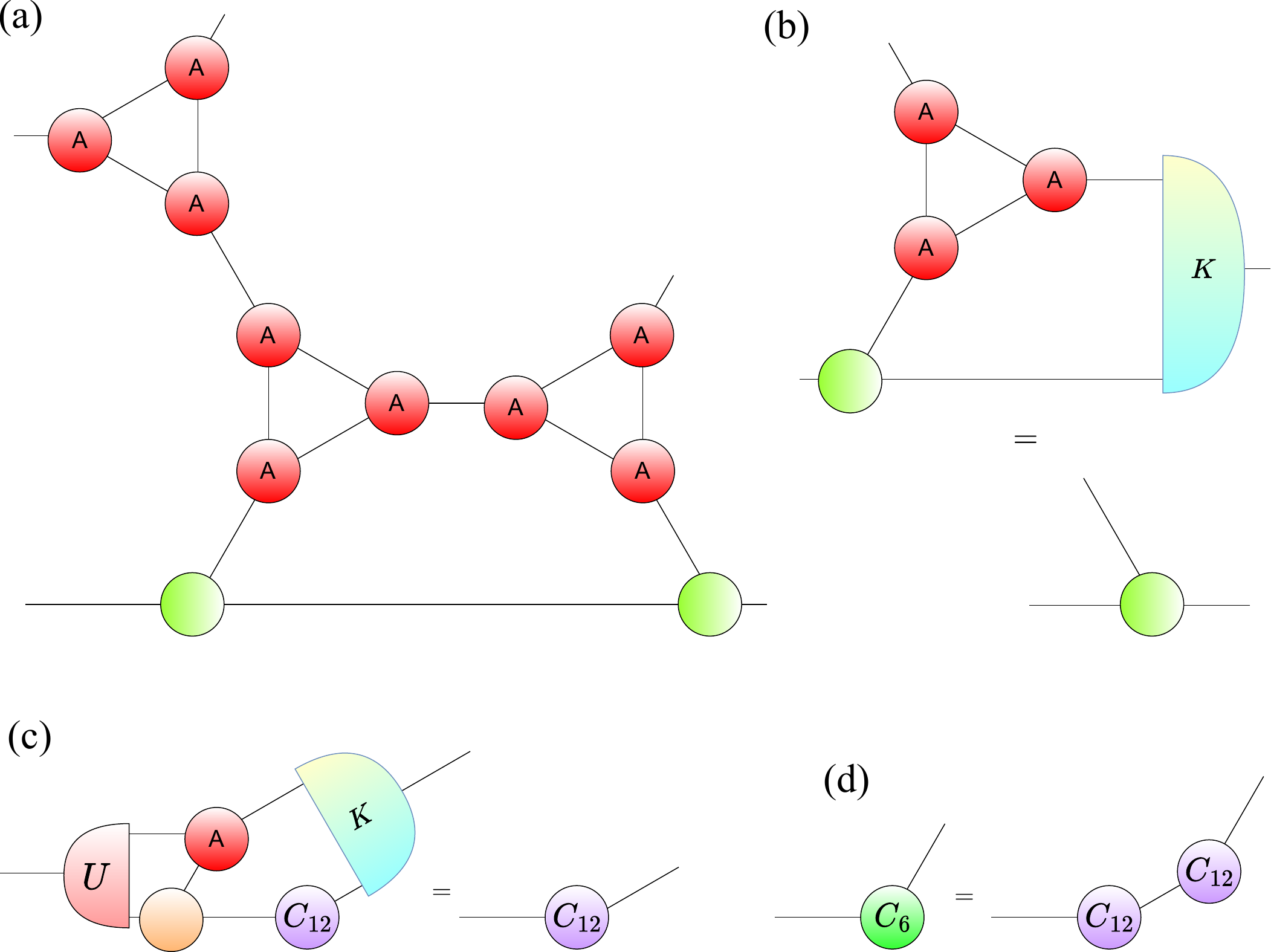} 
 \caption{\label{fig:FigureAddStar}%
   Star lattice: (a) Definition of the second bMPS. (b) Update rule for the boundary tensor, which requires the new isometric projector $K$. (c) The projector can be determined from the new corner matrix $C_{12}$, which is a square root of the previously introduced corner matrix $C_{6}$. Its update rule can be viewed as a modification of the rule in Fig.~\ref{fig:Figure12}(b) , but results here in two different isometries by using SVD of the enlarged matrix $C_{12}$. }
\end{figure}

Let us now discuss the second bMPS.  Its boundary MPS is shown in Fig.~\ref{fig:FigureAddStar}(a). Note that these transfer matrices and bMPS largely mimic the analogous quantities on the honeycomb lattice. We show the update rule for bMPS in Fig.~\ref{fig:FigureAddStar}(b) and note that it is the same as the one for the honeycomb lattice. It should be mentioned that we absorb here all three tensors $A$ within a single operation, since it is more computationally efficient to first contract these three tensors together and then apply the resulting contraction to the boundary tensor.  

Here, we also introduce the isometric projector $K$. To find this projector, we can introduce the corner matrix $C_{12}$, which must be placed on the intersections of the two different kinds of bMPS. The previously defined $C_{6}$ corner matrix was its square. To find the update rule for the new corner matrix $C_{12}$, we modify the rule in Fig.~\ref{fig:Figure12}(b) and replace the eigenvalue decomposition with SVD, which now results in the two isometric projectors $U$ and $K$ simultaneously.

We should also mention the following. One can just use the second bMPS, while the projector can be equivalently determined from the update rule on the honeycomb lattice. This can be beneficial in some situations, since the second bMPS does not contain factorization steps, thus it can be employed without any positivity restrictions. 

\subsection{4-6-12 lattice}
The square-hexagon-dodecahedron (SHD) or 4-6-12 lattice is more complex than the previously studied lattices. It consists of three different polygons, which, in principle, require their own corner matrices. It also has two different types of bMPS. Additionally, the main branch of the bMPS update requires a nonsymmetric factorization step, which we discuss below in detail. However, many other lattices can be viewed as the limiting cases of the 4-6-12 lattice. 
For example, the CTMRG algorithm on the ruby lattice, to be described in the next subsection, is naturally obtained from the CTMRG on the 4-6-12 lattice with a few modifications. 
\begin{figure}
\includegraphics[width= \linewidth]{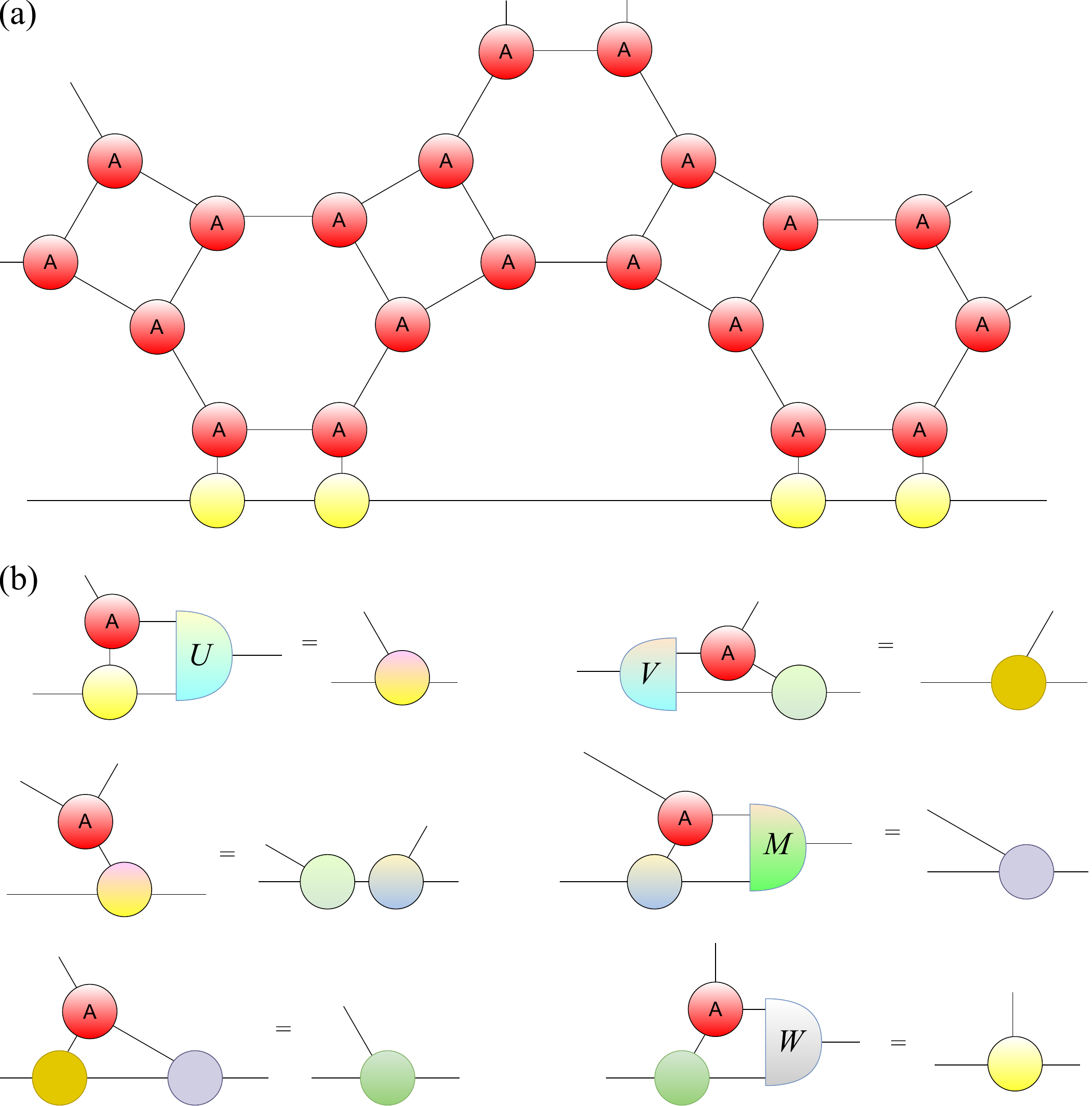} 
 \caption{\label{fig:Figure14}%
   SHD lattice: (a) Definition of the first transfer matrix on the 4-6-12 lattice and the respective bMPS. (b) The local updates of bMPS tensors for the first type of the transfer matrix. }
\end{figure}

We show the transfer matrix for the 4-6-12 lattice and the corresponding bMPS in Fig.~\ref{fig:Figure14}(a). The transfer matrix is much more complex than the previously studied cases. The local update rules are shown in Fig.~\ref{fig:Figure14}(b). These updated rules require four different isometric projectors and a nonsymmetric factorization step. The 'nonsymmetric' means that the rank-4 tensor is factorized into the product of two different rank-3 tensors. This factorization can be performed exactly at the cost of the enlargement of the factorized index dimension. Another option is to truncate this factorized index by means of a certain projector. We discuss the truncation procedure below.

\begin{figure*}
\includegraphics[width= \textwidth]{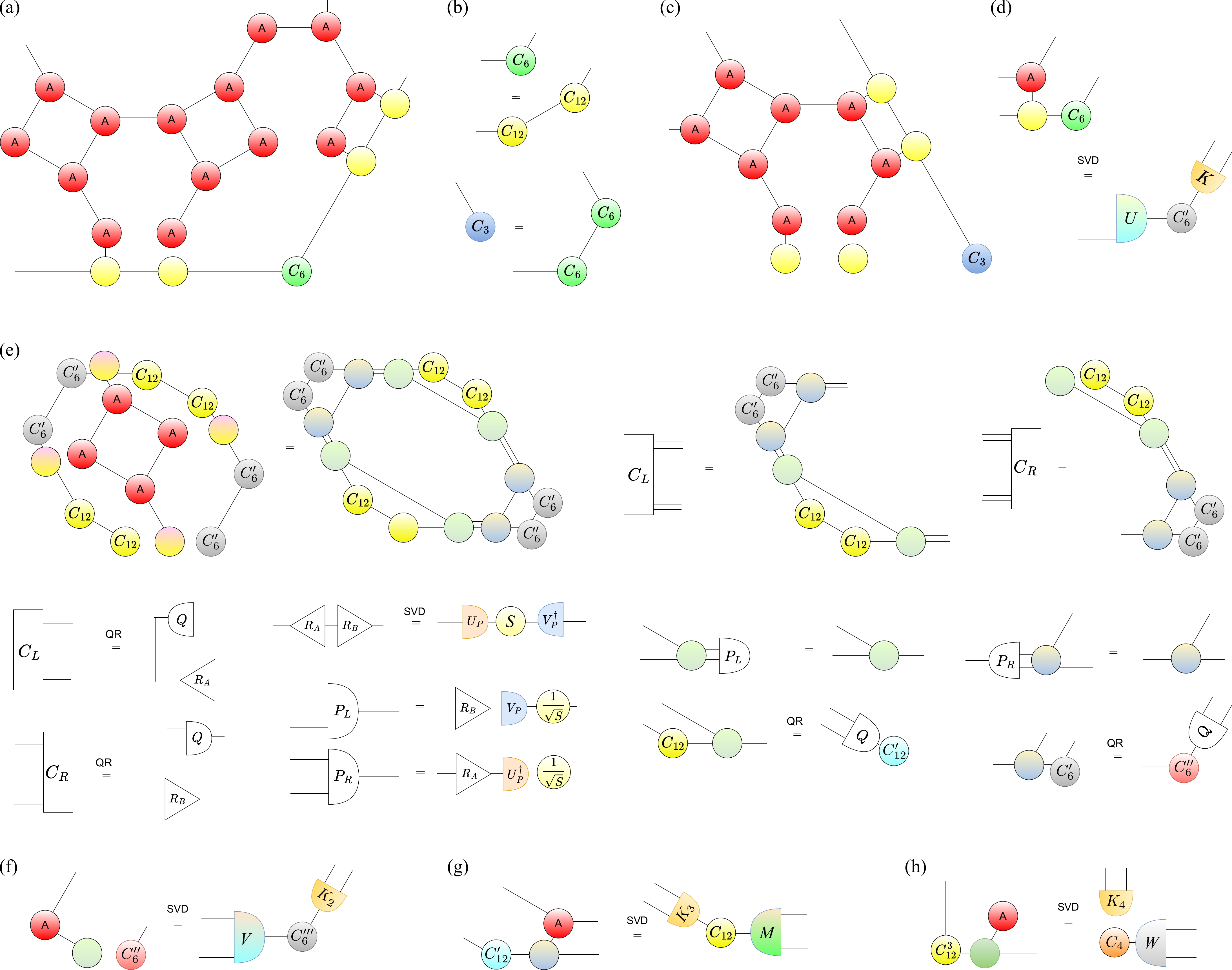} 
 \caption{\label{fig:Figure15}%
   SHD lattice: (a) Definition of the corner matrix $C_{6}$ as an intersection point between two bMPS with the angle $\pi/3$.  (b) We use a representation of the corner matrix $C_{6}$ as a square of the corner matrix $C_{12}$, which can be defined as an intersection of two different types of bMPS. Here, we use $C_{12}$ as a conventional tool in the calculations. We also define the corner matrix $C_{3}$ as a square of the corner matrix $C_{6}$. (c) The illustration of the role of the corner matrix $C_{3}$ as an intersection of two bMPS. (d) We do not work with the matrix $C_{3}$ explicitly, but use its implicit representation in terms of the matrices $C_{6}$. This allows us to find the first isometric projector $U$ and the updated corner matrix ${C}'_{6}$ through SVD of the enlarged matrix $C_{6}$. Note that the isometry $K$ in SVD is auxiliary in this version of the algorithm; we do not employ it hereafter, as all other isometries $K$. (e) The illustration of the nonsymmetric factorization step. First, we can express full lattice in terms of corner matrices and bMPS tensors. Then, we can use exact factorization of the rank-4 tensor, where we do not perform any truncation. The enlarged index is shown with two lines. We express this density matrix, which represents the full lattice as a product of two matrices $C_{L}$ and $C_{R}$. To obtain the projectors, we apply the biorthogonalization procedure to the tensors $C_{L}$ and $C_{R}$. The obtained projectors are then applied to truncate the enlarged factorized indices. We can also update the matrices $C_{12}$ and ${C}'_{6}$. The update is carried out with QR decompositions. Note that the new matrices $C''_{6}$ and ${C}'_{12}$ are not symmetric or diagonal. (f) Another step of the matrix $C''_{6}$ update, which results in the second isometry $V$ and another corner matrix $C'''_3$. (g) Analogous update step for the matrix~${C}'_{12}$ gives back the matrix $C_{12}$ and also the isometrical projector $M$. (h) To obtain the final projector $W$, we construct the corner matrix $C_{4}$ of the squares in the 4-6-12 lattice from the bMPS tensors and the third power of the matrix~$C_{12}$. The SVD of this enlarged corner allows us to obtain the last projector $W$ and also the matrix $C_{4}$, which corresponds to correlations internal to the squares.  
   }
\end{figure*}

In Fig.~\ref{fig:Figure15}(a) we determine the first necessary element of CTMRG: the corner matrices $C_{6}$, which are internal to dodecahedrons. Below, we also obtain the matrices $C_{6}$, which are internal to two types of hexagons and to squares. 
We can define the second type of bMPS for the 4-6-12 lattice. This second bMPS will have the angle $\pi/6$ with the first type of bMPS. On their intersection, it is possible to define the corner matrices $C_{12}$. We do not describe the update iterations for the second type of bMPS in this subsection, but we introduce the corner matrix $C_{12}$ as a convenient tool. Its connection to the matrix $C_{6}$ is shown in Fig.~\ref{fig:Figure15}(b). 

In Fig.~\ref{fig:Figure15}(b) we also introduce the corner matrix $C_{3} = C_{6}^{2}$. Its geometrical meaning is explained in Fig.~\ref{fig:Figure15}(c). Next, we derive the projectors. The first step is to update the matrix~$C_{6}$, as shown in Fig.~\ref{fig:Figure15}(d). After the dimension increase of the matrix $C_{6}$, we perform SVD to truncate it and obtain the matrix ${C}'_{6}$ and also the first isometric projector $U$. To understand why SVD is the appropriate procedure for the truncation, we note that the enlarged matrix ${C}'_{6}$ corresponds to the nonsymmetric angles of the first kind of hexagons. Its truncation must be performed along the same lines, as we described for the square-octagon lattice. 

Next, let us perform the nonsymmetric factorization. First, we can employ the arbitrary factorization method to factorize the rank-4 tensor into the product of two rank-3 tensors. In this exact factorization, a new factorized index emerges, which has, in general, rather large bond dimension. We can deal with this large dimension in two different ways. If the tensor network to be contracted has a small internal index dimension (e.g., for the Ising models $D=2$), then we can use arbitrary exact factorization and proceed without a truncation, since the factorized index dimension does not grow too much (and it is truncated back later using the projectors). The arbitrariness of the decomposition is only a gauge freedom in this factorized index, which does not influence the succeeding calculations. In this case, we can just proceed to the final update steps for the corner matrices at the end of Fig.~\ref{fig:Figure15}(e). 

We can now describe how to perform the truncation of the factorized index. To this end, we write the full lattice contraction in terms of the corner matrices and factorized rank-3 tensors, as shown in Fig.~\ref{fig:Figure15}(e). This construction can be cast in the form of multiplication of two matrices $C_{L} C_{R}$, which together can be viewed as a density matrix. Unfortunately, this matrix is not symmetric, hence, we cannot truncate it using simple eigenvalues. We choose instead to use the biorthogonalization procedure from Ref.~\cite{CTMRG_for_iPEPS_2}. Biorthogonalization steps applied to $C_{L}$ and $C_{R}$ are also illustrated in Fig.~\ref{fig:Figure15}(e). Biorthogonalization results in projectors $P_{L}$ and $P_{R}$, which can be used to truncate the factorized index. After the truncation, we can use the truncated rank-3 tensors to update the matrices $C_{12}$ and ${C}'_{6}$. Note that the update uses QR decomposition and results in nonsymmetric 
new matrices ${C}'_{12}$ and $ C''_{6}$. These matrices represent the correlations inside the squares of the 4-6-12 lattice. By updating the matrices ${C}'_{12}$ and $ C''_{6}$, as shown in Fig.~\ref{fig:Figure15}(f) and \ref{fig:Figure15}(g), we obtain another two isometric projectors $V$ and $M$ and also arrive back at the dodecahedron corner matrix $C_{12}$. In this process, we also obtain the corner matrix $C'''_{6}$, which represents correlations in the second type of hexagons. 

Finally, we determine the last isometrical projector $W$. This projector corresponds to the squares of the 4-6-12 lattice. In this sense, one naturally obtains it from the update rule of the corner matrix $C_{4}$, which appears at the intersection of two types of bMPS. Since we describe here the algorithm with only one type of bMPS, the matrix $C_{4}$ does not naturally appear in our calculations. To obtain the projector, we use a trick and represent $C_{4}$ inside the dodecahedron as its self-consistency condition $C_{4} = C_{12}^{3}$ (note that all corner matrices inside the dodecahedron can be self-consistently constructed from the matrix $C_{12}$). As we show in Fig.~\ref{fig:Figure15}(h), one can write an update step for the matrix $C_{4}$ and truncate it using SVD (SVD must be applied, since the angles of the squares in the 4-6-12 lattice are not symmetric under reflections). SVD results in the new diagonal matrix $C_{4}$ (internal to the squares, but not to the dodecahedrons) and also in two isometries. The first isometry is precisely the projector~$W$. The second one is not employed in our version of the algorithm, but it is necessary for the update of the second type bMPS, if included in the iteration. 

\subsection{Ruby lattice}

In this subsection, we discuss the tensor network contraction on the ruby (also, bounce or 3-4-6) lattice with rank-4 tensor $A$ placed on all the nodes of the lattice. We require from the tensor $A$ only one reflection symmetry. As we show below, the tensor $A$ also needs to be positive with respect to factorization across the reflection axis. 

The ruby lattice can be obtained from the 4-6-12 lattice by contracting certain edges of hexagons. As a result, the CTMRG on the ruby lattice is similar to the 4-6-12 lattice. To highlight the similarity of the two algorithms, we use the same conventions on the figures of analogous tensors or update steps. Note that we can explicitly map the algorithm on the ruby lattice to the 4-6-12 lattice CTMRG by employing the symmetric factorization of the rank-4 tensor $A$ on the ruby lattice. 

\begin{figure}
\includegraphics[width= \linewidth]{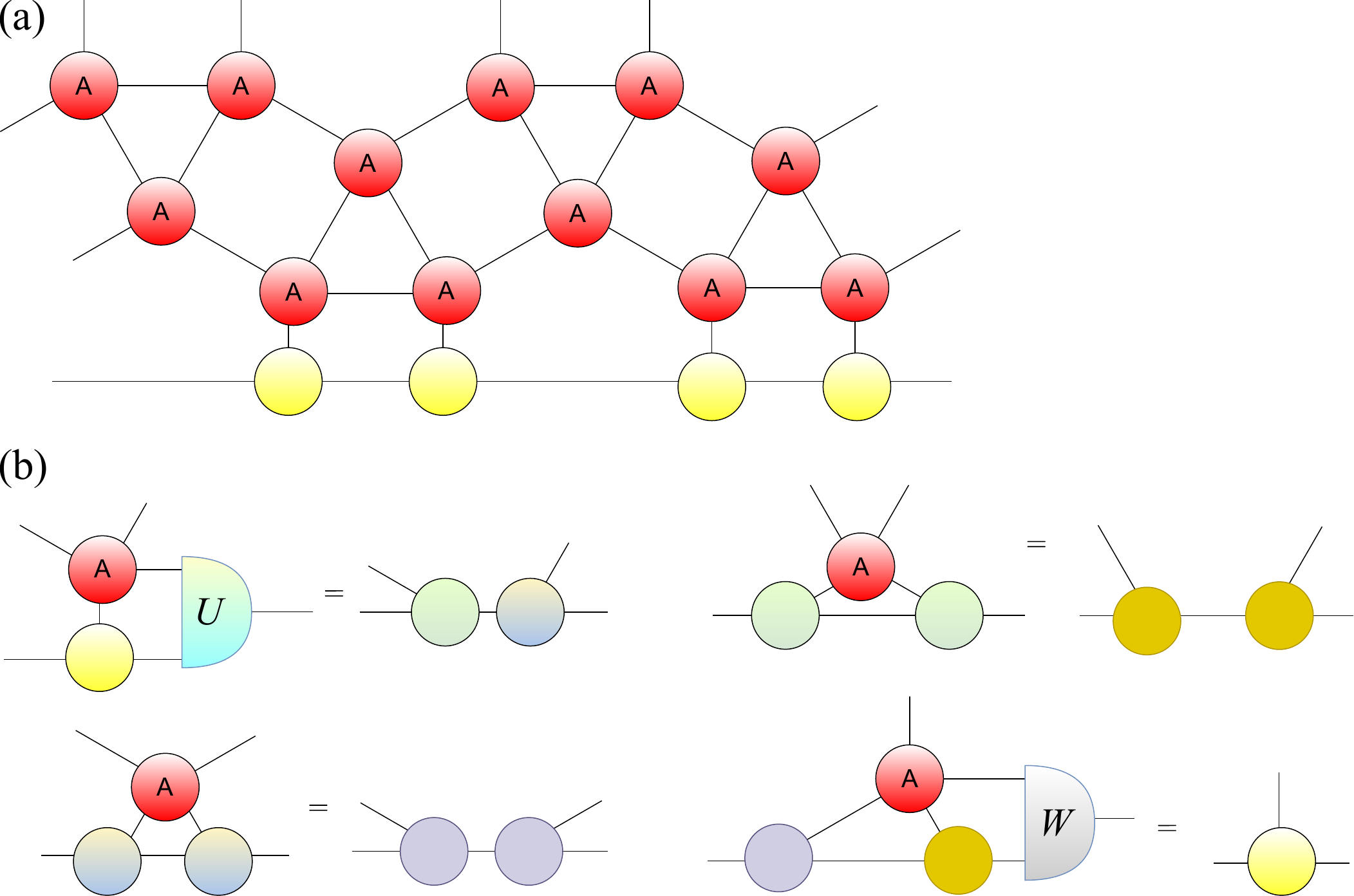} 
 \caption{\label{fig:Figure16}%
   Ruby lattice: (a) Definition of the first transfer matrix and the respective bMPS. (b) The local updates of bMPS tensors for the first type of transfer matrix. Note the similarity of the updates and conventions with Fig.~\ref{fig:Figure14}.  }
\end{figure}

The ruby-lattice transfer matrix is shown in Fig.~\ref{fig:Figure16}(a). The corresponding local update steps are shown in Fig.~\ref{fig:Figure16}(b). Compared to the 4-6-12 lattice, we find that a pair of projecting steps turn into factorization steps, and now one needs only two different isometric projectors $U$ and $W$. We also observe that the nonsymmetric factorization step remains a part of the calculation. 

\begin{figure*}
\includegraphics[width= \textwidth]{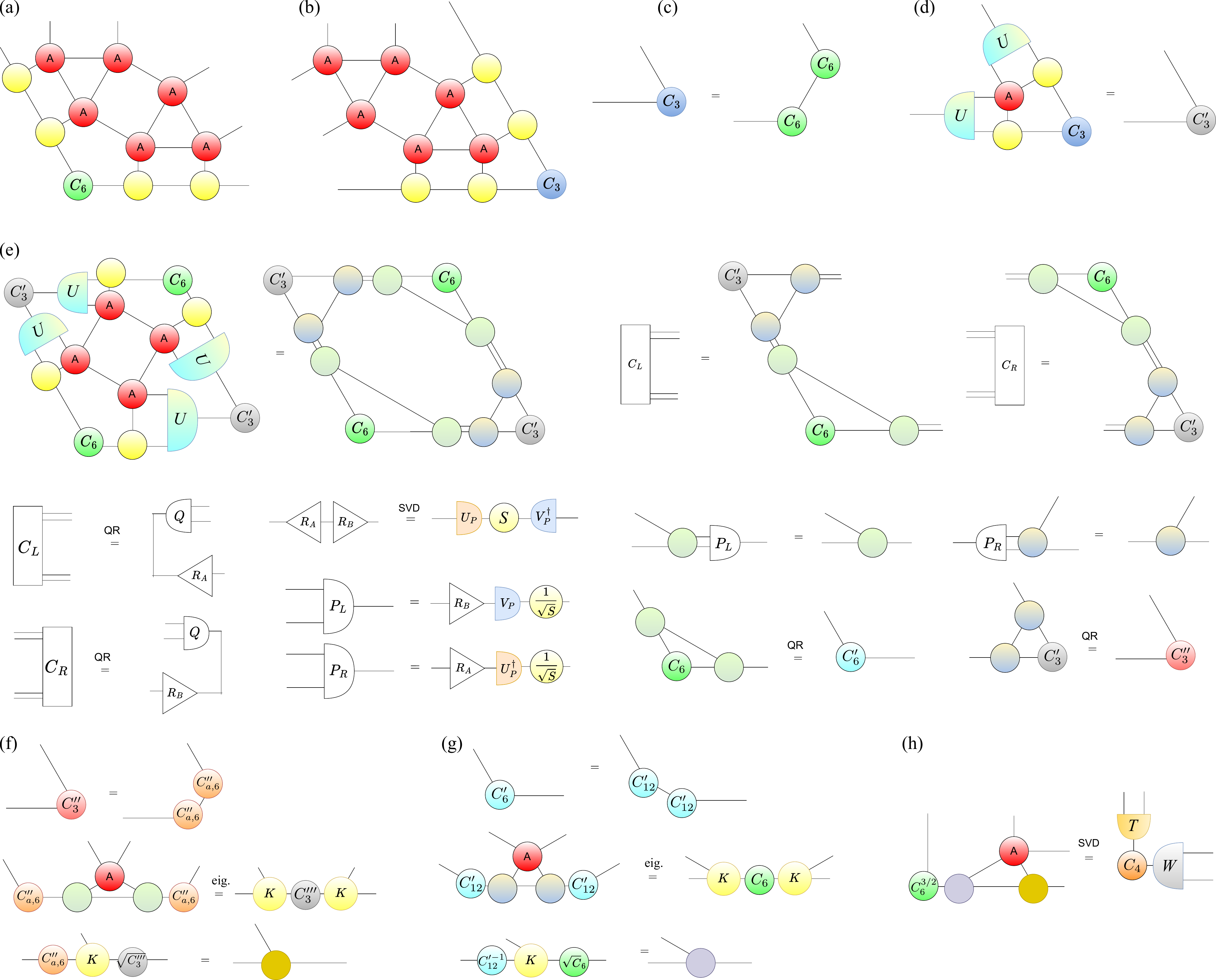} 
 \caption{\label{fig:Figure17}%
   Ruby lattice: (a) Definition of the corner matrix $C_{6}$ inside the hexagon as an intersection point between two bMPS with the angle $\pi/3$.  (b)  The definition of the corner matrix $C_{3}$ inside the hexagon as an intersection point between two bMPS with the angle $2\pi/3$. (c) The consistency condition between the corner matrices $C_{3}$ and $C_{6}$. (d) The first update step of the corner matrix $C_{3}$ allows us to find the first isometric projector $U$ and the updated corner matrix ${C}'_{3}$. (e) The illustration of the nonsymmetric factorization step. First, we can express the lattice in terms of the corner matrices and bMPS tensors. Then, we employ the exact factorization of the rank-4 tensor, where we do not perform any truncation. The enlarged index is shown with two lines. Next, we write this density matrix, which represents the lattice as a product of two matrices $C_{L}$ and $C_{R}$. To obtain the projectors we apply the biorthogonalization procedure to the tensors $C_{L}$ and $C_{R}$. The obtained projectors are then applied to truncate the enlarged factorized indices. We can also update the matrices $C_{6}$ and ${C}'_{3}$.  Note that the new matrices ${C}''_{3}$ and ${C}'_{6}$ are not diagonal. (f) The first factorization step is performed in a similar way to factorizations on different lattices. The main difference is that the matrix ${C}''_{3}$ is no longer diagonal. To proceed further, we factorize this matrix with the Cholesky decomposition. This decomposition works only for the positive matrices, which is a major limitation for this type of algorithm. After the decomposition, we can perform a factorization step. (g) Another factorization step, which is performed analogously to the previous point. The only difference is that we obtain a new $C_{6}$ matrix as a byproduct of factorization. (h) The final projector can be found from the SVD of the effective corner matrix $C_{4}$. This decomposition requires $C_{6}^{3/2}$, thus sets an additional positivity condition.   }
\end{figure*}

Let us now discuss the calculation of projectors and factorizations. To this end, we introduce the corner matrices. 
The corner matrices for the different intersections of the bMPS are shown in Figs.~\ref{fig:Figure17}(a) and \ref{fig:Figure17}(b), and the consistency condition between them is shown in Fig.~\ref{fig:Figure17}(c). Figure~\ref{fig:Figure17}(d) illustrates the update step for the $C_{3}$ corner matrix, which results in the corner matrix ${C}'_{3}$, representing the correlations inside the triangles. The isometry $U$ is chosen to diagonalize and truncate the corner matrix ${C}'_{3}$. Note that the positivity of ${C}'_{3}$ depends on the positivity of the tensor $A$ across the reflection line. 

Next, we perform the nonsymmetric factorization step, as in Fig.~\ref{fig:Figure17}(d). The procedure is the same as for the 4-6-12 lattice, thus we do not discuss it here in detail. The main difference is only the update of the corner matrices, which results in the new matrices ${C}'_{6}$ and $ {C}''_{3}$, which are symmetric, but generally not diagonal, and possibly not positive. For the positive matrices we can perform their Cholesky decomposition, as in Figs.~\ref{fig:Figure17}(f) and \ref{fig:Figure17}(g), and to use blocks from the Cholesky decomposition in the factorization steps. The factorization steps also result in the new corner matrix $C_{6}$. 

The final step is to determine the second isometric projector~$W$ from the update of the effective matrix $C_{4}$, which is partially mimicked by the power $3/2$ of the matrix $C_{6}$. This power can be computed only for the positive matrix $C_{6}$, hence, its positivity becomes a necessary condition. The process of the update for the matrix $C_{4}$ is shown in Fig.~\ref{fig:Figure17}(h).

The algorithm described above depends substantially on the positivity of the matrices $C_{6}$, $ {C}'_{6}$, and $C''_{3}$. The positivity is preserved by the majority of the update steps. The only exceptions are the update in Fig.~\ref{fig:Figure17}(d) and the factorization in Fig.~\ref{fig:Figure17}(g). If the tensor $A$ is positive with respect to symmetric factorization, then these two steps are also guaranteed to preserve positivity, and the method works. We also discuss the positivity issues in more detail in Sec.~\ref{subsec:Positivity} and in Appendix~\ref{app:A}.

\subsection{Dice lattice}

The dice lattice is an example of a non-Archimedean lattice, which has two different types of vertices. We construct a tensor network consisting of rank-6 rotationally symmetric tensors $A$ and rank-3 rotationally symmetric tensors $B$ placed on the respective nodes of the dice lattice. The algorithm requires changes with respect to previously discussed lattices, but the main idea remains the same. 

\begin{figure}
\includegraphics[width= \linewidth]{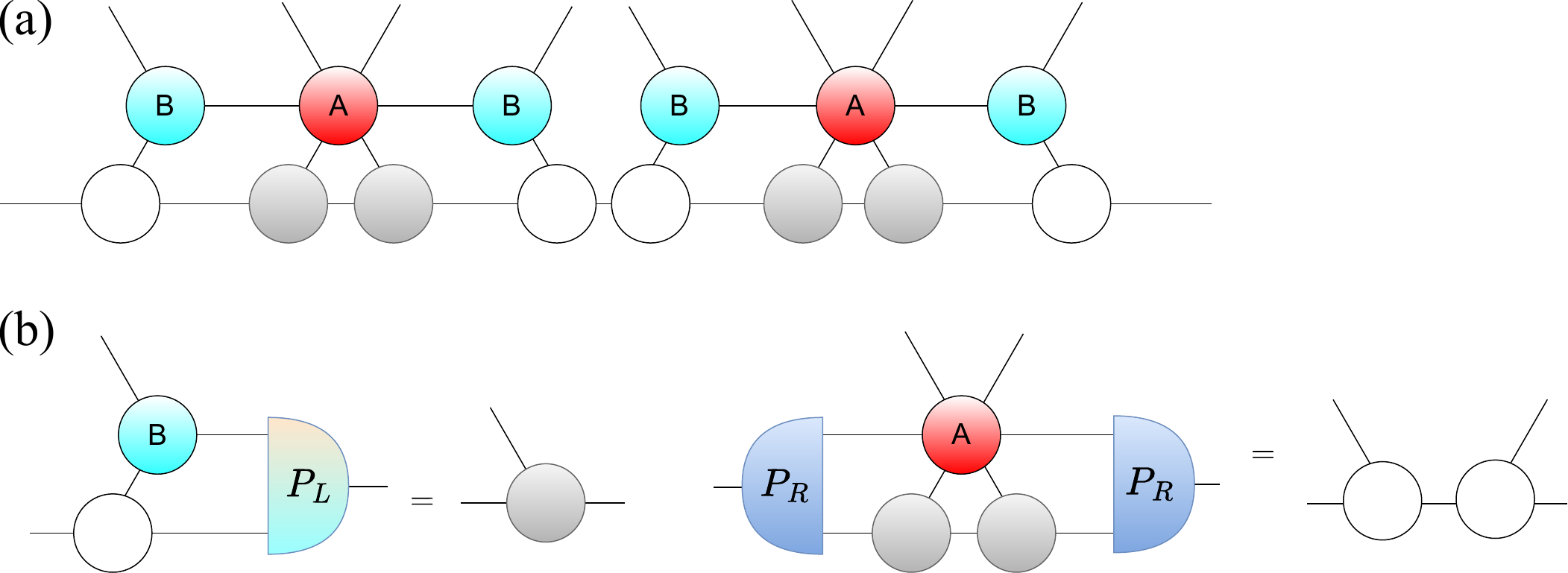} 
 \caption{\label{fig:Figure18}%
   Dice lattice: (a) Definition of the transfer matrix on the dice lattice and the respective bMPS. (b) The local updates of bMPS tensors.  }
\end{figure}

First, in Fig.~\ref{fig:Figure18}(a) we show the transfer matrix and the respective bMPS on the dice lattice. Note that the bMPS consists of two different types of tensors. This is necessary due to two different types of bulk tensors $A$ and $B$. The updates of the local tensors are shown in Fig.~\ref{fig:Figure18}(b). For the update, we need biorthogonal (not isometric) projectors $P_{L}$ and $P_{R}$ with $P_{L}P_{R} = 1$, and also to perform one symmetric factorization step. To define the projectors we introduce the corner matrices and the respective updates. 

\begin{figure*}
\includegraphics[width= \linewidth]{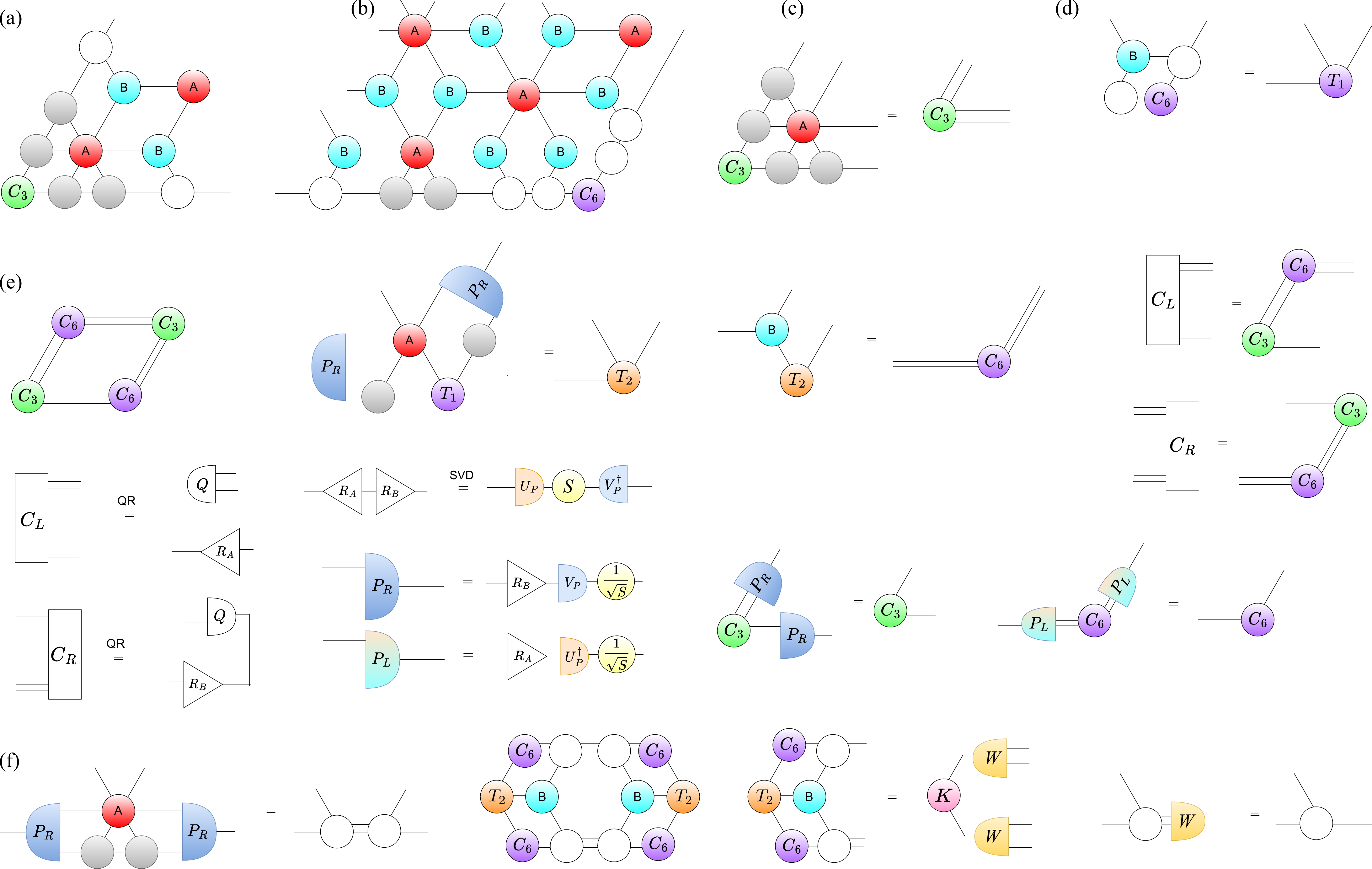} 
 \caption{\label{fig:Figure19}%
   Dice lattice: (a) Definition of the corner matrix $C_{3}$. (b) Definition of the corner matrix $C_{6}$. (c) The growth step for the corner matrix $C_{3}$. (d) The update steps for the corner matrix $C_{6}$, which result in the new matrix $C_{6}$ with a larger dimension. (e) The contracted tensor network in terms of the larger matrices $C_{3}$ and $C_{6}$, which can be written as a product of the matrices $C_{L}$ and $C_{R}$. We apply the biorthogonalization to $C_{L}$ and $C_{R}$ to obtain the projectors $P_{L}$ and $P_{R}$. (f) The symmetric factorization step. Initially, we perform a factorization without truncating the factorized index (indicated with two lines). We can write a contracted tensor network in terms of factorized tensors and corner matrices. The index can be truncated using the isometry obtained from the eigendecomposition of the density matrix (or its half denoted as $K$).   }
\end{figure*}

The corner matrices $C_{6}$ and $C_{3}$ are shown in Figs.~\ref{fig:Figure19}(a) and \ref{fig:Figure19}(b). We do not have consistency relations between these matrices, sine the dice lattice does not consist of symmetric polygons. We introduce two separate updates for these matrices, which are shown in Figs.~\ref{fig:Figure19}(c) and \ref{fig:Figure19}(d). As a result of these updates, we obtain the enlarged corner matrices, which we truncate by means of biorthogonal projectors. To determine these projectors, we express the contracted tensor network in terms of these corner matrices $C_{3}$ and $C_{6}$, as shown in Fig.~\ref{fig:Figure19}(e). Equivalently, this contraction can be written as a product of two matrices $C_{L}$ and $C_{R}$. To find the projectors, we biorthogonalize the matrices $C_{L}$ and $C_{R}$. Finally, we perform the symmetric factorization. In Fig.~\ref{fig:Figure19}(f) we first perform an exact factorization without the truncation. The enlarged factorized index is depicted with two lines. We can now rewrite the whole tensor network in terms of these factorized tensors, corner matrices, and bulk tensors $B$. Using this representation, we take the density matrix for the enlarged factorized index and truncate the index according to its spectrum. Equivalently, one can diagonalize half of the density matrix, as is shown in Fig.~\ref{fig:Figure19}(f). This diagonalization results in the isometry $W$, which can be used to truncate the factorized index. 

\subsection{Remarks on the positivity and other lattices}\label{subsec:Positivity}

In the introduced CTMRG approaches, many times we faced the necessity for the symmetric factorization step. Unfortunately, not every symmetric tensor can be decomposed symmetrically. The necessary condition for this decomposition relies on the positivity of the tensors. In particular, its eigenvalues around the symmetric bipartition must be all positive. For positive bulk tensors, which usually appear in classical statistical mechanics applications, these conditions hold automatically. For more general tensors, which appear, e.g., in the iPEPS calculations, the bulk tensors can be both positive and negative,  thus the simple versions of the described algorithms can face convergence issues. Note that the CTMRG algorithms on honeycomb, square-octagon, 4-6-12, and square lattices do not contain symmetric factorizations, hence, these can be straightforwardly applied to arbitrary bulk tensors without any positivity restrictions.

For nonpositive bulk tensors and lattices with symmetric factorizations, one can introduce certain algorithmic modifications. First, we should not assume that the bMPS consists of one type of symmetric tensor. 
Instead, we can assume now that the bMPS contains several different types of tensors. In this case, the symmetric factorizations turn into nonsymmetric factorizations, which can be performed in the same way, as we described for the 4-6-12 lattice. We discuss these generalizations to nonpositive bulk tensors by taking example of triangular lattice in Appendix~\ref{app:A}. 

In this section, we studied the dice lattice and various Archimedean lattices. Still, there are three Archimedean lattices with the coordination number $z=5$, which we have not covered: maple-leaf, Shastry-Sutherland, and trellis lattices. We believe that the CTMRG approach can be extended to these lattices as well, but their natural anisotropy makes it difficult to devise a simple truncation procedure. In fact, the density matrices on these lattices are generally not symmetric, thus one needs to apply biorthogonalizations and nonsymmetric factorizations all the time.

\section{Benchmarks and results}\label{sec:Benchmarks}

\subsection{Classical lattice models and tensor networks}

Infinite tensor networks on various lattices can appear in different problems, e.g., in calculations with the wave-function variational ansatzes as iPEPS, calculations of contractions of infinite circuits, or in certain models of classical statistical mechanics on the lattice. In this section, we focus on the latter and study certain well-known statistical mechanics models on different lattices. First, we describe these models, their physics and observables, and their mappings into the tensor network problems. This mapping is not unique, since one can map the statistical mechanics model into the tensor networks on different lattices. It is useful to briefly discuss the possible transformations between certain lattices, because this allows us to study the same problem with different CTMRG algorithms and cross-check the results. 

Our main focus of interest is the classical Ising model~\cite{kramers1941statistics,onsager1944crystal, IsingIntro, mccoy1973two}. This model is formulated on the general lattice as follows: we place a ``spin'' variable $\sigma_{i}$ on every site $i$ of the lattice. This variable is classical and can take only two values: $\sigma_{i} = +1$ or $\sigma_i =- 1$.  We can now define the energy of the system as follows:
\begin{equation}
    E = -J \sum_{\langle ij \rangle} \sigma_{i} \sigma_{j} - B \sum_{i} \sigma_{i},
\end{equation}
where $\langle ij \rangle$ denotes all the nearest-neighbor pairs of sites on the lattice and the second sum is taken over all lattice sites. We set the coupling $J$ to $+1$ or to $-1$ for the ferromagnetic or antiferromagnetic Ising model, respectively.  The quantity $B$ corresponds to the amplitude of the external (magnetic) field, which is generally set to zero, but for certain lattices we also study characteristics in the nonzero field. 

Now, we can express the partition function $Z(\beta) = \sum_{\sigma_{i}} \exp{[-\beta E(\sigma)]}$, where $\beta = 1/T$ is the inverse temperature in units of $k_{\rm B}=1$, as well as the magnetization $\langle \sigma_{i} \rangle = \sum_{\sigma} \sigma_{i} \exp{[-\beta E(\sigma)]} /Z(\beta)$. 
At low temperature, the system described by the ferromagnetic Ising model undergoes phase transition to the state with a nonzero spontaneous magnetization, which corresponds to the nonzero value of $\langle \sigma_{i} \rangle$, while at higher temperature the system is in the disordered phase with $\langle \sigma_{i} \rangle=0$. 
In the antiferromagnetic models, one can also observe a phase transition between the antiferromagnetic and paramagnetic phases driven by thermal fluctuations and the external field. 

The partition function can be written as follows:
\begin{equation}
    Z(\beta) = \sum_{\sigma_{i}} \prod_{i} \exp(\beta B \sigma_{i}) \prod_{\langle ij \rangle} W_{\sigma_{i} \sigma_{j}},
\end{equation}
where $W_{\sigma_{i}\sigma_{j}} = \exp(\beta J \sigma_{i} \sigma_{j}) \equiv W_{ij}$ is the bond matrix. We can now introduce the tensor $x_{ij...n} = \exp(\beta B \sigma_{i}) \delta_{ij...n}$, where the rank of the tensor is equal to the connectivity $z$ of the lattice (if the lattice contains sites with different connectivities, then one must introduce several different tensors $x$ for each type of sites), each individual index $i,j,...,n$ takes two magnetization values $\pm1$, and $\delta_{ij...n}$ is a type of the Kronecker $\delta$ tensor, with the component equal to one, if all indices have the same value ($\sigma_i=\sigma_j=...=\sigma_n$), and zero otherwise. 

Let us add more comments on the tensor $x$. First, in the absence of the external magnetic field $B$, it reduces to the $\delta$-tensor.  Second, this tensor depends on the single value $\exp(-2\beta B)$. Below, we denote these values somewhat loosely as $x = \exp(-2\beta B)$, which can be restricted to the interval $0 < x < 1$. We assume that it is usually clear from the context, if we mean the tensor or the corresponding constant.  Besides that, we also introduce the constant $a = \exp(-2\beta)$. 

\begin{figure}
\includegraphics[width= \linewidth]{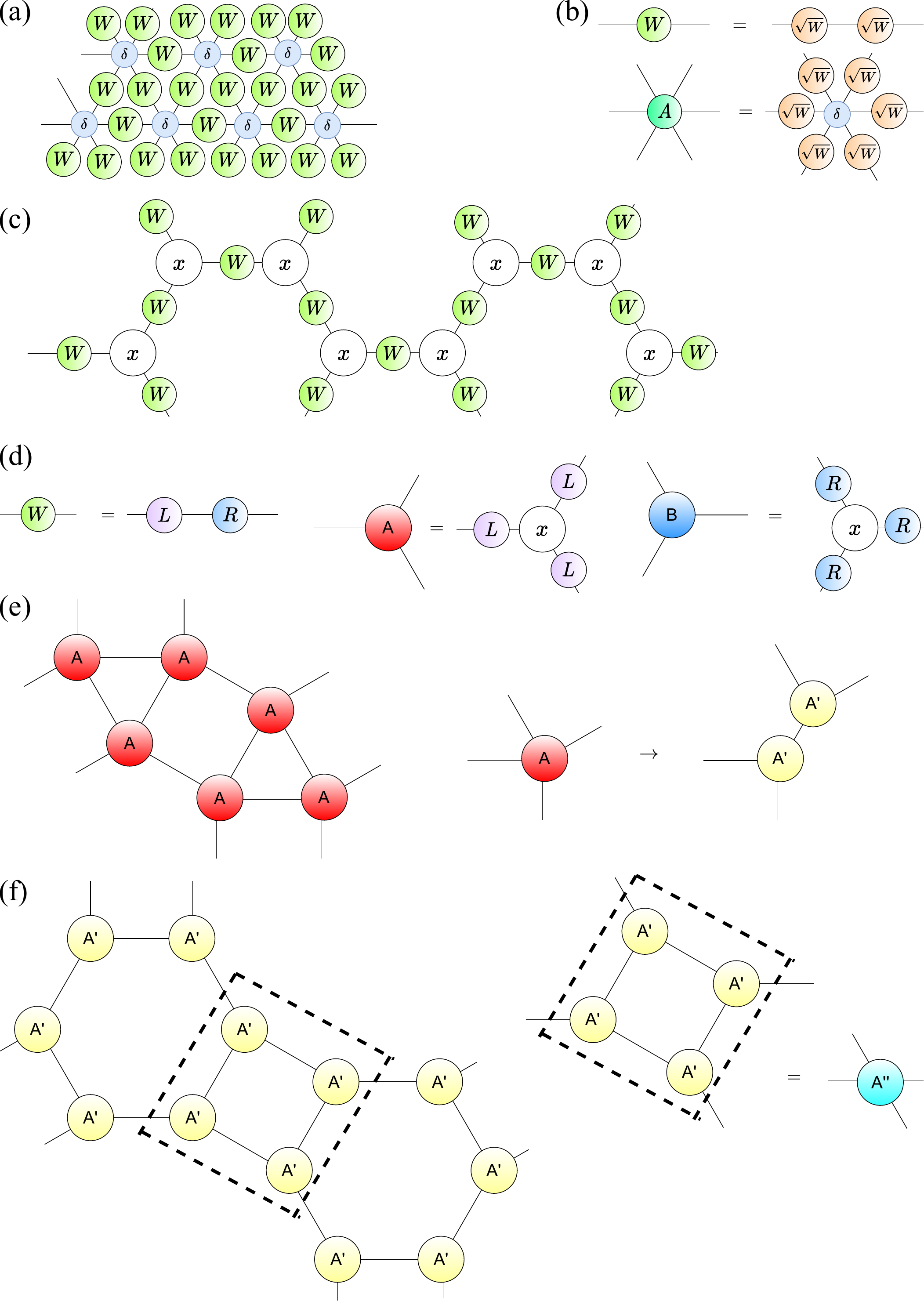} 
 \caption{\label{fig:TensorsMaps}%
   (a) The general construction of the tensor network on the lattice (here, triangular) by placing tensors $\delta$ on the sites and matrices $W$ on the bonds. (b) For the ferromagnetic models we can find the unique positive symmetric square root of the bond matrix $W$. These square roots can be absorbed into the tensors $\delta$, forming the tensor network of identical rotationally and reflection-symmetric tensors $A$.  (c) For the bipartite antiferromagnetic model in the external field, the partition function can be determined by placing identical tensors $x$ on the lattice sites and bond matrices $W$ on the links. (d) For the antiferromagnetic model we do not have a unique symmetric decomposition of the matrix $W$, but we can choose an arbitrary one (e.g., based on SVD) and then absorb the resulting matrices $L$ and $R$ into different site tensors, forming the tensor network with two-site unit cell. (e) Example of the transformation mapping the ruby lattice into the SHD one. Note that here we assume the symmetric factorization of the tensor $A$ on the ruby lattice. If this factorization is not available, then one can employ a nonsymmetric factorization, but the mapping results in the two-site unit cell on the SHD lattice. (g) The mapping from the SHD to kagome lattice.}
\end{figure}

With the introduced tensors, we can define the tensor network for the partition function $Z(\beta)$ by placing the tensors~$x$ on all lattice sites and the matrices~$W_{ij}$ on the corresponding links. This network is illustrated in Fig.~\ref{fig:TensorsMaps}(a) for the triangular lattice in the absence of the external field (i.e., we can use $\delta$-tensors instead of $x$-tensors) and in Fig.~\ref{fig:TensorsMaps}(c) for the  honeycomb-lattice Ising model in the external field. 

The introduced form of tensor network does not match exactly the ones discussed in the previous section, since it contains additional matrices $W_{ij}$ on the links of the lattice. It looks possible to generalize the CTMRG approach directly to these models, similar to the bond-type statistical mechanics models on the square lattice. However, we are interested in mapping the tensor network directly into one the forms from the previous section. 

Let us first discuss the ferromagnetic case. For the ferromagnetic model, the matrix $W$ is positive and symmetric, thus one can define the positive symmetric square root $q = \sqrt{W}$. This allows for the decomposition, which is shown in Fig.~\ref{fig:TensorsMaps}(b), and the square roots can be absorbed into the tensors $x$ (or $\delta$ without external field), transforming to the new tensor $A$. \change{This construction is also related to the Fisher superexchange Ising model~\cite{Fisher} with additional spins on the bonds of the original lattice. In this representation, the indices of the tensor~$A$ take values in these bond spins, while $\sqrt{W}$ is a new bond matrix with the redefined $\beta$. }The new tensor network consists of identical tensors $A$ on all lattice sites, where we can directly apply the algorithms from the previous section. 

For the antiferromagnetic model, we cannot define the positive square root $q$, thus we introduce an arbitrary nonsymmetric decomposition $W = L \times R$, which is shown in Fig.~\ref{fig:TensorsMaps}(d).  This decomposition can be obtained, e.g., from SVD. Then, we absorb the matrices $L$ and $R$ into different tensors $x$, as in Fig.~\ref{fig:TensorsMaps}(d). Note that this construction works only on the bipartite lattices, as the honeycomb one, where we obtain the two-site unit cell tensor network with two different tensors on different sublattices of the original lattice. For other not bipartite frustrated lattices (e.g., triangular or kagome), the more complex tensor network encoding is required, as is discussed in Refs.~\cite{vanhecke2021frustrated, colbois2022partial, song2022tensor, colbois2021artificial, song2023unified}. 

Finally, let us briefly discuss different mappings of the models into the tensor networks. It is possible to map tensor networks on different lattices into each other, as we show in Figs.~\ref{fig:TensorsMaps}(e) and \ref{fig:TensorsMaps}(f) for the cases of ruby, SHD and kagome lattices. These mappings allow cross-checking different CTMRG algorithms, since they can be applied to the same lattice model.

\subsection{Triangular lattice}

We begin the benchmark analysis from the ferromagnetic Ising model on the triangular lattice. This model is integrable and its transition temperature is exactly known. The model can be represented as a tensor network contraction on the triangular lattice, with the tensors being both completely symmetric and positive. In this subsection, we analyze several observables to determine the transition temperature and compare it with the exact values. We consider the following observables: the onsite magnetization $\langle \sigma \rangle$, which appears in the low-temperature ferromagnetic phase, and various corner Hamiltonian characteristics. Up to multiplicative normalization, the corner (or entanglement) Hamiltonian is defined as follows:
\begin{equation}
    H_{c} = - \log( C_{3}).
\end{equation}

In integrable models, these corner Hamiltonians have remarkable properties: they can be connected to the boost operators of the integrable spin chains, which are obtained in the extreme anisotropic limit of the integrable model~\cite{Corner_spectra_1,Corner_spectra_2, Orus_corner_universality, Entanglement_spectra_1, Entanglement_spectra_2}.  These boost operators, in turn, generate integrals of motion of the integrable chain and they are the basis of the recent approaches to finding the new integrable models \cite{Leew_Yang_Baxter_Boost}. The boost operator in integrable models also has integer eigenvalues. It means that the spectrum of $H_{c}$ has a form: $E = E_{0} + \Delta n$, where $E_{0}$ is a normalization-dependent factor, $n$ is integer, and $\Delta$ is an entanglement gap (or a Schmidt gap). 

Note that the Hamiltonian spectrum is degenerate, and generally there are many eigenstates that correspond to the same $n$. We can characterize the Hamiltonian with its gap $\Delta$ and its degeneracies $d(n)$. Since $d(n)$ are integer numbers, they cannot change without phase transition, where the gap $\Delta$ vanishes. As a result, we can study the phase transitions in integrable models by analyzing the gap $\Delta$~\cite{Schmidt_gap}, and characterize the phases with their respective degeneracies $d(n)$. For certain models, it was found that the degeneracies are related to the representation theory of the quantum deformed Kac-Moody algebras \cite{jimbo1991combinatorics, jimbo1994algebraic, davies1993diagonalization}, but we do not pursue this characterization here. For the Ising model on the square or triangular lattices, the spectra are known exactly.  These have free fermionic form: $E = \sum_{l} \epsilon_{l} n_{l}$, where $n_{l} = 0, 1$, and $\epsilon_{l}$ can be obtained according to the following rule~\cite{Corner_spectra_2}:
\begin{equation}
    \epsilon_{l} = (2l+1)\Delta,
\end{equation}
for the disordered phase, where $l = 0,1,2,3,\ldots$, and for the ordered phase
\begin{equation}
    \epsilon_{l} = l\Delta,
\end{equation}
where $l = 1,2,3,\ldots$. From these spectra, we can find the degeneracies by computing the partition function of the entanglement spectrum in terms of $q = \exp(-\Delta)$,  which in the disordered phase has a form:
\begin{multline}
    Z = \prod_{l = 0}^{\infty} (1 + q^{2l+1}) = 1 + q + q^{3} + q^{4} + q^{5} + q^{6} + q^{7} + \\ + 2 q^{8} + 2 q^{9}  +  2 q^{10} + 2 q^{11} + 3 q^{12} + \ldots .
\end{multline}
This formula defines the degeneracies in the disordered phase $d=\{1,1,0,1,1,1,1,1,2,2,2,2,3,\ldots\}$.  

In turn, in the ordered phase
\begin{equation}
    Z = \prod_{l = 1}^{\infty} (1 + q^{l}) = 1 + q + q^{2} + 2 q^{3} + 2 q^{4} + 3 q^{5} + 4 q^{6} + 5 q^{7} + \ldots .
\end{equation}
This corresponds to the degeneracy pattern $d=\{1,1,1,2,2,3,4,5,\ldots\}$.  Below, we observe that exactly this spectrum appears in the Ising models on triangular, kagome, square-octagon, star, and dice lattices, while for other lattices they appear in the vicinity of the phase transition, which is always governed by the Ising conformal field theory (see Ref.~\cite{Spectra_near_transition} for the spectra universality near the phase transition).  We also observe that on ruby and SHD lattices, where we do not observe the exact degeneracies, the spectra are still free fermionic, which still leads to a large number of nontrivial relations between different entanglement eigenvalues. 

In general, we analyze the following observables as a check of our CTMRG scheme: we measure the magnetization and entanglement gap to determine the transition temperature, and we explore the closeness of the entanglement spectrum to the degenerate integer values as an additional check.  It should be mentioned that for integrable models it is possible to develop a ``logarithmic'' CTMRG, which works directly with the corner Hamiltonian \cite{okunishi2005real}. 

\begin{figure}
\includegraphics[width= \linewidth]{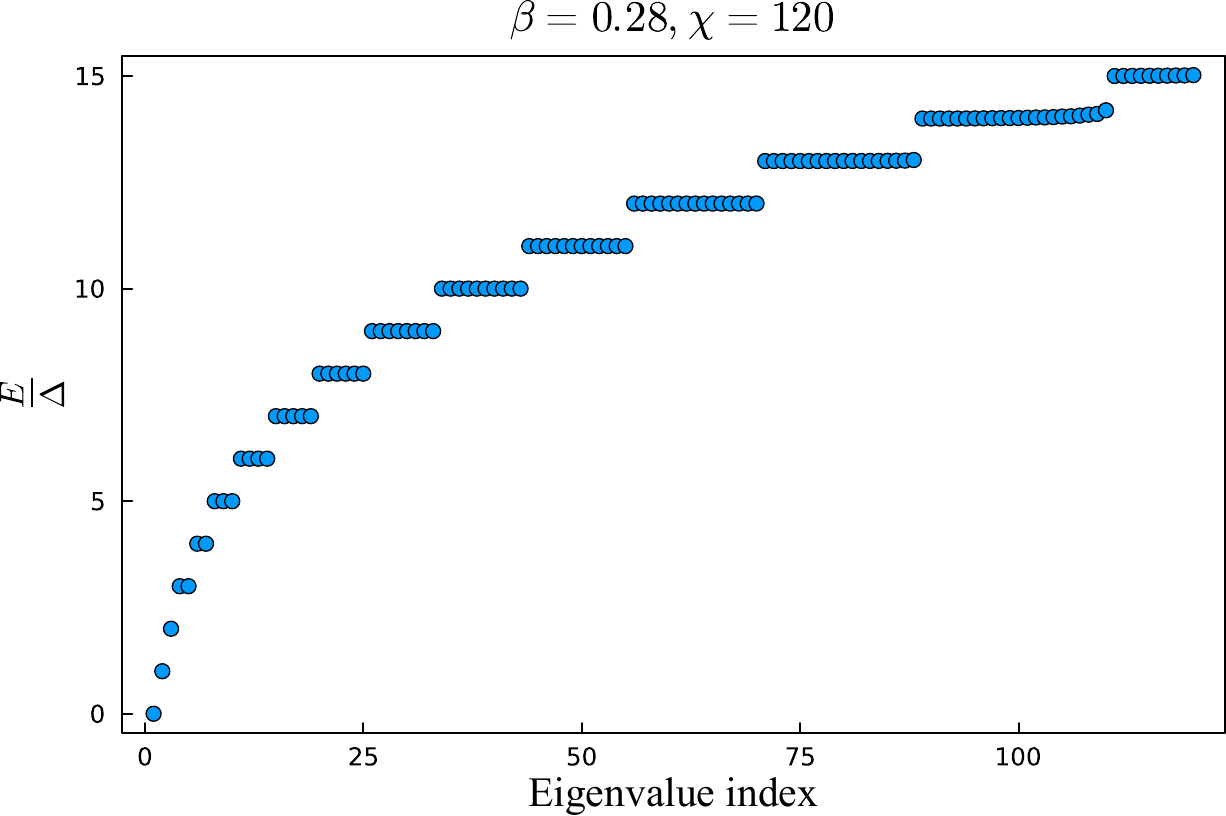} 
 \caption{\label{fig:Figure20}%
   Triangular lattice: The corner matrix spectrum in the ordered phase of the Ising model, normalized by the gap $\Delta$ between the first and second eigenvalues. The bond dimension of the corner matrix  $\chi=120$ and the inverse temperature $\beta = 0.28$.  }
\end{figure}

We show the corner spectrum in the ordered phase ($\beta = 0.28$) in Fig.~\ref{fig:Figure20}, where we normalize the eigenvalues by the gap~$\Delta$. It is clear that with this normalization the eigenvalues form degenerate multiplets with integer eigenvalues. The degeneracies form a sequence $d=\{1,1,1,2,2,3,4,...\}$. This sequence is independent of the precise value of $\beta$ inside the ordered phase. For the first eigenvalues, the closeness to the integer holds within a precision $10^{-11}$.  For the higher eigenvalues, we notice distortions of the spectrum from the integer values and from the exact degeneracy. These deviations originate from the finite values of the bond dimension~$\chi$ and one can systematically improve the accuracy by the increase of $\chi$. Note that the corner eigenvalues $E$ are logarithmic quantities, thus their closeness to integers is still very impressive. The figure proves that our CTMRG algorithm captures the main properties of the corner spectra in integrable models. 

\begin{figure}
\includegraphics[width= \linewidth]{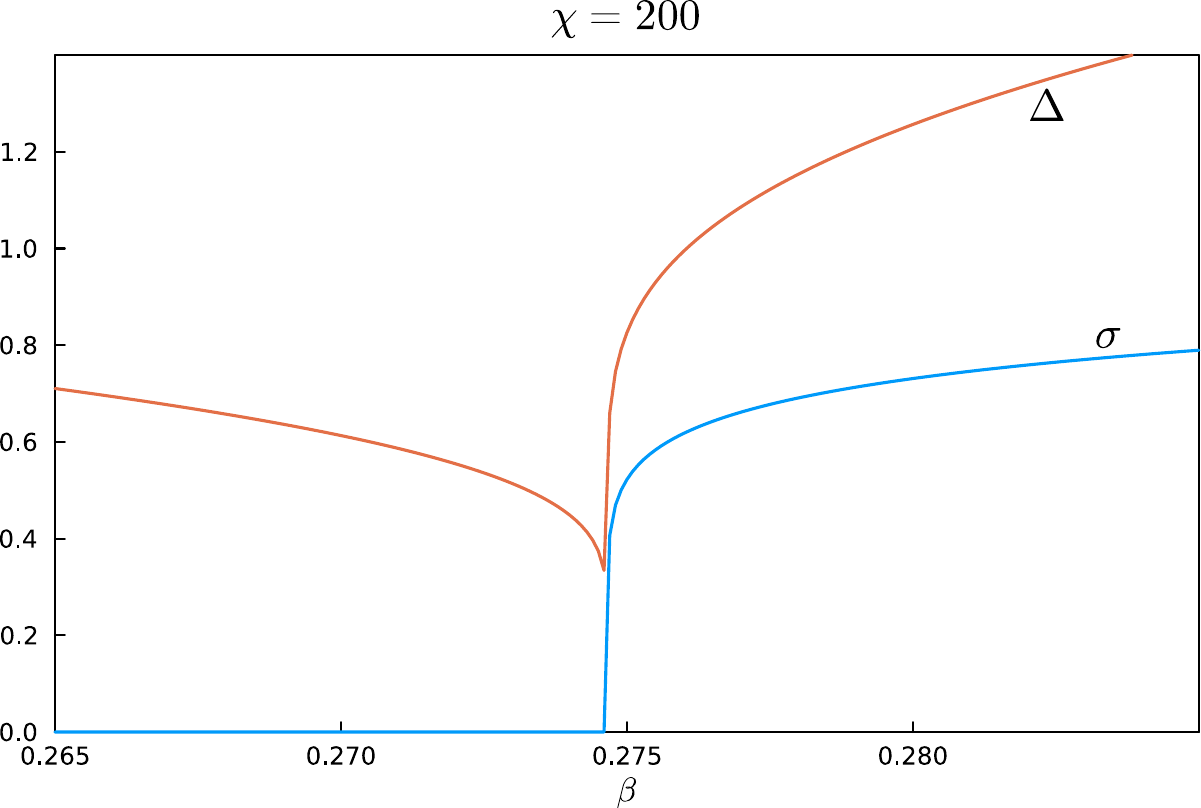} 
 \caption{\label{fig:Figure21}%
   Triangular lattice: Dependence of the magnetization~$\sigma$ and entanglement gap $\Delta$ on the inverse temperature $\beta$ at $\chi = 200$.  }
\end{figure}

In Fig.~\ref{fig:Figure21} we show the dependence of the magnetization and entanglement gap $\Delta$ on the inverse temperature $\beta$. The behavior of the magnetization indicates the second-order phase transition with the estimated $\beta_{c} \approx 0.2746$. This agrees well with the exact critical value $\beta_c^{(\rm ex)} = {\ln{(3)}}/{4} \approx 0.274653$. The behavior of the gap $\Delta$ agrees with the magnetization, indicating a clear minimum in the vicinity of the phase transition. In principle, the gap must vanish at the transition point. However, the convergence in the critical region is slow and to show this explicitly, one needs to further increase $\chi$, since the corner spectrum becomes continuous at the transition point.

\subsection{Kagome lattice}

\begin{figure}
\includegraphics[width= \linewidth]{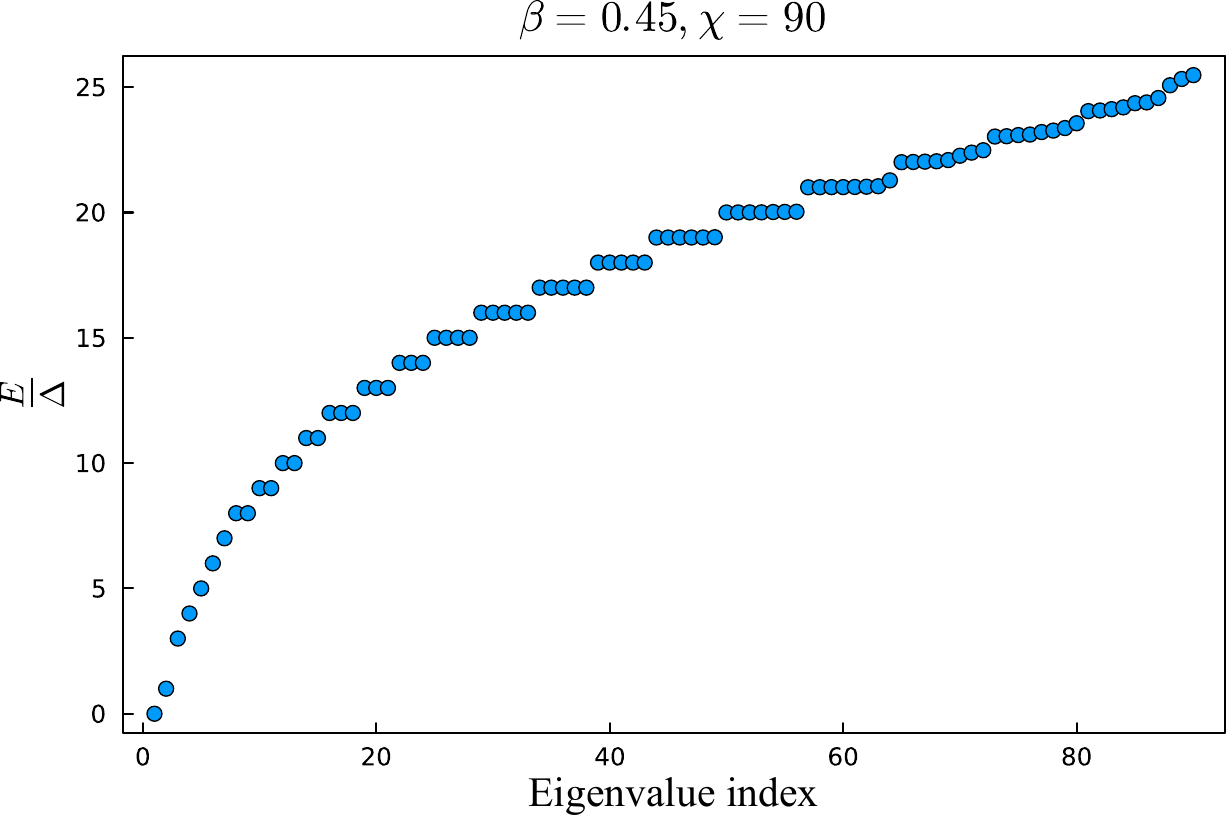} 
 \caption{\label{fig:Figure22}%
   Kagome lattice: The corner matrix spectrum in the disordered phase of the Ising model, normalized by the gap $\Delta$ between the first and second eigenvalues. The bond dimension $\chi=90$ and the inverse temperature $\beta = 0.45$, which is close to the transition point. }
\end{figure}

Next, we study the same ferromagnetic Ising model on the kagome lattice. Analogously to the triangular lattice, we analyze the on-site magnetization and entanglement (corner) spectrum, which we now define as the eigenvalues of the logarithm of corner matrix $C_{6}$ (this is just a normalization convention). We also define the entanglement gap $\Delta$ as a difference between the first and second multiplets of the entanglement spectrum.  The model is integrable, and we obtain the same integer-level spacings with the nearly exact degeneracies. In Fig.~\ref{fig:Figure22} we show the spectrum in the disordered phase on the kagome lattice.  It is clear that the levels are integer-valued and form a degeneracy pattern in the disordered phase of the form $d=\{1,1,0,1,1,1,1,1,2,2,2,2,3,\ldots\}$.

It is also interesting to analyze the corner matrix ${C}_{3}'$, which is defined inside the triangles (see also Fig.~\ref{fig:Figure6}) and its entanglement spectrum. We note that generally this spectrum is not integer-valued and is not exactly degenerate, but the spectra follow the free fermionic pattern and become integers only in the vicinity of the phase transition. This is a common trait for many corner matrices of different Ising models. For a majority of models, we observe that only one of the corner matrices exhibits the integer spectrum, while other corner matrices are only free fermionic. For certain lattices, e.g., the ruby or SHD, we find that all corner matrices are only free fermionic. 
\begin{figure}
\includegraphics[width= \linewidth]{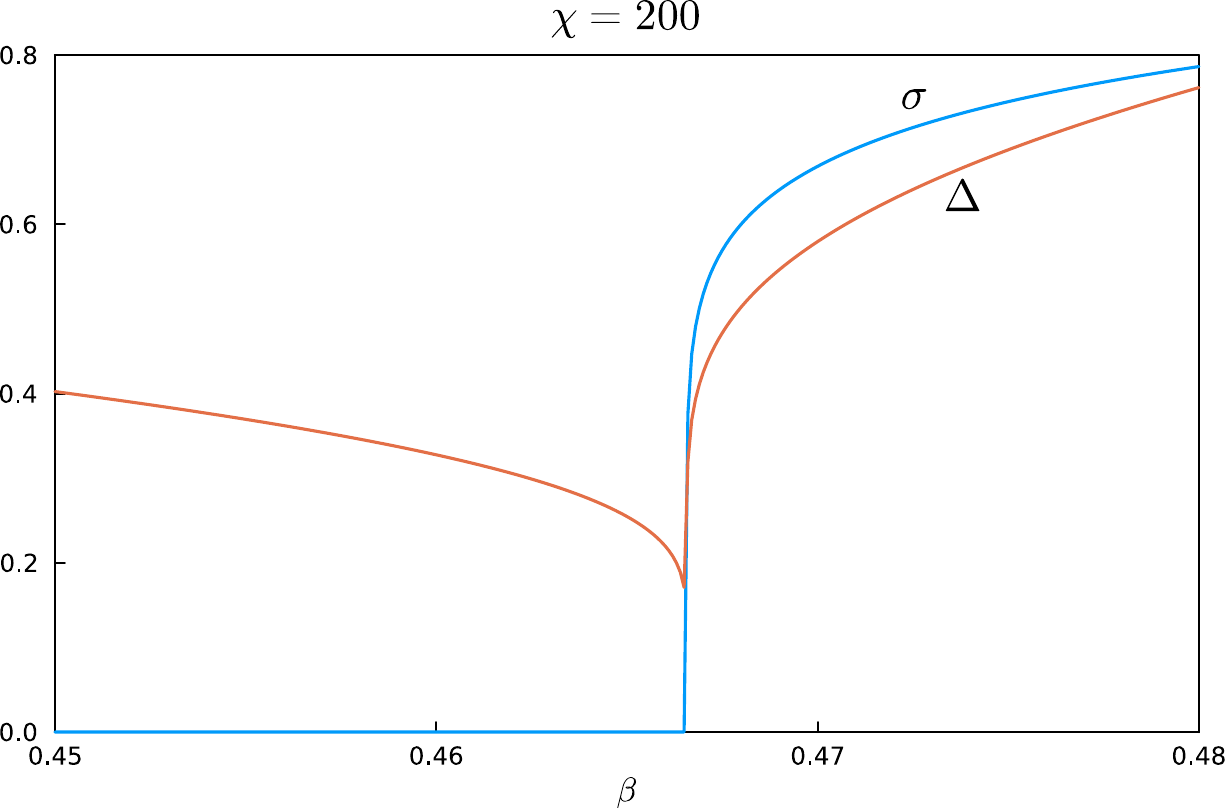} 
 \caption{\label{fig:Figure23}%
   Kagome lattice: Dependence of the magnetization $\sigma$ and entanglement gap $\Delta$ on the inverse temperature $\beta$ at $\chi = 200$. The exact critical value $\beta_{c} = 0.46657$, while our estimate is $\beta_c = 0.4666(1)$.}
\end{figure}

In Fig.~\ref{fig:Figure23} we show the dependence of the magnetization and entanglement gap on the inverse temperature $\beta$. The exact critical value $\beta_{c}^{(\rm ex)} = \ln(3+2\sqrt{3})/{4} \approx 0.466566$, which agrees with our result $\beta_{c} = 0.4666(1)$. Note that this result is obtained purely from the computation of CTMRG on a mesh with spacing $\Delta\beta = 0.0001$, which defines the accuracy of the estimates. In principle, one can improve accuracy by additional fitting and finer mesh in the vicinity of the phase transition. 
\begin{figure}
\includegraphics[width= \linewidth]{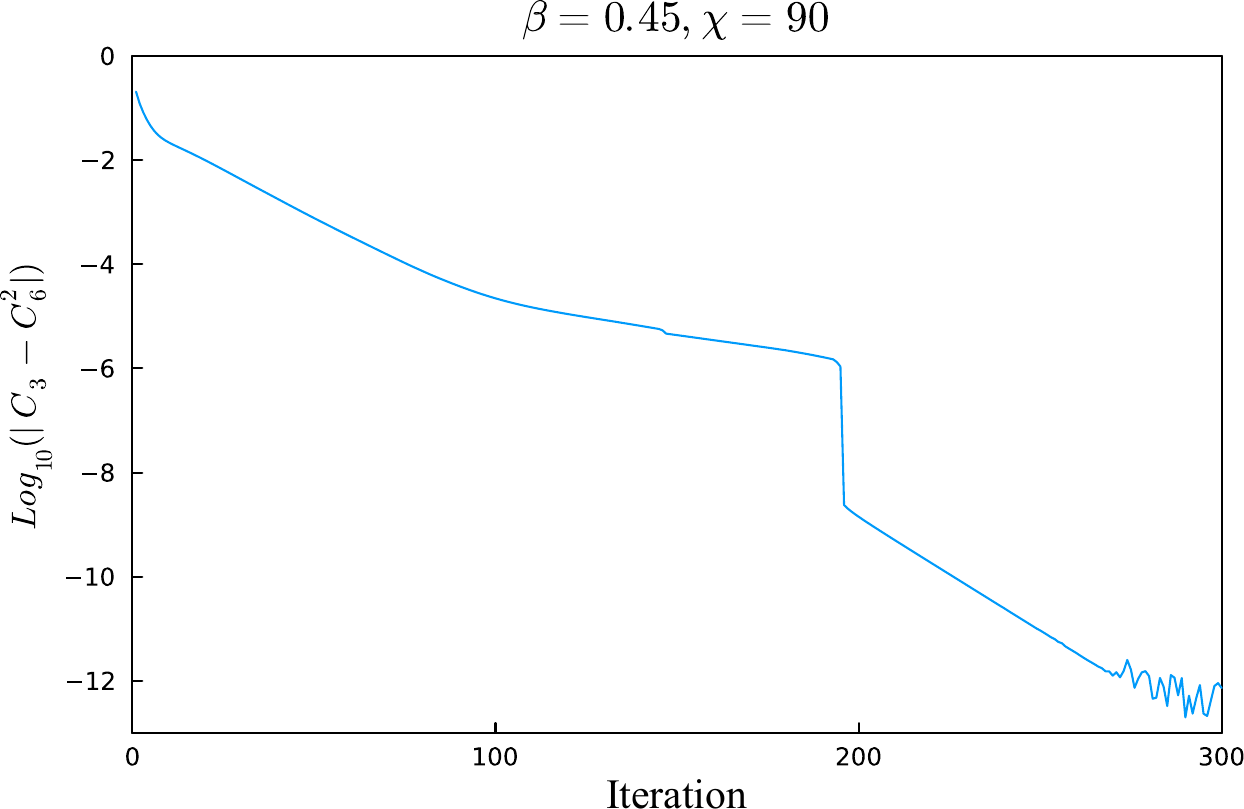} 
 \caption{\label{fig:Figure24}%
   Kagome lattice: 
   The logarithm of the Frobenious norm of the difference between two matrices $C_3$ and $C_6^2$ (preliminary normalized). The parameters are $\chi = 90$ and $ \beta = 0.45$.}
\end{figure}

Finally, as we specified in Sec.~\ref{subsec:Alg.kagome}, the corner matrix~$C_{3}$ inside the hexagons can be obtained in two different ways. In the converged state, these two definitions must be identical. Hence, the agreement must indicate the convergence and self-consistency of the approach. In Fig.~\ref{fig:Figure24} we show the logarithm of the norm of the difference between two definitions of the matrix $C_{3}$. It is clear that the difference vanishes (up to machine precision errors) after a sufficient number of iteration steps.  This confirms the consistency of the proposed scheme and also proves that various procedures to converge the CTMRG environments ultimately lead to the same results.

\subsection{Square-octagon lattice}
The square-octagon lattice does not introduce any additional difficulties compared to the previous lattices. We observe the same integer entanglement spectrum and the dependence of the magnetization and entanglement gap is also analogous to the previous results. 

The possible difference from the previously discussed cases is the presence of two different bMPS, which are rotated by the angle $\pi/4$. Note that the quantities as the magnetization can be computed with both types of bMPS. We computed the magnetizations with both types of bMPS and observed that the results agree up to machine-level precision (of the order of $10^{-13}$). 

\begin{figure}
\includegraphics[width= \linewidth]{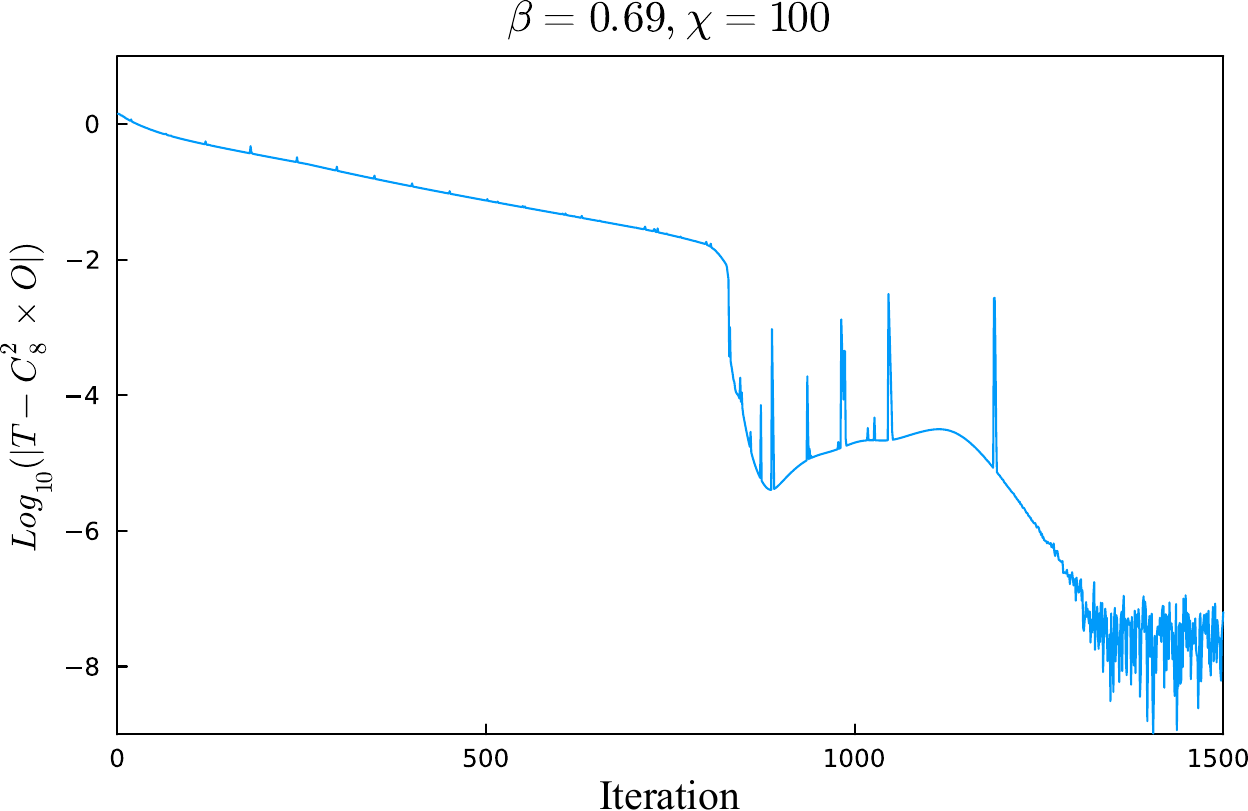} 
 \caption{\label{fig:Figure25}%
   Square-octagon lattice: Convergence of the norm of the difference between the tensors $T$ and $C_{8,i}^{2} O_{ijk}$ (preliminarily normalized). The parameters are $\chi = 100$ and $\beta = 0.69$. The norm of the difference converges to the value around $10^{-8}$.}
\end{figure}

Another possible check of consistency does not employ two different bMPS, but operates in the same bMPS algorithmic loop (see Fig.~\ref{fig:Figure11}), where we also introduce the symmetric tensor $T$. At the same time, we can always replace the tensor~$T$ with the first boundary MPS tensor~$O$ [shown in yellow in Fig.~\ref{fig:Figure10}(a)] and two corner matrices $C_{8}$, which account for the remaining angle $\pi/2$ in the tensor $T$. For the consistency of all calculations, these two ways to represent the corner in any computation of observables must agree. Hence, the consistency condition can be written as $T_{ijk} = C_{8,i}^{2} O_{ijk}$. We show the convergence of the Frobenious norm of the tensor difference in Fig.~\ref{fig:Figure25}. 

\begin{figure}
\includegraphics[width= \linewidth]{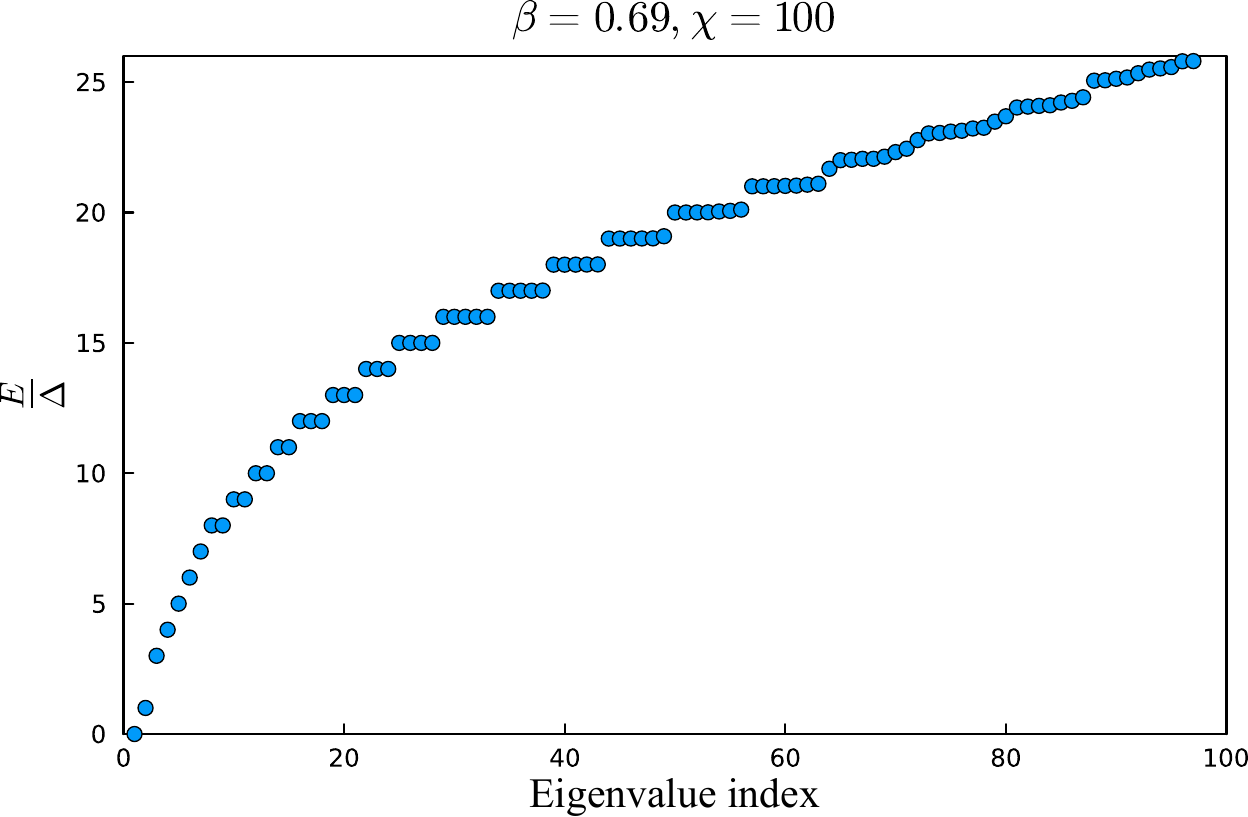} 
 \caption{\label{fig:Figure26}%
   Square-octagon lattice: The corner matrix spectrum in the disordered phase of the Ising model, normalized by the gap $\Delta$ between the first and second eigenvalues. The parameters are $\chi=100$ and $\beta = 0.69$. }
\end{figure}

\begin{figure}
\includegraphics[width= \linewidth]{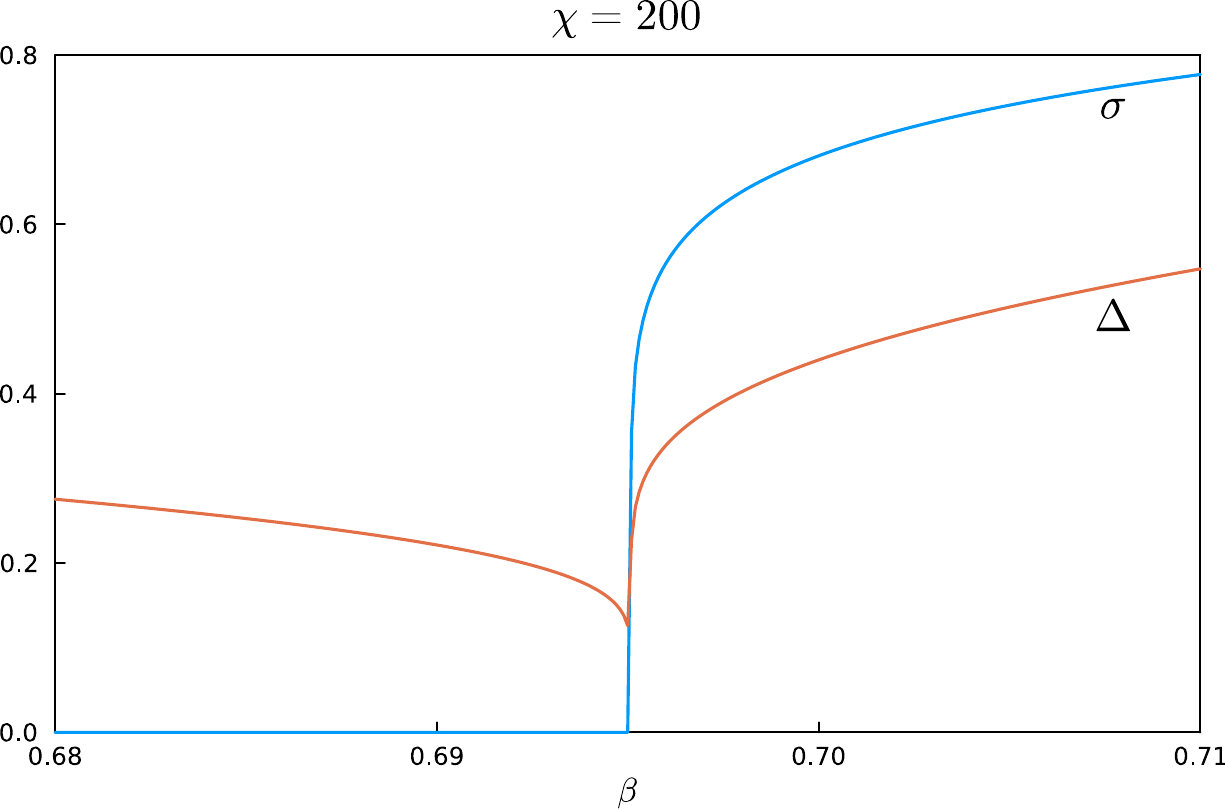} 
 \caption{\label{fig:Figure27}%
   Square-octagon lattice: Dependence of the magnetization $\sigma$ and entanglement gap $\Delta$ on the inverse temperature $\beta$ at $\chi = 200$. The estimated critical value $\beta_c = 0.6950(1)$.}
\end{figure}

We show the entanglement spectrum in the disordered phase in Fig.~\ref{fig:Figure26}. This spectrum is obtained from the logarithm of the corner matrix $C_{8}$. Note that the spectral degeneracies are identical to the ones of the kagome lattice with the corresponding degeneracy patterns.
In turn, the results for the magnetization and for the entanglement gap are shown in Fig.~\ref{fig:Figure27}. The critical inverse temperature estimated in our analysis $\beta_{c} = 0.6950(1)$ agrees well with the exact result,  $\beta_{c}^{(\rm ex)} = {\ln(1+\sqrt{2}/2 + \sqrt{10+8\sqrt{2}}/2)}/{2} \approx 0.6950741$. The entanglement gap is calculated from the matrix $\log{C_{8}}$. 

\subsection{Star lattice}

\begin{figure}
\includegraphics[width= \linewidth]{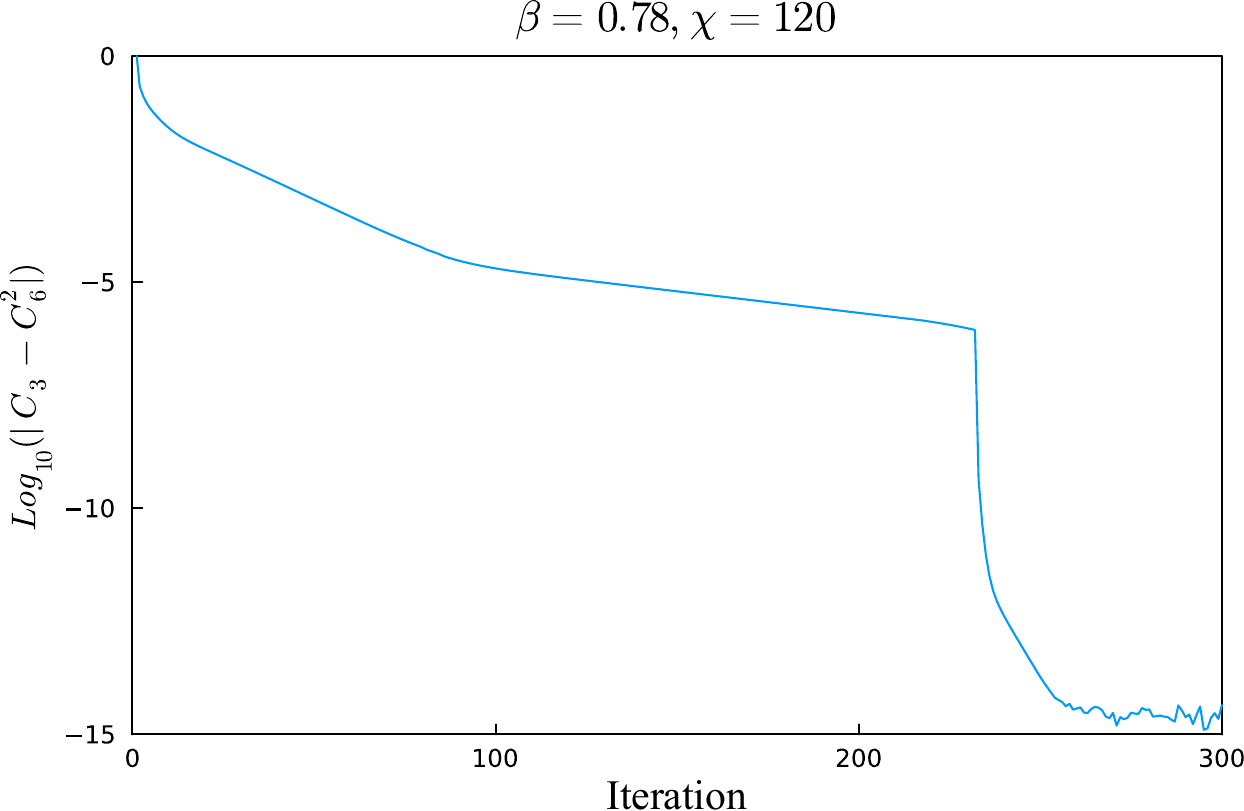} 
 \caption{\label{fig:Figure28}%
   Star lattice: Convergence of the norm of the difference between two definitions of the matrix $C_{3}$ (preliminarily normalized). The parameters are $\chi = 120$ and $\beta = 0.78$.}
\end{figure}

The star lattice has a lot in common with the kagome lattice. In fact, the kagome lattice can be obtained from the star lattice with a contraction of one of the indices, hence, the CTMRG algorithms on both lattices are very similar. Here, we analyze the same quantities as for the kagome lattice. In particular, we compute the same consistency condition, which corresponds to the difference between two different ways how the matrix $C_{3}$ inside the dodecahedrons can be defined.  We show the convergence of the difference between two matrices in Fig.~\ref{fig:Figure28}. The difference converges almost to the machine precision. 

\begin{figure}
\includegraphics[width= \linewidth]{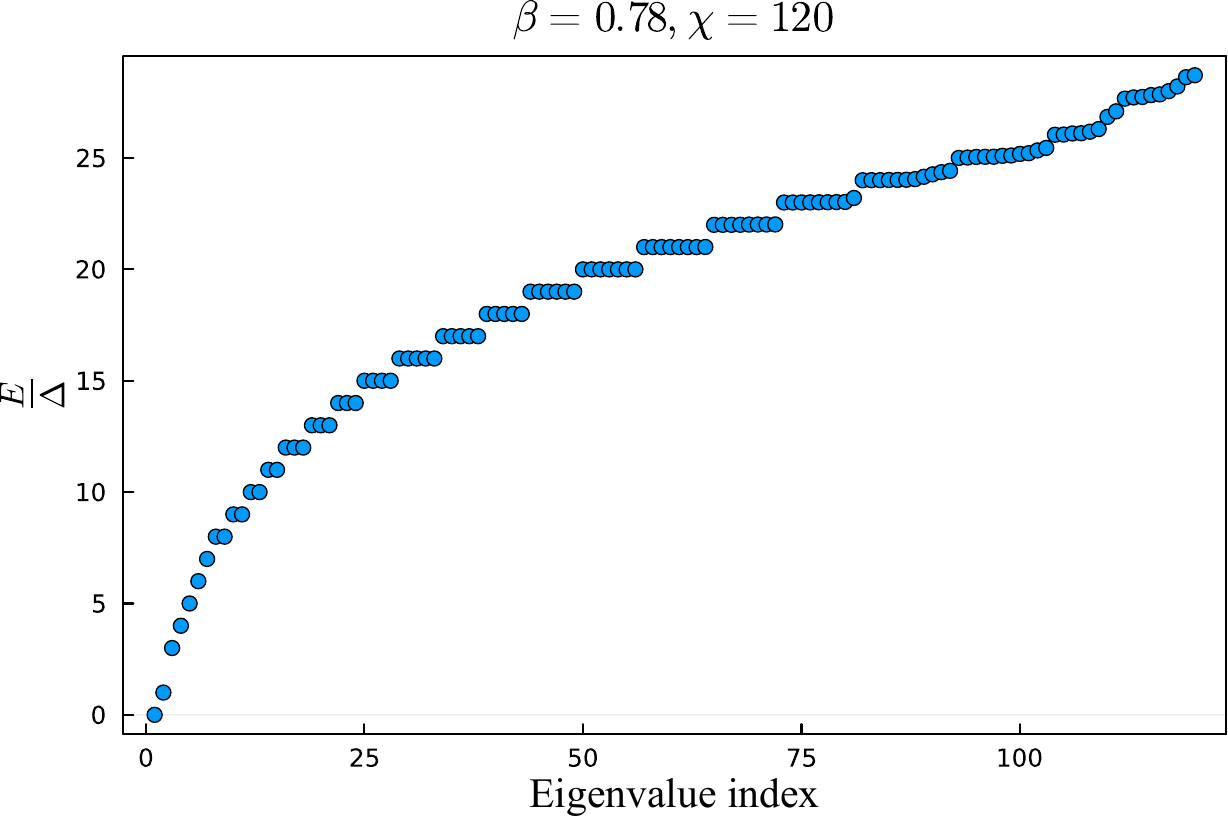} 
 \caption{\label{fig:Figure29}%
   Star lattice: The corner matrix spectrum in the disordered phase of the Ising model, normalized by the gap $\Delta$ between the first and second eigenvalues. The parameters are $\chi=120$ and $\beta = 0.78$. }
\end{figure}

\begin{figure}
\includegraphics[width= \linewidth]{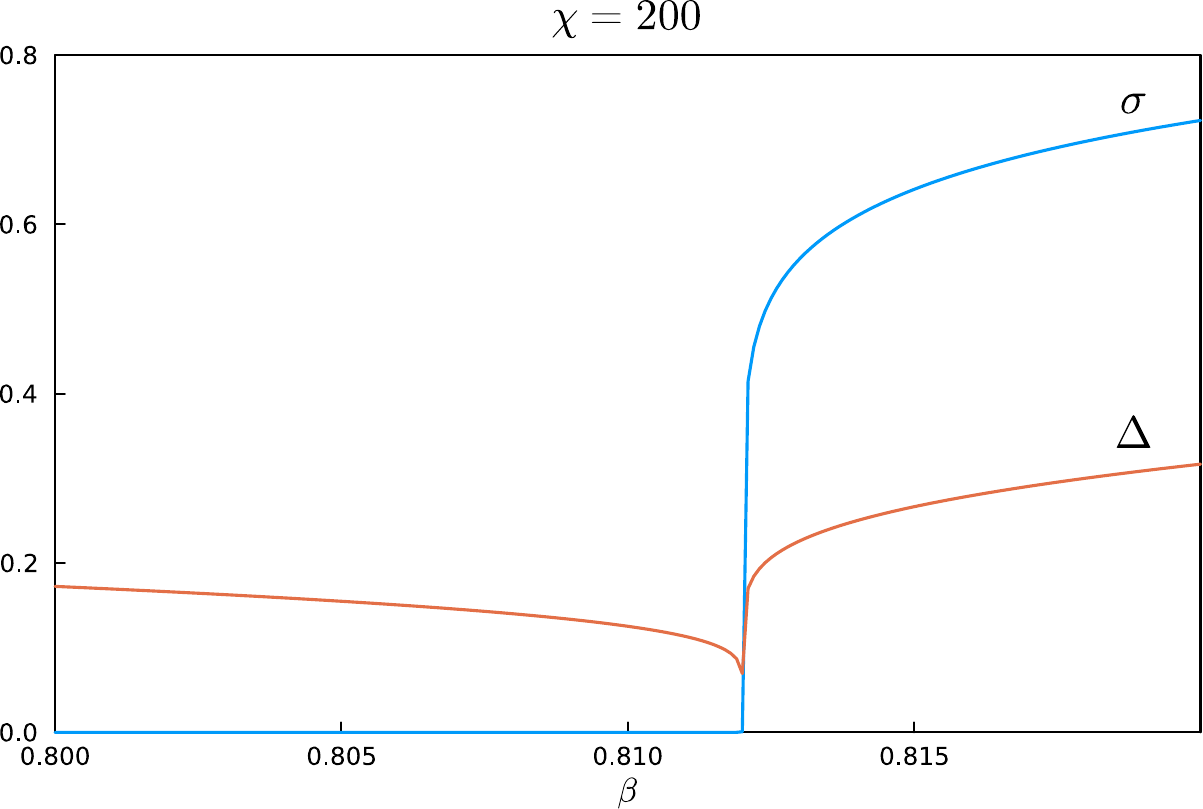} 
 \caption{\label{fig:Figure30}%
   Star lattice: Dependence of the magnetization $\sigma$ and entanglement gap $\Delta$ on the inverse temperature $\beta$ at $\chi = 200$. The estimated critical value $\beta_c = 0.8120(1)$.}
\end{figure}

Next, we discuss the entanglement spectrum and observables on the star lattice. These are shown in Figs.~\ref{fig:Figure29} and \ref{fig:Figure30}, respectively. The entanglement spectrum is the same as for the square-octagon and kagome lattices, while the critical temperature also agrees well with the exact result, $\beta_{c}^{\rm(ex)} = {\ln{(3/2+\sqrt{3}/2 + \sqrt{12+10\sqrt{3}}/2)}}/{2} \approx 0.8120101$. Note that here we obtain the entanglement spectrum and the entanglement gap from the corner matrix~$C_{12}$. 

\subsection{Ruby lattice}
On the ruby lattice, it is necessary to apply nonsymmetric factorizations. These require a long procedure of finding projectors $P_{L}$ and $P_{R}$, which is illustrated in Fig.~\ref{fig:Figure17}.  This procedure employs the inversion of the square root of the matrix $S$, with $S$ requiring the contraction of all corner matrices. It means that $S \propto C_{6}^{6}$, where the proportionality means the order of magnitude estimate.  Now, imagine the system is far from criticality, where the eigenvalues of the matrix~$C_{6}$ quickly decrease to zero. For these corner matrices, the eigenvalues of $S$ decrease much faster. This can lead to machine-precision errors since the eigenvalues of $S$ can become smaller than $10^{-16}$. In this case, the intermediate steps in the calculation of $S$ seem inaccurate, and the resulting projectors $P_{L}$ and $P_{R}$ can be wrong. This leads to a possible algorithm instability, which appears only at high values of $\chi$.  

In this study, we choose the following scheme to mitigate the potential numerical instability: We start the CTMRG scheme with a small $\chi$ and then gradually increase it step by step, with convergence ensured at each step. We stop increasing $\chi$, if it has reached the maximal value or if the eigenvalues of $S$ become lower than the predefined threshold, e.g., $10^{-14}$. It is possible that this problem can be solved with the introduction of more precise schemes for finding projectors, which were proposed on the square lattice in Ref.~\cite{fishman2018faster}.  Note that in the vicinity of the critical point, the instability appears at much higher values of $\chi$ than our typical maximal values. 

\begin{figure}
\includegraphics[width= \linewidth]{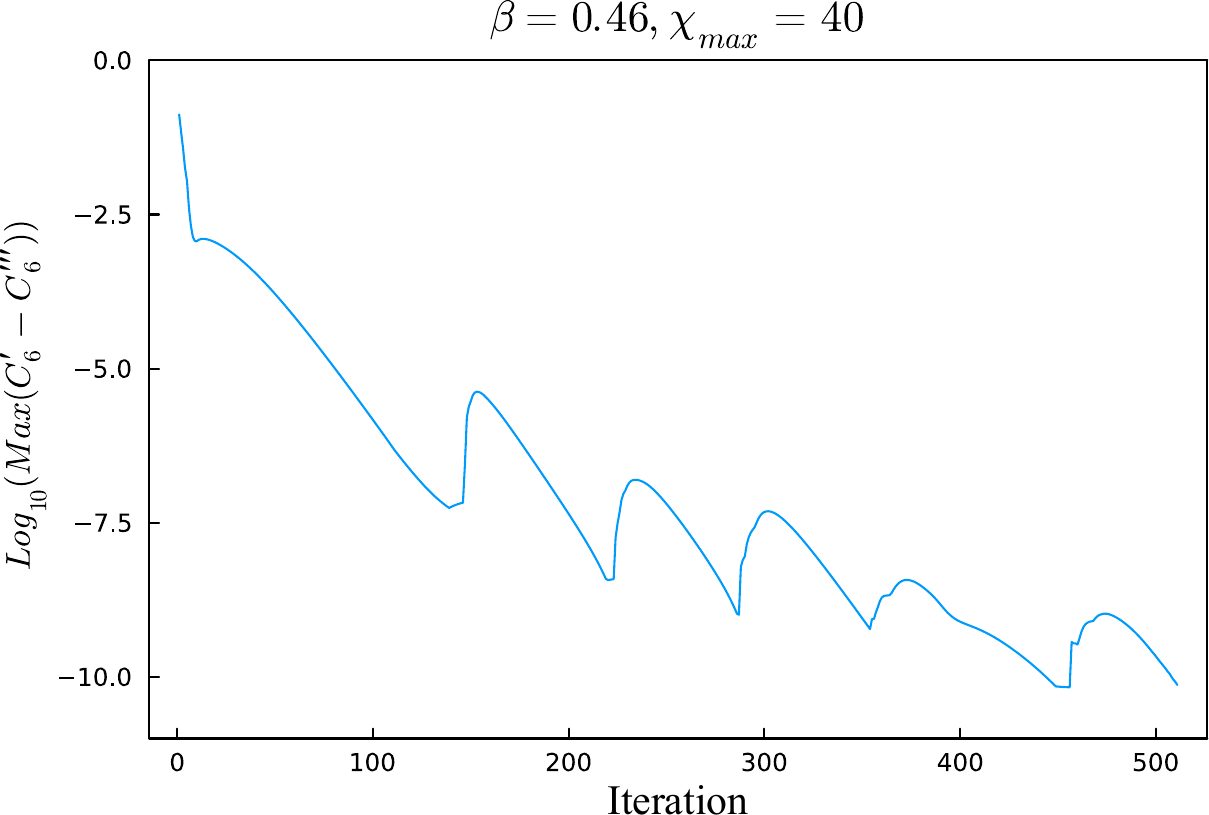} 
 \caption{\label{fig:Figure31}%
   Ruby lattice: Convergence of the maximal difference between the ten largest eigenvalues of the corner matrices ${C}'_{6}$ and $C'''_{6}$ (preliminary normalized). The parameters are $\chi = 40$ and $\beta = 0.46$.  The fluctuations on the plot correspond to the consecutive increases of $\chi$. }
\end{figure}

We can now describe certain consistency checks on the ruby lattice. Note that the ruby lattice contains two types of hexagons with the corresponding corner matrices ${C}'_{6}$ and $C'''_{6}$. These two matrices appear on the different stages of the algorithmic loop, hence, in principle, they are not connected. But from the geometrical arguments, we can make a statement that these matrices must be the same.  Indeed, this also corresponds to our numerical observations. The first three eigenvalues converge to the precision of about $10^{-10}$, while the smallest eigenvalues can have larger differences up to $10^{-7}$. In Fig.~\ref{fig:Figure31} we show the convergence of the maximal difference between the ten largest eigenvalues of the respective matrices.  

Next, we describe the entanglement spectrum on the ruby lattice. Unfortunately, the spectra (for all types of corner matrices) do not exhibit a clear integer degeneracy pattern. Still, these spectra remain free fermionic. In particular, we are able to reorganize the first two dozen eigenvalues in the form of the free fermionic Hamiltonian. Another clear sign of the free fermionic nature of the spectra is that all eigenvalues can be grouped in pairs of the form $\{E, E+\Delta\}$, which we also observe in all cases.  We believe that the free fermionic nature of the spectra is due to the possibility of the exact fermionization of the Ising models on the planar lattices, which is a basis of the Pfaffian method of the Ising model solution~\cite{mccoy1973two, Pfaffian_ruby, Pfaffian_SHD} (see also Ref.~\cite{wille2023topological} for a tensor network introduction to the duality). 

\begin{figure}
\includegraphics[width= \linewidth]{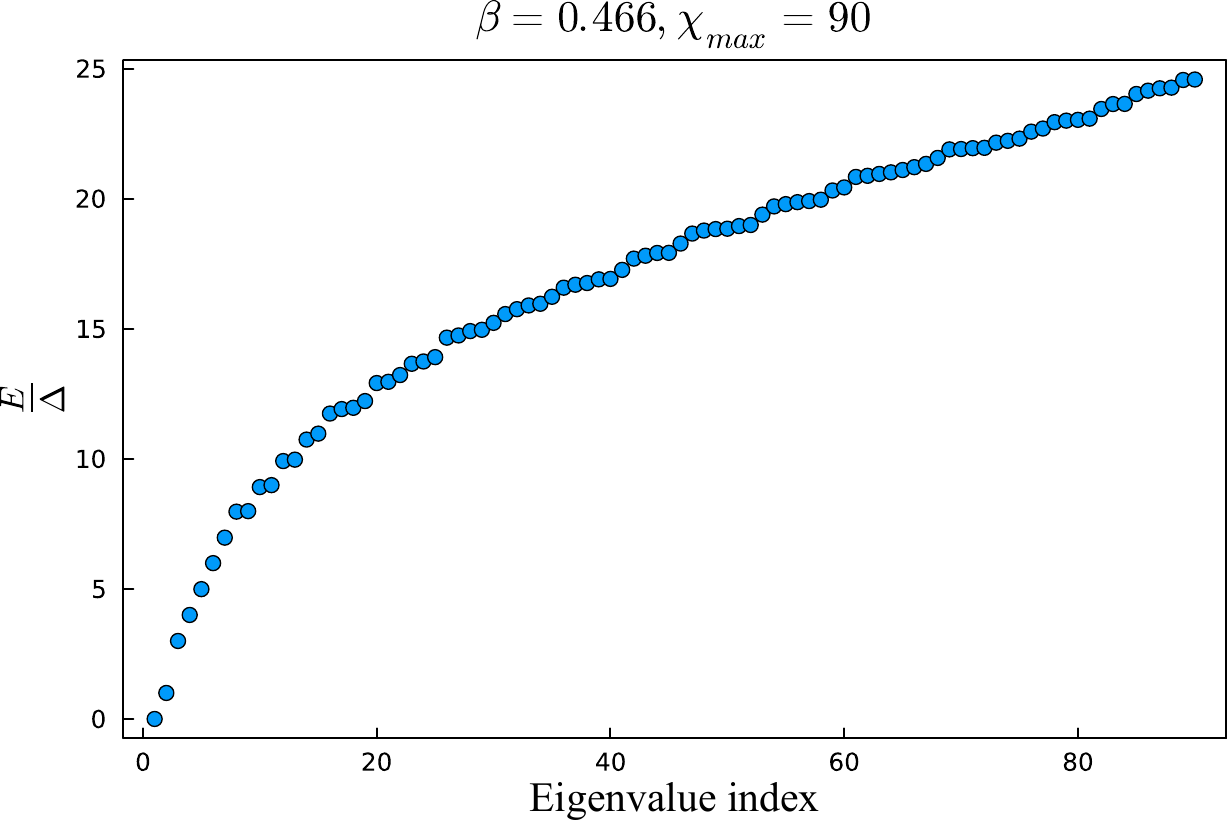} 
 \caption{\label{fig:Figure32}%
   Ruby lattice: The corner matrix spectrum in the disordered phase of the Ising model, normalized by the gap $\Delta$ between the first and second eigenvalues. The parameters are $\chi=90$ and $\beta = 0.466$, which is close to the critical point. }
\end{figure}

The exact integer degeneracy appears only in the vicinity of the phase transition due to universality~\cite{Spectra_near_transition}. For all other values of the inverse temperature~$\beta$, the spectra deviate from the exactly degenerate integer form. To check this property, we mapped the tensor network on the ruby lattice to another tensor network on the kagome lattice, where we applied the kagome-lattice CTMRG and obtained identical not-exactly integer spectra. We show the spectrum in the vicinity of the critical point in Fig.~\ref{fig:Figure32}. The first eigenvalues are nearly integer (with a difference of the order of $10^{-3}$) and nearly degenerate, but the higher eigenvalues show disordered behavior. 

\begin{figure}
\includegraphics[width= \linewidth]{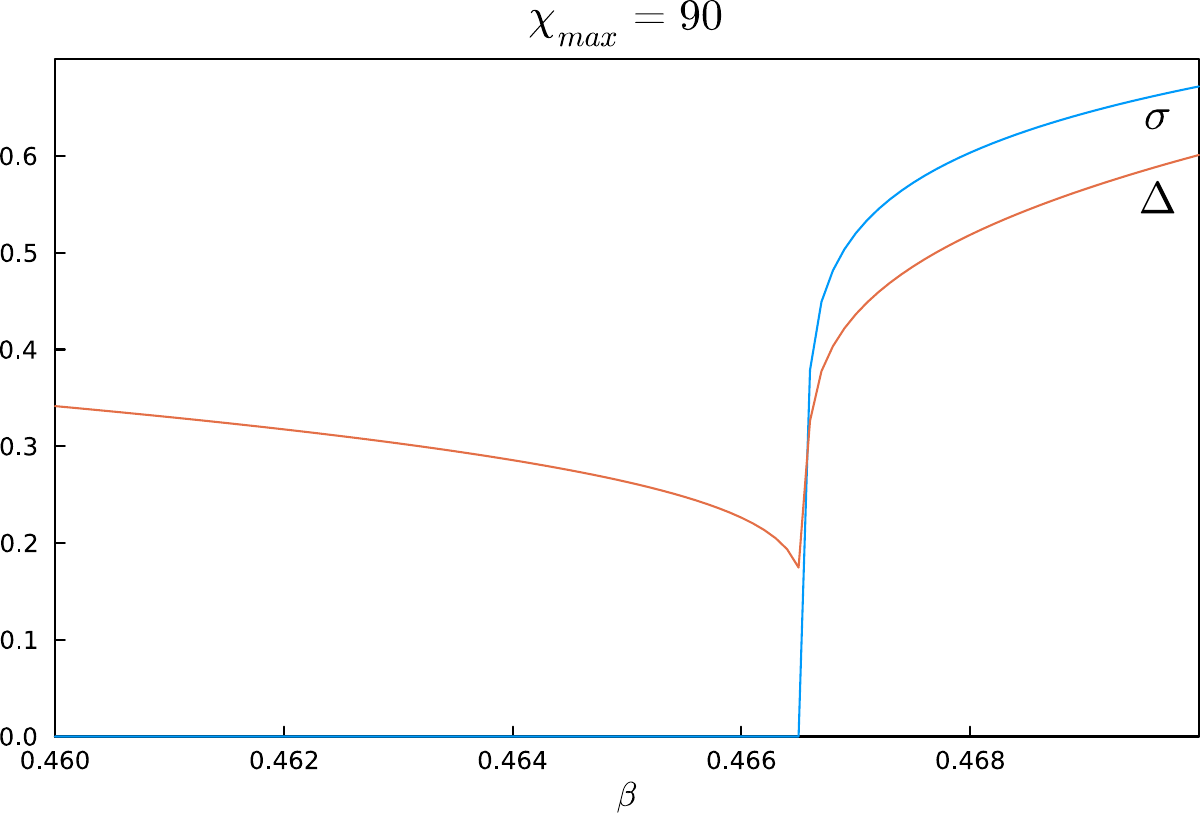} 
 \caption{\label{fig:Figure33}%
   Ruby lattice: Dependence of the magnetization $\sigma$ and entanglement gap $\Delta$ on the inverse temperature $\beta$ at $\chi_{\max} = 90$. The estimated critical value is $\beta_c = 0.4666(1)$.}
\end{figure}

Finally, we analyze the magnetization and entanglement gap on the ruby lattice. The exact result for the critical temperature is given by $\beta_{c}^{\rm(ex)} = \ln{(3 + 2\sqrt{3})}/4 \approx 0.466566$. This agrees well with our estimates in Fig.~\ref{fig:Figure33}, where we show the dependence of the magnetization and entanglement gap on the inverse temperature $\beta$. 

\subsection{SHD lattice}

\begin{figure}
\includegraphics[width= \linewidth]{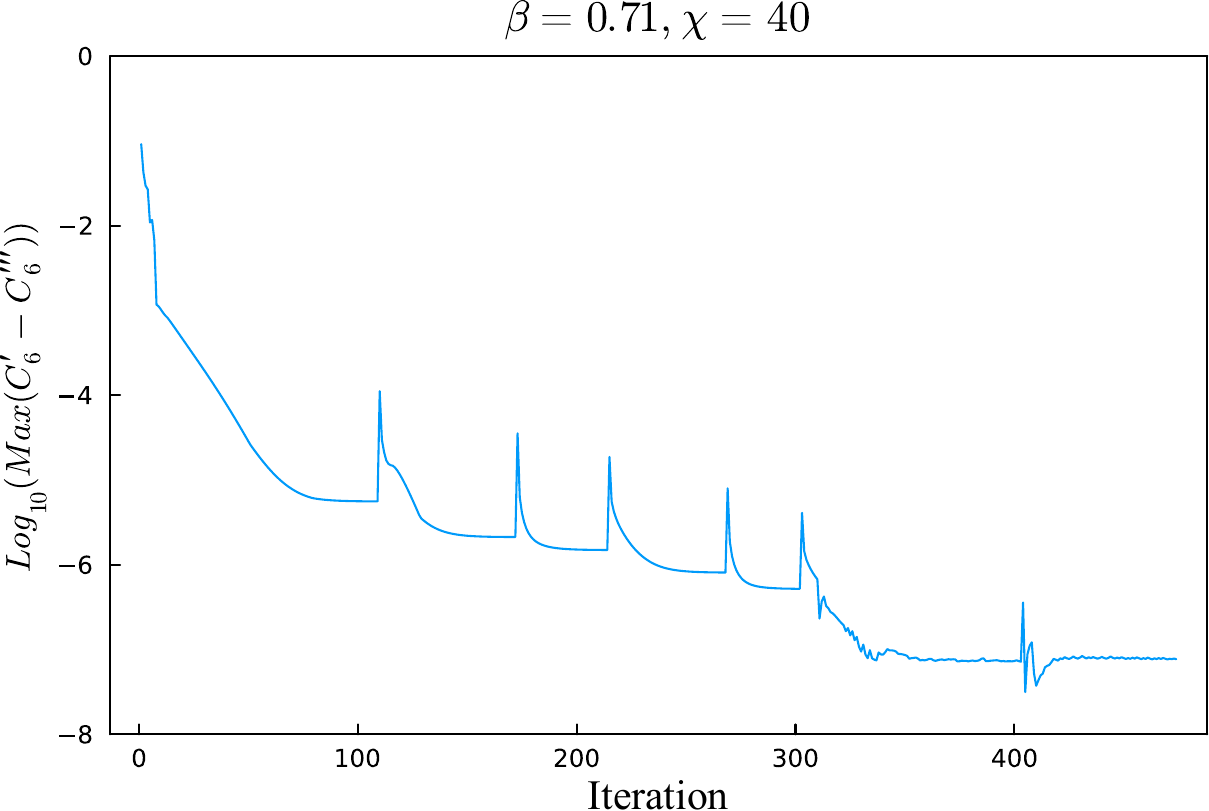} 
  \caption{\label{fig:FigureSHD1}%
   SHD lattice: Convergence of the maximal difference between the ten largest eigenvalues of the corner matrices ${C}'_{6}$ and $ {{{C}'''_{6}}}$ (preliminary normalized). The parameters are $\chi = 40$ and $\beta = 0.71$.  The fluctuations on the plot correspond to the consecutive increases of $\chi$. }
\end{figure}

The case of SHD lattice is completely analogous to the ruby lattice in its algorithmic realization with all the corresponding discussions. In particular, there exists the potential numerical instability at high $\chi$. 
As a check of convergence of the CTMRG scheme on the SHD lattice, we analyze the difference between the spectra of two different hexagon corner matrices shown in Fig.~\ref{fig:FigureSHD1}. The results are somewhat worse than for the ruby lattice, which is a general observation for the SHD lattice. 

\begin{figure}
\includegraphics[width= \linewidth]{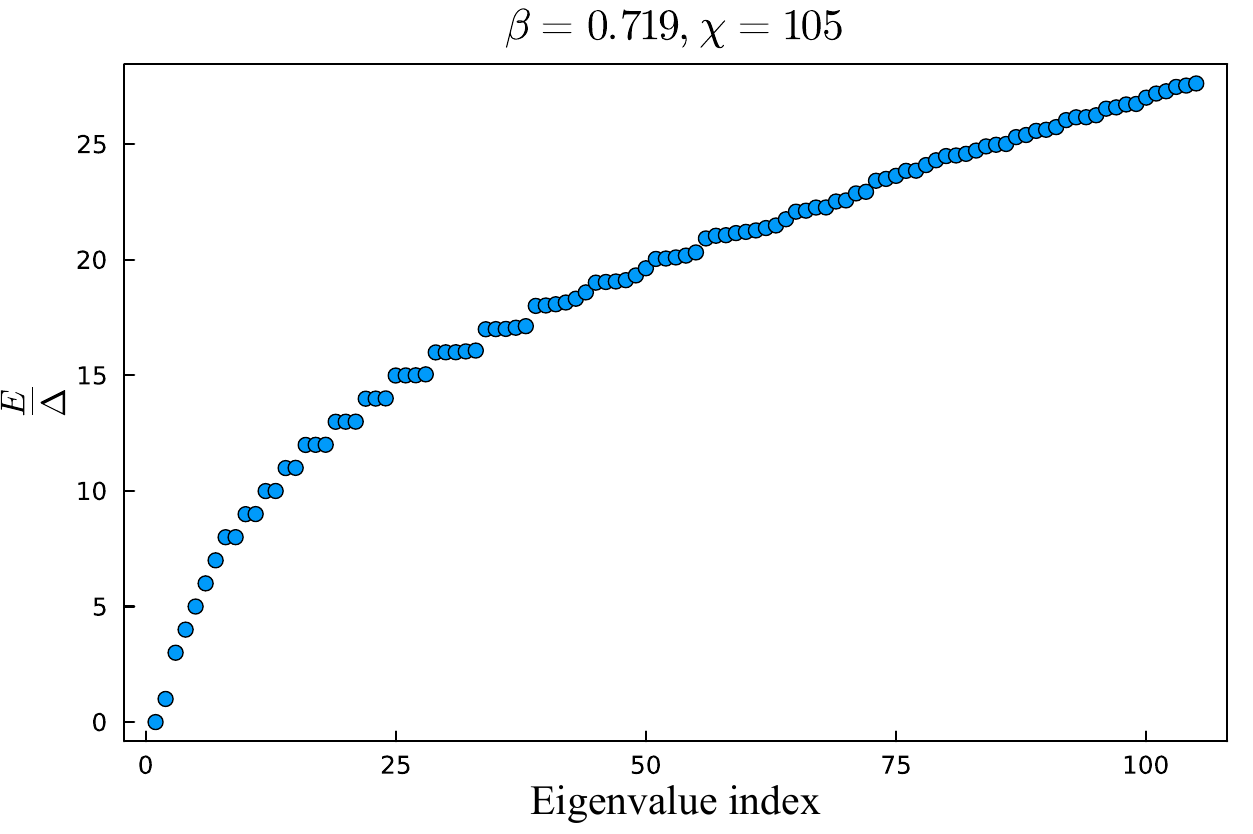} 
 \caption{\label{fig:FigureSHD2}%
   SHD lattice: The corner matrix spectrum in the disordered phase of the Ising model, normalized by the gap $\Delta$ between the first and second eigenvalues. The parameters are $\chi=105$ and $\beta = 0.719$.}
\end{figure}

The entanglement spectra on the SHD lattice are also not exactly integer, though they converge to the integer values near the phase transition. To confirm this result, we also mapped the tensor network to the kagome lattice, where we applied the kagome-lattice CTMRG. The results of the kagome-lattice CTMRG are in agreement with the SHD-lattice calculations. The behavior of the spectrum near the critical point (in the disordered phase) is shown in Fig.~\ref{fig:FigureSHD2}.  The eigenvalues are close to the ones obtained for the square-octagon or star lattices, but this holds only in the vicinity of the phase transition. 

\begin{figure}
\includegraphics[width= \linewidth]{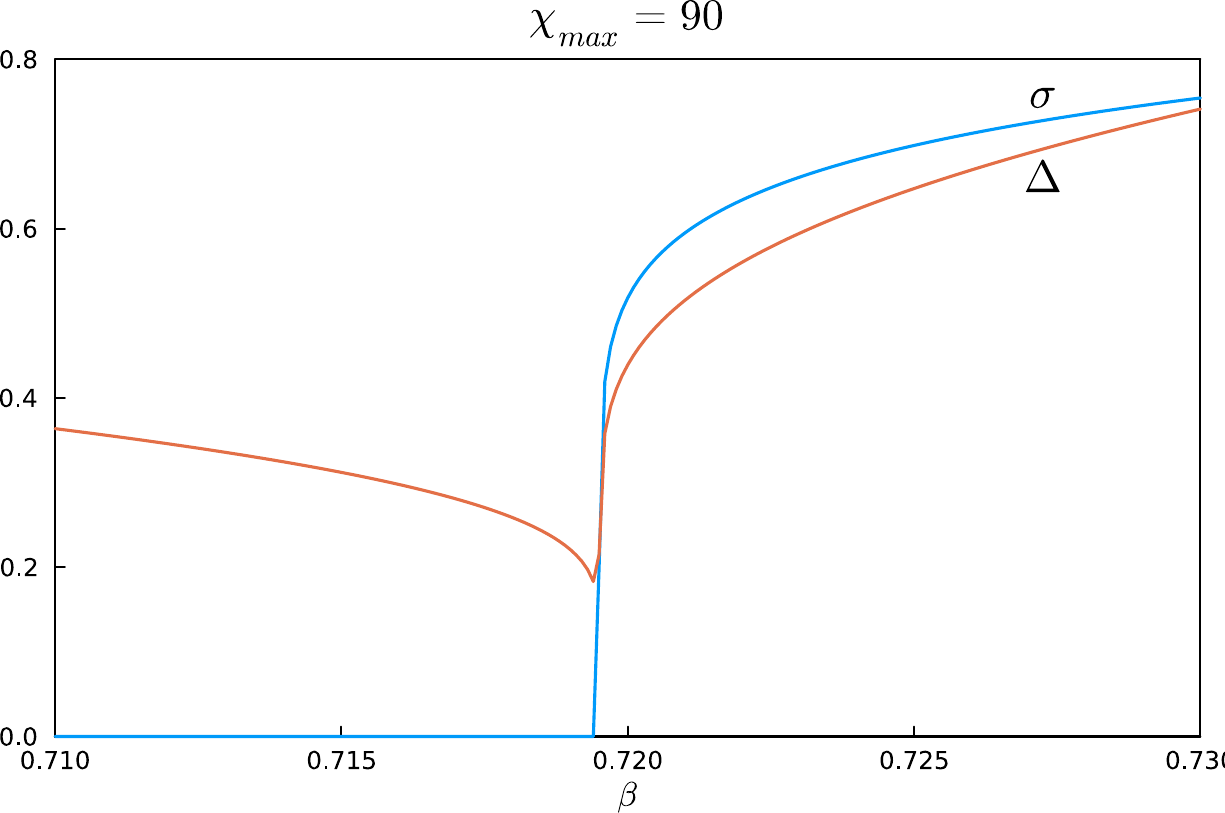} 
 \caption{\label{fig:FigureSHD3}%
   SHD lattice: Dependence of the magnetization $\sigma$ and entanglement gap $\Delta$ on the inverse temperature $\beta$ at $\chi_{\max} = 90$. The exact value of the inverse critical temperature $\beta_{c}^{\rm(ex)} \approx 0.71951019$, while according to our estimates $\beta_c = 0.7195(1)$.}
\end{figure}

Finally, we analyze the magnetization and the entanglement gap defined from $\log{(C_{6})}$, where $C_{6}$ is the corner matrix inside the dodecahedrons. The corresponding observables and the estimate of the critical temperature are shown in Fig.~\ref{fig:FigureSHD3}. 

\subsection{Dice lattice}

The dice-lattice CTMRG has similar peculiarities as the ruby and SHD lattices, since the corresponding algorithm includes the nonsymmetric factorization. Due to this, all comments regarding possible numerical instabilities at large $\chi$ hold here as well. To proceed, we also introduce the maximal value of $\chi$ for this lattice, which depends on the spectrum of $S$ in the nonsymmetric factorization. 

\begin{figure}
\includegraphics[width= \linewidth]{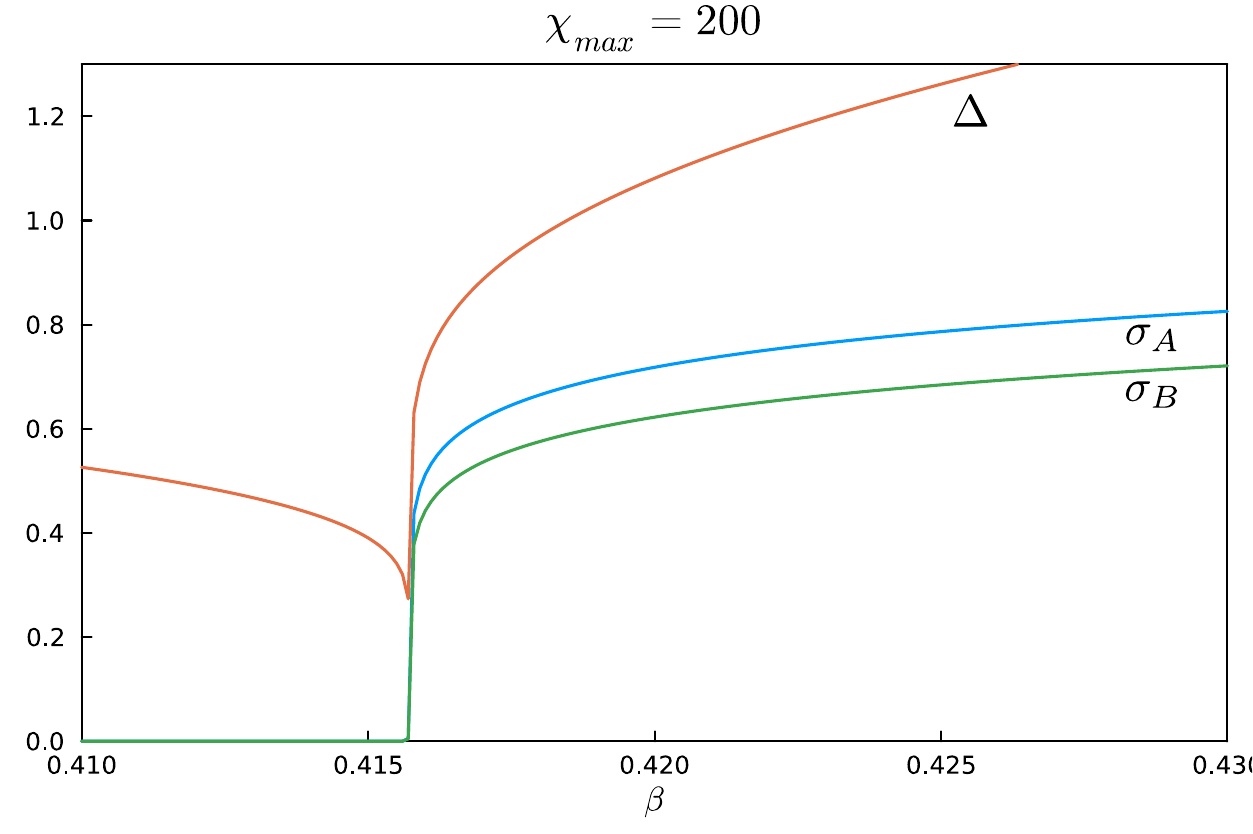} 
 \caption{\label{fig:Figure34}%
   Dice lattice: Dependence of the local magnetizations $\sigma_{A,B}$ on two types of lattice sites $A,B$ and entanglement gap $\Delta$ on the inverse temperature $\beta$ at $\chi_{\max} = 200$. The estimated critical value of the inverse temperature is $\beta_c = 0.4157(1)$.}
\end{figure}

Note that we also have additional differences in the case of dice lattice. First, it does not contain a single type of diagonal corner matrix with the eigenvalues related to the entanglement spectrum. In contrast, we have two different corner matrices $C_{6}$ and $ C_{3}$, which are symmetric. In general, these two matrices cannot be simultaneously diagonalized.  Still, the behavior of their eigenvalues contains information on the criticality of the system. In particular, we define the entanglement gap $\Delta$ as the difference of logarithms of the second and first eigenvalues of the matrix $C_{3}$.  We employ this as a definition of the entanglement gap for the dice-lattice Ising model. Second, the dice lattice has two types of sites: trivalent (type $B$) and six-valent ones (type $A$), which can have different on-site magnetizations. The entanglement gap and the magnetizations on two types of sites are shown in Fig.~\ref{fig:Figure34}.  The critical temperature for the dice lattice can be obtained from its duality to the kagome lattice, thus $\beta_{c}^{\rm(ex)} \approx 0.415721472$. Our result for the critical temperature is close to the specified exact value, in particular, $\beta_c = 0.4157(1)$. 

As we stated, the spectra of the corner matrices $C_{3}$ and $C_{6}$ cannot be defined simultaneously and do not give us useful information, in particular, the integer spectrum with exact degeneracies. To obtain these, we employ the duality with the kagome lattice as a guiding principle. According to the calculations on the kagome lattice, we know that the kagome-lattice Ising model exhibits integer and exactly degenerate corner spectra on the hexagons. Surely, the CTMRG approach to the dual-lattice Ising model (which is exactly equivalent to the original one up to some redefinitions of the inverse temperature) must also contain these integer degenerate levels but in a more complex way.  

The original kagome-lattice corner matrices correspond to the corners of the hexagons, which are dual to the six-valent vertices of the dice lattice. The six-valent tensors are now surrounded by six tensors $T_{1ijk}$ (see also Fig.~\ref{fig:Figure19}). We also take the matrices $q$ (defined as square roots of the bond matrices of the Ising model, $q = \sqrt{W}$) and reabsorb them into the tensors $T_{1}$ as follows: $T'^{\sigma}_{i k} = T_{1ijk}q_{j}^\sigma$, where $\sigma = \{+1, -1\}$. 
With these new tensors, the partition function is proportional to $\sum_{\sigma} \Tr [(T'^{\sigma})_{ij}^{6}] $, where we take $T'^{\sigma}_{ij}$ as matrices (rank-2 tensors) in indices $i$ and $j$ for the fixed $\sigma$, perform exponentiation of these matrices to the sixth power for each $\sigma$, and then take a trace with respect to the indices $i,j$ and sum over $\sigma$.

\begin{figure}
\includegraphics[width= \linewidth]{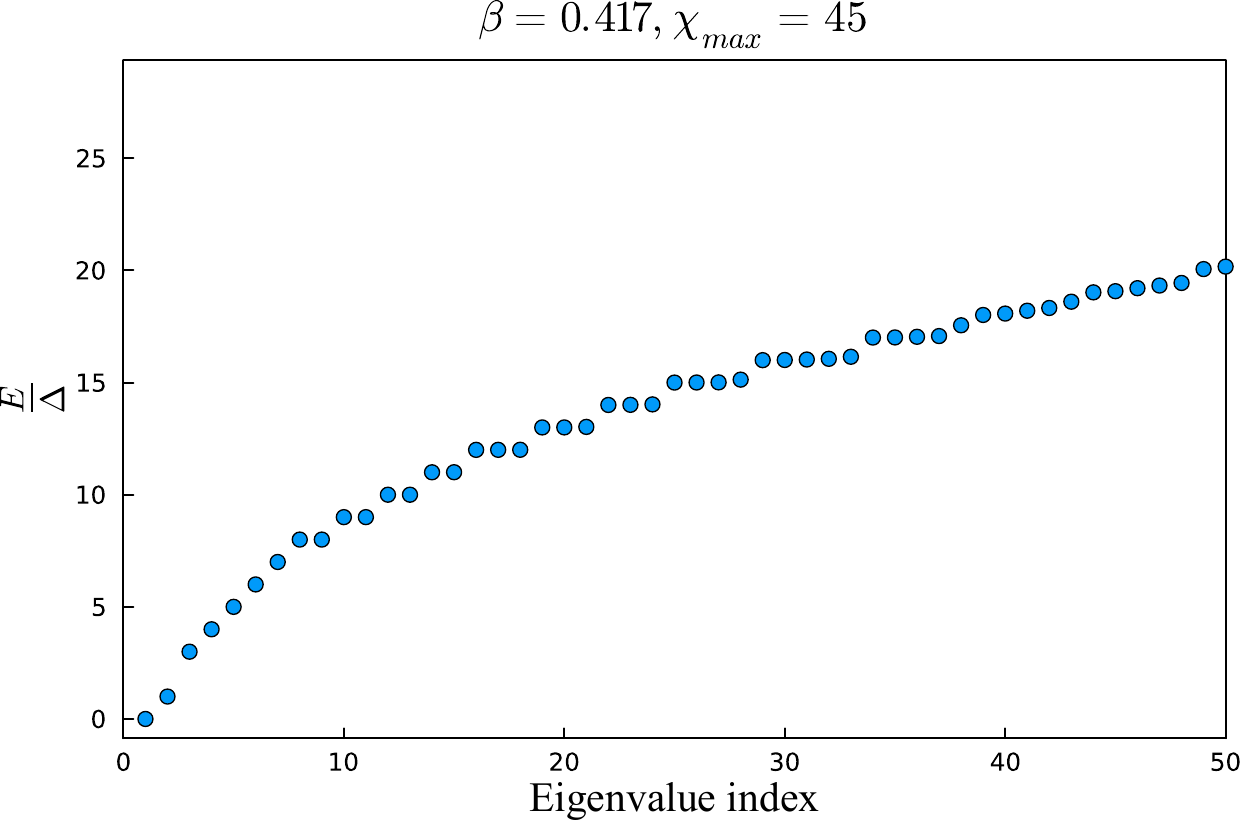} 
 \caption{\label{fig:Figure35}%
   Dice lattice: The $T'$-tensor spectrum in the ordered phase of the Ising model, normalized by the gap $\Delta$ between the first and second eigenvalues. The parameters are $\chi=45$ and $\beta = 0.417$. }
\end{figure}

We define the entanglement spectrum as logarithms of the eigenvalues of $T'^{\sigma}_{ij}$ for both $\sigma$. We find that this spectrum is integer-valued and contains the same degeneracies as the kagome-lattice corner spectrum. The differences are in the reverse role of the phases: the disordered phase on dice lattice corresponds to the ordered phase on kagome lattice and vice versa.  We also have additional degeneracy in the disordered phase of the dice lattice, since the spectra for $\sigma = \pm1$ are the same in this phase.  We show the eigenvalue pattern of the tensor $T'$ in the ordered phase in Fig.~\ref{fig:Figure35}. The degeneracy pattern is $d=\{1,1,0,1,1,1,1,1,2,2,2,2,3,3,3,4,5,\ldots\}$,  which is identical to the spectral degeneracies of kagome, triangular, square-octagon, and star lattices in the disordered phase. Note that here we combine the spectra of both $\sigma$. 
This is the reason for the number of depicted eigenvalues exceeding the chosen maximal $\chi$. 

\begin{figure}
\includegraphics[width= \linewidth]{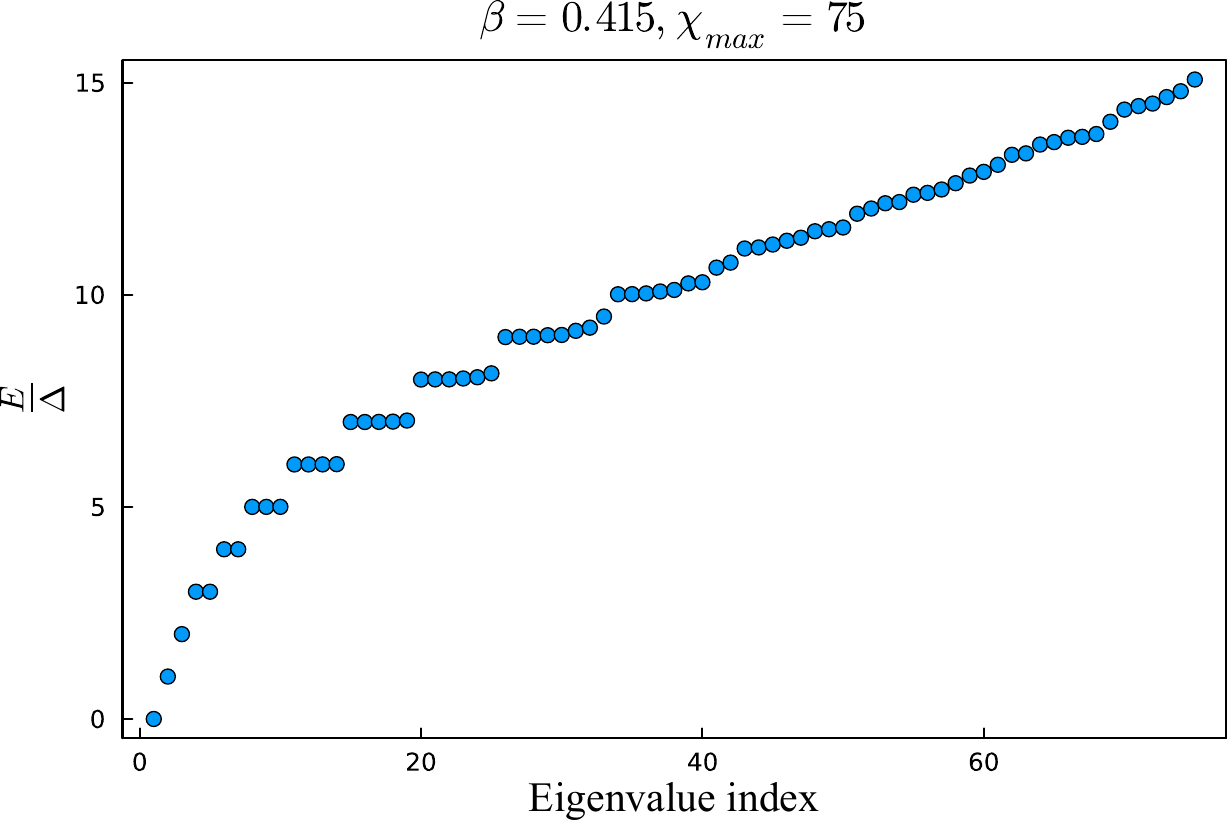} 
 \caption{\label{fig:Figure36}%
   Dice lattice: The $T'$-tensor spectrum in the disordered phase of the Ising model, normalized by the gap $\Delta$ between the first and second eigenvalues. The parameters are $\chi=75$ and $\beta = 0.415$. }
\end{figure}

The spectrum of the matrix $T'$ in the disordered phase of the dice-lattice Ising model is shown in Fig.~\ref{fig:Figure36} (with $\sigma = +1$, since $\sigma=-1$ results in the identical spectrum).  The degeneracy pattern is now $d=\{1,1,1,2,2,3,4,5,6,...\}$, which coincides with the degeneracy pattern of the triangular- and kagome-lattice Ising models in the ordered phase. 
%

\subsection{Honeycomb lattice}

Finally, let us discuss certain aspects of the honeycomb-lattice CTMRG.  The  CTMRG approach on the honeycomb lattice was introduced and successfully applied in two separate studies~\cite{Lukin2023ctmrg_honeycomb, nyckees2023critical}.  In this subsection, we aim to show that our extension of the honeycomb-lattice CTMRG to the two-site unit cell indeed converges, obeys certain consistency checks, and agrees with the scheme proposed in Ref.~\cite{nyckees2023critical}.  

\begin{figure}
\includegraphics[width= \linewidth]{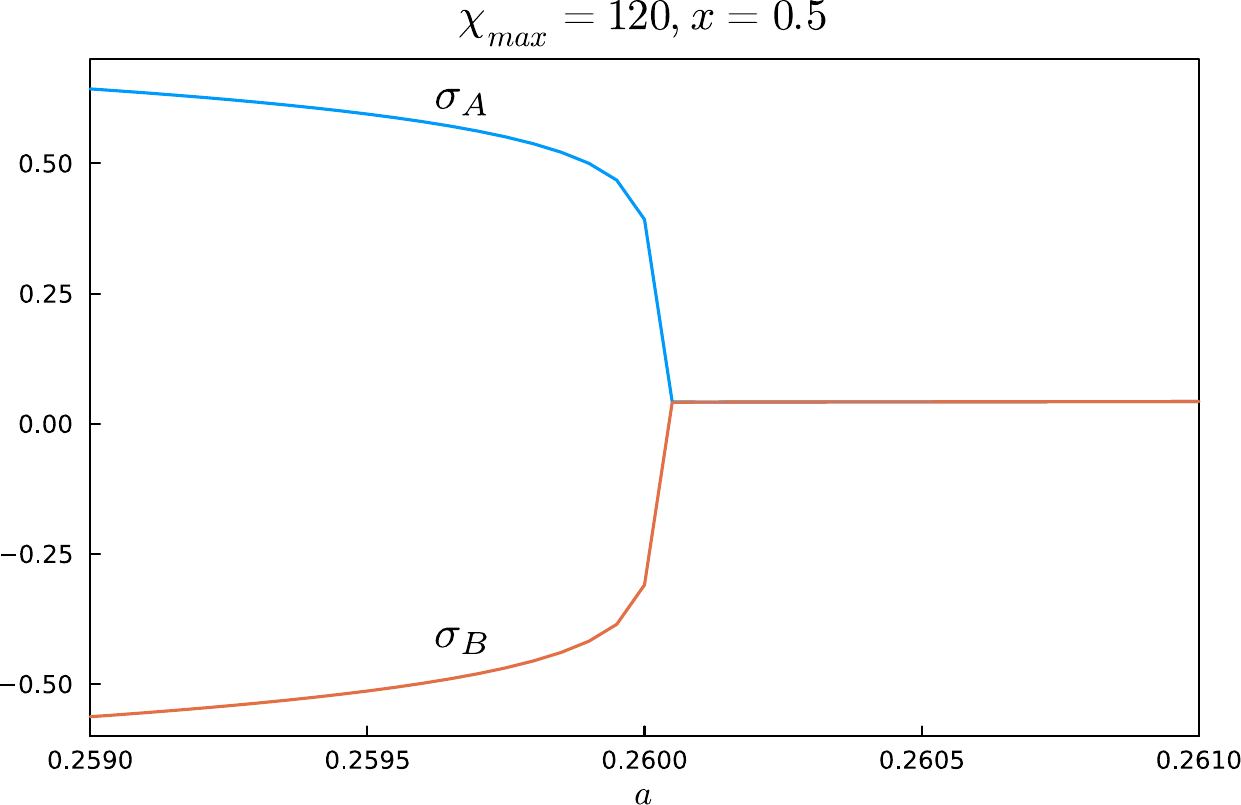} 
 \caption{\label{fig:Figure37}%
   Honeycomb lattice: Dependence of the local magnetizations $\sigma_A$ and $\sigma_B$ on the inverse temperature $\beta$ at $x=0.5$ and $\chi_{\max} = 120$. }
\end{figure}

To check the two-site algorithm convergence on the honeycomb lattice, we need a simple model with a two-site unit cell. A natural suggestion is the honeycomb-lattice antiferromagnetic Ising model in the external field. This model is not exactly solvable, but there are numerical results for the critical point~\cite{KIM2006245, wu1989critical}. As suggested above, we use the parameter $x = \exp(-2\beta B)$ to characterize the strength of the external field and the parameter $a = \exp(-2 \beta )$ to characterize the temperature. For the comparison, we choose $x = 0.5$, where the published numerical results suggest a critical point $a_c = 0.260013$~\cite{KIM2006245}. Our results in Fig.~\ref{fig:Figure37} are in good agreement and suggest the critical point $a_c = 0.26005(5)$. 

Let us comment on the convergence of the algorithm. 
In Sec.~\ref{subsec:Alg.honeycomb}, we suggested that it is sufficient to employ only one corner matrix~$C_{6}$ to obtain the projectors. In our numerical analysis, we confirm this suggestion, if the system is far from criticality.  In the vicinity of the phase transition, it is sometimes necessary to employ two or all three corner matrices to obtain the environments $Q_{L}$ and $Q_{R}$ for the corresponding projectors (these are defined analogously to the dice-lattice algorithm shown in Fig.~\ref{fig:Figure19}).

\begin{figure}
\includegraphics[width= \linewidth]{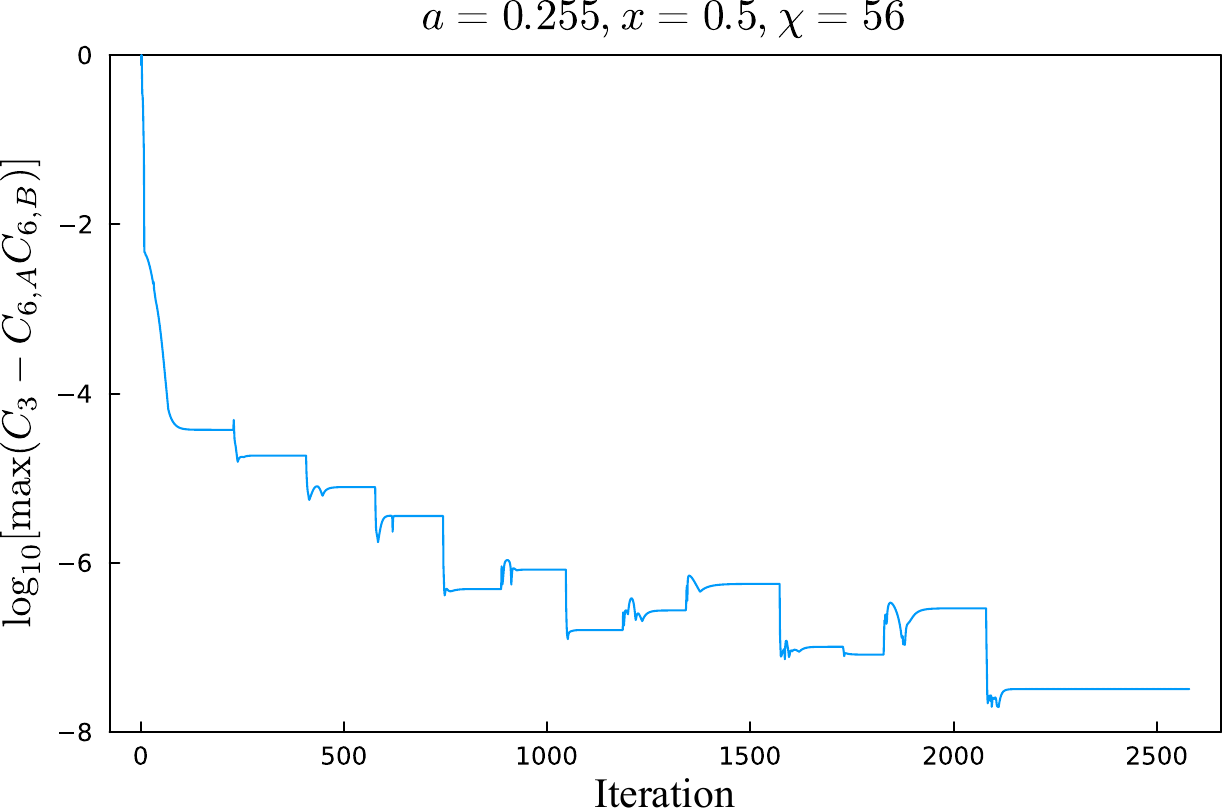} 
  \caption{\label{fig:Figure38}%
   Honeycomb lattice: Convergence of the maximal difference between $C_3=(R_AR_B)(C_{6,A}C_{6,B})$ and $C_{6,A}C_{6,B}$. Both products are preliminary normalized by their respective maximal elements. The parameters are $\chi = 56$ and $a = 0.255$.  }
\end{figure}

Next, we turn to the consistency checks. The first check concerns the matrix $C_{3}$, which was used in Ref.~\cite{nyckees2023critical}. As we suggested in Sec.~\ref{subsec:Alg.honeycomb} [see Figs.~\ref{fig:Figure8}(e) and \ref{fig:Figure8}(g)], the matrix~$C_{3}$ must be identical to $C_{6}^2$ or, in the case of the two-site unit cell, it must be equal to the corresponding matrix product $C_{6,A} C_{6,B}$.  For the latter, the equation in Fig.~\ref{fig:Figure8}(e) 
must be viewed as a consistency condition on the tensors $C_{6,A}$, $C_{6,B}$, $R_{A}$, and $R_{B}$ [note that in Fig.~\ref{fig:Figure8}(e) additional projectors and bulk tensors $A$ appear, but they can be grouped into just two tensors $R_{A}$ and $R_{B}$]. 
We show the convergence of this equation in Fig.~\ref{fig:Figure38}. It is important to note that the converged values of the order of $10^{-7}$ is the maximal difference, which usually occurs for the last eigenvalues, while the differences in the first matrix elements are identical to the machine precision. We conclude that the equation in Fig.~\ref{fig:Figure8}(e) holds for our converged environments. Hence, the corner environments of Ref.~\cite{nyckees2023critical} can be obtained from ours, and the algorithms are equivalent up to the choice of projectors. 

\begin{figure}
\includegraphics[width= \linewidth]{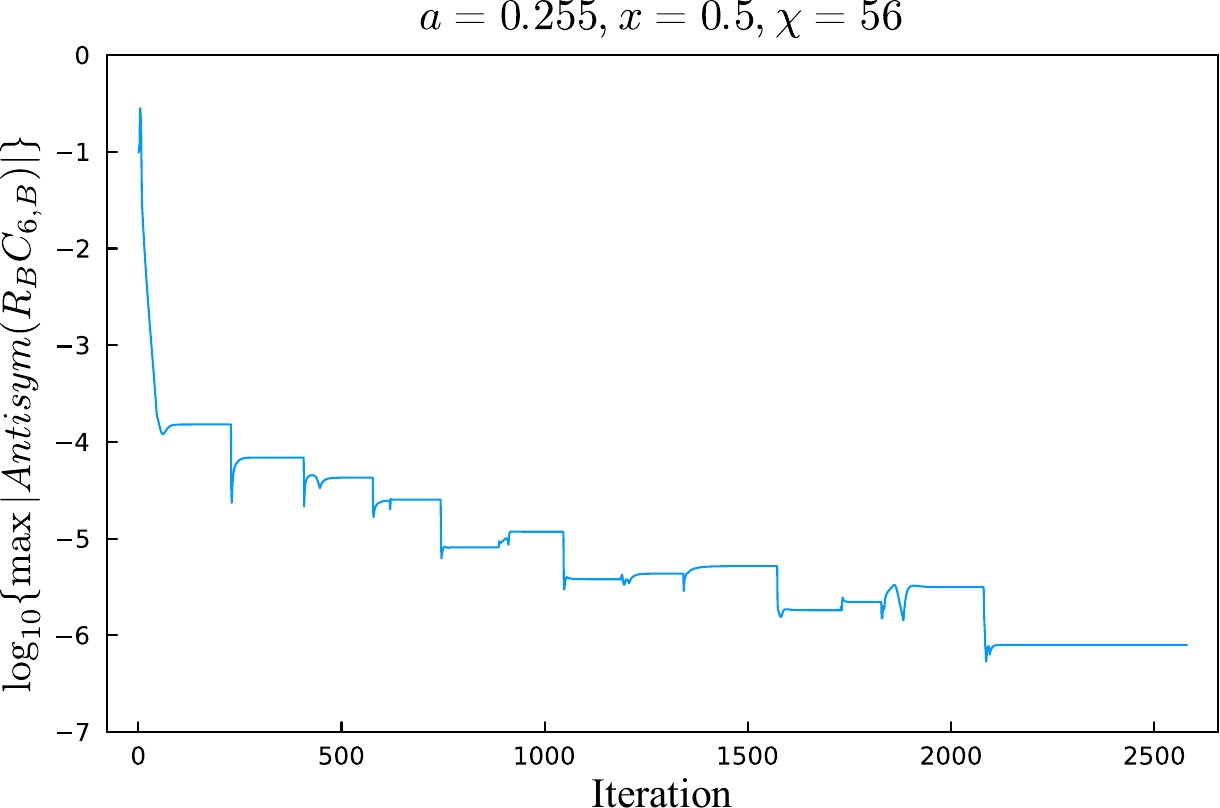} 
  \caption{\label{fig:Figure39}%
   Honeycomb lattice: Convergence of the maximal absolute value of the antisymmetrization of the tensor  $R_{B} C_{6,B}$  (both $R_{B}$ and $C_{6,B}$ are normalized by their respective norms). The parameters are $\chi = 56$ and $a = 0.255$.  }
\end{figure}

We can now discuss the second consistency check. The rank-3 tensor $R_{B}  C_{6,B}$ must be symmetric in its first and last auxiliary indices.  We show the evolution of the maximal absolute value of the antisymmetrization of $R_{B} C_{6,B}$ in Fig.~\ref{fig:Figure39}. As in the previous consistency check, the maximal difference corresponds to the smallest eigenvalues of the corner matrices, while the first tensor elements after antisymmetrization vanish up to machine precision.
We also probed the same consistency test with several types of directional update method to obtain the projectors and usually observed much poorer convergence of the consistency checks. 

It should be mentioned that we also briefly studied the ferromagnetic Ising model on the honeycomb lattice without an external field. In this case, we used the simplest isotropic and homogeneous CTMRG ansatz, with projectors determined either from the corner matrices $C_{6}$ or $C_{3}$. We observed that the converged results are generally independent of the scheme and that $C_{3}$ computed from the converged corner matrix $C_{6}$ is the same as the converged matrix $C_{3}$ from the $C_{3}$-based CTMRG up to machine precision.

\begin{figure*}
\includegraphics[width=\textwidth]{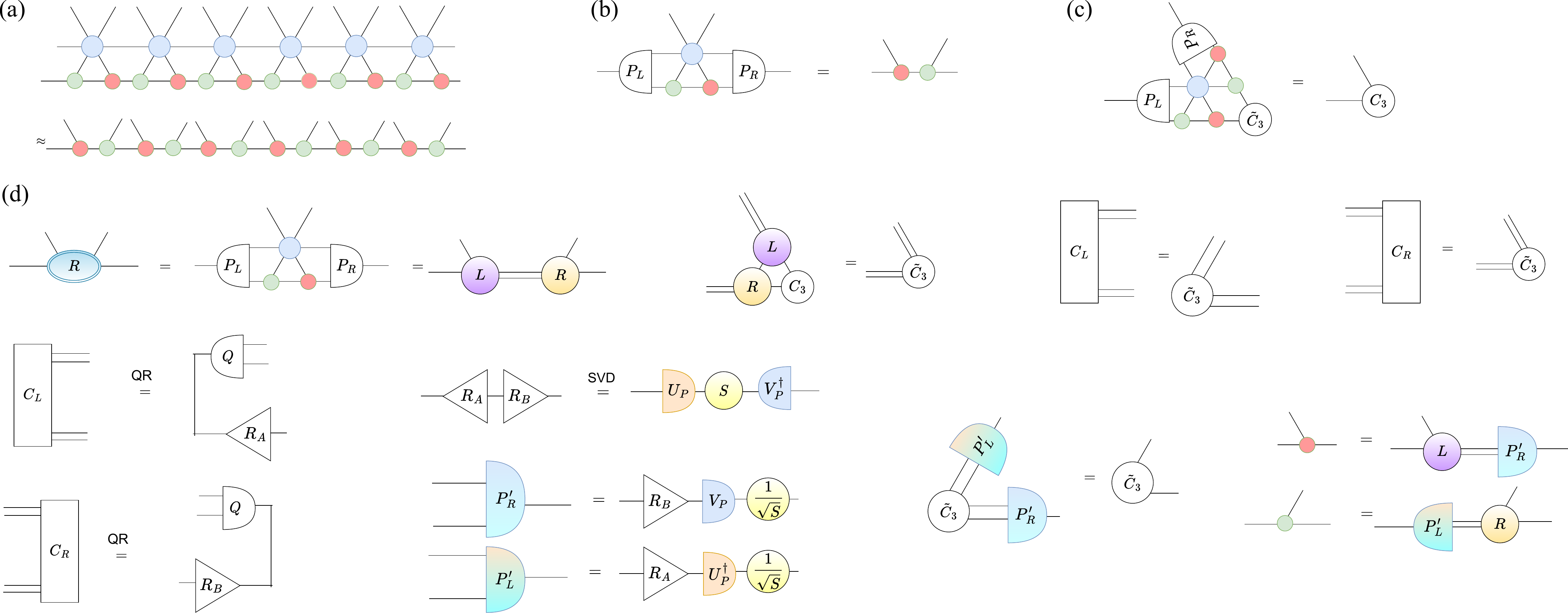} 
  \caption{\label{fig:Figure40}%
   Triangular lattice: (a) Transfer matrix for the nonpositive bulk tensor and the corresponding bMPS, which consists of two different boundary tensors. (b) The update rule for the boundary tensors, which mimics the steps shown in Fig.~\ref{fig:Figure1}. The difference is that the factorization is nonsymmetric and projectors are not isometric. (c) The modified update step for the corner matrix~$C_{3}$ (which is also no longer symmetric). The projectors $P_{L}$ and $P_{R}$ can be obtained, e.g., from the biorthogonalization of the environments with enlarged $C_{3}$. (d) The nonsymmetric factorization step. First, as in the positive case, we define the tensor~$R$. Then, we perform an arbitrary factorization of this tensor into two rank-3 tensors $L$ and $R$. We do not perform truncation in this step, which is illustrated with the double-line links for the enlarged not-truncated factorization  index.  The tensors $L$ and $R$ are then absorbed into the corner matrix. The new corner matrix has an enlarged bond dimension, which is truncated back with the projectors $P'_{L}$ and $ P'_{R}$, which we also obtain from biorthogonalization. The same projectors can be also used to truncate the factorization index in $L$ and $R$ back to the bMPS bond dimension.
   }
\end{figure*}

\section{Conclusions and Outlook}\label{sec:Conclusions}
In this study, we developed the CTMRG algorithms on triangular, kagome, square-octagon, ruby, SHD, star, and dice lattices. 
The procedure to construct the CTMRG loop is rather general and can be summarized as follows: 
(i) define all unique bMPS on the lattice and find how the individual tensors of the bMPS are updated during the absorption of the bulk tensors into the bMPS; 
(ii) define the corner matrices as intersections of different bMPS; 
(iii) find the updates of corners from the updates of bMPS local tensors; 
(iv) employ the corner tensors to define environments, which allow finding optimal truncations for the local bMPS tensors.  
All together, this forms the self-consistent scheme, which allows us to find the CTM environments on many different lattices. 

We conclude that the procedure is general and can be extended to the lattices not covered in this study, including the Shastry-Sutherland, maple-leaf, trellis, square-kagome lattice, etc.  It can also be extended to anisotropic tensor networks or the ones with the enlarged unit cells.  Another potential direction is to dualize these algorithms to define CTMRG for the interaction-round-a-face models on the dual lattice. The interaction-round-a-face models can also be studied with the CTMRG methods \cite{CTM_intro_2} (see, e.g., Refs.~\cite{hyperbolic_2, hyperbolic_3, hyperbolic_4, hyperbolic_6, hyperbolic_7, hyperbolic_8} for the corresponding discussion on triangular and hyperbolic lattices).

Other relevant research directions include the completely variational formulation of the proposed algorithms~\cite{liu2022variational},  its comparison with VUMPS~\cite{VUMPS_CTMRG}, and the development of the modified convergence schemes in line with Ref.~\cite{fishman2018faster}. 
Potential practical applications encompass the variational optimization of the iPEPS wave function with the proposed CTMRG approaches as a method to compute correlation functions~\cite{liao2019differentiable, Lukin2023ctmrg_honeycomb}, studies of the frustrated lattice models~\cite{nyckees2023critical, song2023unified}, as well as quantum thermal systems~\cite{vanhecke2023quantum_thermal}. 
Currently, we are also investigating possible consistency conditions between the CTM environment tensors and projectors (to some extent, in line with Ref.~\cite{fishman2018faster}) with the aim of employment of additional structure appearing in projectors for the computation of arbitrary two- and three-point long-range correlation functions.

\acknowledgements
The authors thank Mari Carmen Ba\~nuls and Natalia Chepiga for helpful comments during the manuscript preparation.
The authors acknowledge support by the National Research Foundation of Ukraine under the call ``Excellent science in Ukraine'' (2024--2026)
and support by the Office of Naval Research Global (ONRG), Grant \#~N62909-23-1-2088 (program manager Dr.~Martina Barnas).

\appendix

\section{The CTMRG on the triangular lattice with nonpositive bulk tensors }\label{app:A}

By taking an example of triangular lattice, let us discuss how to generalize the algorithms to the nonpositive bulk tensors, which inevitably appear, e.g., in the iPEPS calculations. First, according to Fig.~\ref{fig:Figure1}(c), we see that the positivity of the bulk tensor guarantees the existence of symmetric factorization of the tensor in the left part of the equation in Fig.~\ref{fig:Figure1}(c).  If the bulk tensor is not positive, then we conclude that the factorization can be nonsymmetric. This leads us to introduce the more general nonsymmetric bMPS ansatz, which is shown in Fig.~\ref{fig:Figure40}(a).

This new ansatz consists of two different boundary tensors. The update rule for these tensors is illustrated in Fig.~\ref{fig:Figure40}(b). This is a trivial generalization of the update rule from Fig.~\ref{fig:Figure1}(c), though we employ the nonsymmetric factorization instead of symmetric one, and we also replace the isometric projector $P$ with a more general biorthogonal projectors $P_{L}$ and $ P_{R}$. The corner matrix $\tilde{C}_{3}$ is defined in the same way as in Fig.~\ref{fig:Figure2}(a), but the update also includes biorthogonal projectors, as we show in Fig.~\ref{fig:Figure40}(c). Note that the new corner matrix is no longer symmetric or diagonal. Still, we can find the projectors from it by applying biorthogonalization of the environments built from this enlarged matrix $C_{3}$ (or, alternatively, with some form of the directional update).  

Next, let us discuss the factorization step. We define the rank-4 tensor $R$ to be factorized, as shown in Fig.~\ref{fig:Figure40}(d).  We factorize it into two rank-3 tensors $L$ and $R$, where the factorization is performed without truncation. We employ SVD for this factorization and absorb the tensors $L$ and $R$ into $C_{3}$ to obtain the matrix $\tilde{C}_{3}$, but with the enlarged bond dimensions. To truncate it back to the original bond dimension, we find the new biorthogonal projectors $P'_{L}$ and $P'_{R}$. These projectors can be obtained again from the biorthogonalization, but this time of the environments built from the matrix $\tilde{C}_{3}$. The projectors are applied to truncate the matrix $\tilde{C}_{3}$ and also to obtain the new bMPS tensors from $L$ and $R$.

We tested the method on random symmetric (but nonpositive) bulk tensors, finding the quickly convergent CTMRG environments, which also obey the simplest consistency checks on various tensors of CTM.  To define these consistency checks, we used combinations of the matrices $C_{3}$, on-site tensors $O$, and corner tensor $T$, which can also be general to the nonpositive case. 

Note that the method can be extended to triangular lattices with partly or fully broken rotational or reflection symmetries. It is also interesting that all the steps in this nonpositive algorithm have already appeared in various CTMRG algorithms in the main text.

\change{
\section{Additional tests on the $q$-Potts model}\label{app:B}
To confirm a high potential and universality of the developed approach, we also performed a few additional tests of the method on the models with the larger bond dimension~$D$ of the  tensor network. For this purpose, we take the $q$-Potts model~\cite{Potts_1952, Wu_Potts}, which has the following Hamiltonian:
\begin{equation}
    H = - \sum_{\langle ij \rangle_{r}} J_{r} \delta_{s_{i} s_{j}},
\end{equation}
where the sum is taken over nearest neighbours $\langle ij \rangle_{r}$ with the direction $r$ and $s_{i}$ takes values in a set of $q$-roots of the identity $\exp{[i2\pi k/q]}$. $J_{r}$ are coefficients for different directions of bonds $r=\{1,2,3\}$. Hence, the model can be anisotropic. 

The tensor network for this model is constructed in a complete analogy with the simplest Ising model (which corresponds to $q=2$) and has the bond dimension $D = q$. On the triangular lattice, the transition point is exactly known for $q > 3$ and is defined as a root of the equation~\cite{Hintermann1978, Wu_Potts} 
\begin{equation}
    x_{1} x_{2} x_{3} + x_{1} x_{2} + x_{2} x_{3} + x_{1} x_{3}  - q = 0,
\end{equation}
where $x_{r} = \exp{[J_{r}\beta]} - 1$. 

\begin{figure}[t]
\includegraphics[width= \linewidth]{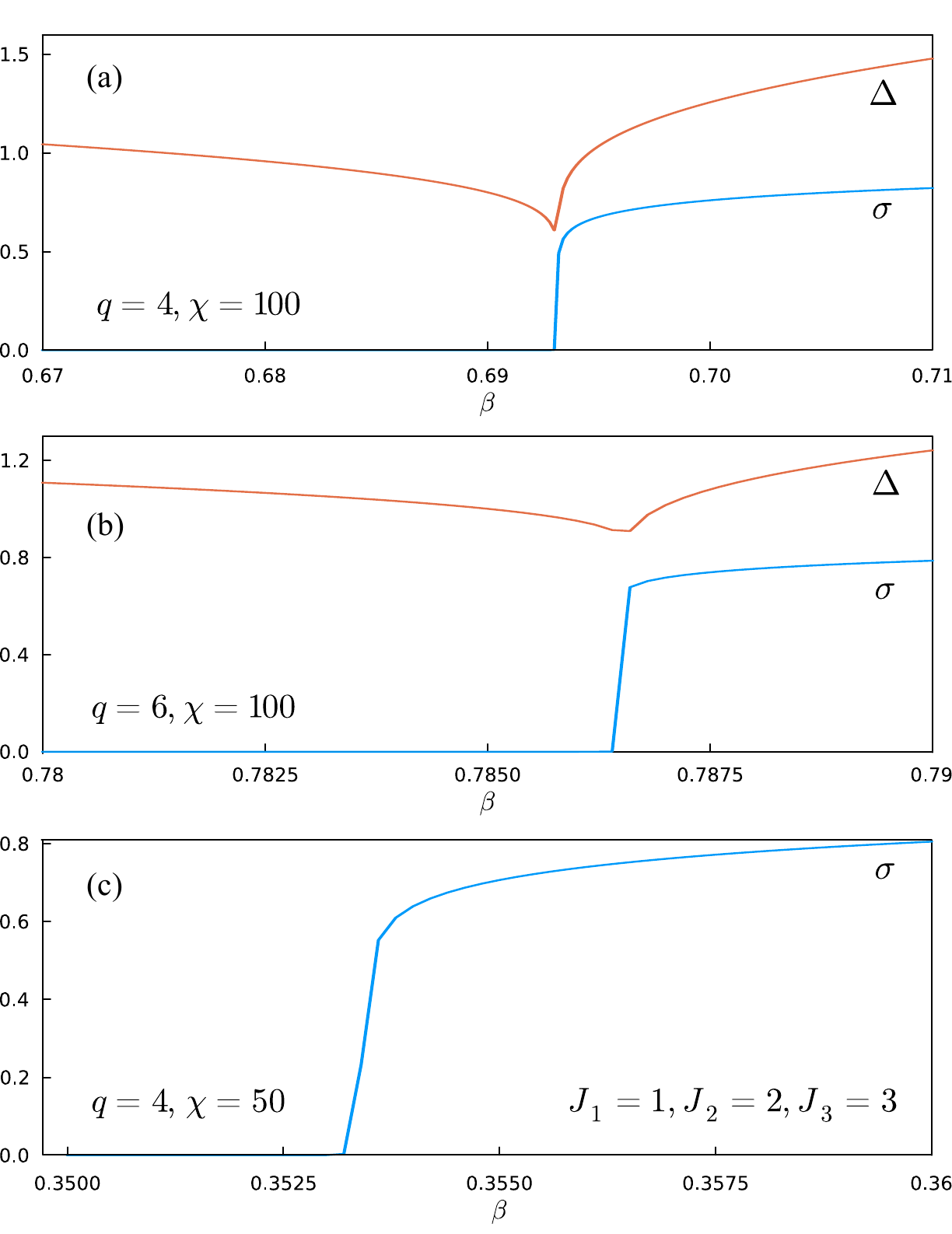} 
 \caption{\label{fig:47}%
   Dependence of the magnetization $\sigma$ and entanglement gap $\Delta$ on the inverse temperature $\beta$ for the $q$-Potts model on triangular lattice. The parameters and exact values of the inverse critical temperature are: (a) isotropic case with $J_{1} = J_{2} = J_{3} = 1, q=4$, $\chi = 100$, and $\beta_{c} \approx 0.6931$; (b) isotropic case with $J_{1} = J_{2} = J_{3} = 1, q=6$,  $\chi = 100$, and $\beta_{c} \approx 0.7866$ (note that the transition is of the first order for $q>4$); 
   (c) anisotropic case with $q=4$, $\chi = 50$, $J_1=1$, $J_2=2$,  $J_3=3$, and $\beta_{c} \approx 0.3535$.
   }
\end{figure}
We perform calculations with $q = 4$ and $q=6$ in the vicinity of the phase transition and show the results in Fig.~\ref{fig:47}  (the inverse temperature~$\beta$ on the $x$-axis is shifted according to the theoretical value of the transition point $\beta_c$). The method allows us to estimate the transition point with the accuracy within the mesh size. The results are shown for both isotropic and anisotropic cases. Note that for the simulation of the anisotropic case, we employ the modification of the algorithm above with three different bMPS and corner matrices. All other details of the update are completely analogous.}


\bibliography{CTMRG}
\end{document}